\documentclass[12pt]{article}
\usepackage{amsmath, amssymb, amsfonts}
\usepackage[dvips]{graphicx}
\def\nn{\nonumber}
\def\dfrac{\displaystyle\frac}
\def\numt#1#2{#1 \times 10^{#2}}
\def\etal{{\it et al.}}
\def\etc{{\it etc.~}}
\def\eg{{\it e.g.,~}}
\def\PR#1#2#3{Phys. Rev. {\bf #1}, #2 (#3)}
\def\PRL#1#2#3{Phys. Rev. Lett. {\bf #1}, #2 (#3)}
\def\PL#1#2#3{Phys. Lett. {\bf #1}, #2 (#3)}
\def\NP#1#2#3{Nucl. Phys. {\bf #1}, #2 (#3)}
\def\PTP#1#2#3{Prog. Theor. Phys. {\bf #1}, #2 (#3)}
\def\EPJ#1#2#3{Eur. Phys. J. {\bf #1}, #2 (#3)}

\def\PTP#1#2#3{Prog. Theor. Phys. {\bf #1}, #2 (#3)}
\def\EPJ#1#2#3{Eur. Phys. J. {\bf #1}, #2 (#3)}

\def\JHEP#1#2#3{JHEP{\bf #1}, #2 (#3)}
\def\l{\left}
\def\r{\right}
\def\gsim{~{\rlap{\lower 3.5pt\hbox{$\mathchar\sim$}}\raise 1pt\hbox{$>$}}\,}
\def\lsim{~{\rlap{\lower 3.5pt\hbox{$\mathchar\sim$}}\raise 1pt\hbox{$<$}}\,}
\def\cerenkov{$\check{\rm C}$erenkov~}

\def\sss{\scriptscriptstyle}
\def\atm{\sss{\rm ATM}}
\def\rct{\sss{\rm RCT}}
\def\sun{\sss{\rm SOL}}
\def\dmns{\delta_{\sss{\rm MNS}}}

\def\dm#1#2{\delta m^2_{#1#2}}
\def\ssun#1{\sin^2 #1\theta_{\sss{\rm SOL}}}
\def\satm#1{\sin^2 #1\theta_{\sss{\rm ATM}}}
\def\srct#1{\sin^2 #1\theta_{\sss{\rm RCT}}}
\def\csun#1{\cos^2 #1\theta_{\sss{\rm SOL}}}
\def\catm#1{\cos^2 #1\theta_{\sss{\rm ATM}}}

\def\Erec{E_{\rm rec}}
\def\effe{\varepsilon_e}
\def\effm{\varepsilon_\mu}

\begin{document}
% ------------------------------------------------------------------------ %
\title{Re-evaluation of the T2KK physics potential\\
with simulations including backgrounds}
\author{{Kaoru Hagiwara$^{1,2}$} and 
{Naotoshi Okamura$^{1}$}\thanks{E-mail address : naotoshi@post.kek.jp} \\
{\small \it $^{1}$KEK Theory Division, Tsukuba, 305-0801, Japan}\\
{\small \it $^{2}$Sokendai, Tsukuba, 305-0801, Japan}}
\date{January 12, 2009}
\maketitle
% ------------------------------------------------------------------------ %
\vspace{-9.5cm}
\begin{flushright}
\hspace*{3ex}
KEK-TH-1268
\end{flushright}
\vspace{9.5cm}
\vspace{-2.0cm}
% ======================================================================== %
% abstract
% ------------------------------------------------------------------------ %
\begin{abstract}
The Tokai-to-Kamioka-and-Korea (T2KK)
neutrino oscillation experiment under examination can have
a high sensitivity to determine the neutrino mass hierarchy 
for a combination of relatively
large ($\sim 3.0^\circ$) off-axis angle beam
at Super-Kamiokande (SK) 
and 
small ($\sim 0.5^\circ$) off-axis angle 
at $L \sim 1,000$~km in Korea.
We elaborate previous studies by taking into account smearing of
reconstructed neutrino energy due to finite resolution of electron or
muon energies, nuclear Fermi motion and resonance production,
as well as the neutral current $\pi^0$ production background to the 
$\nu_\mu \to \nu_e$ oscillation signal.
It is found that the mass hierarchy pattern can still be determined
at $3\sigma$ level 
if $\srct{2}\equiv4|U_{e3}|^2(1-|U_{e3}|^2)\gsim0.08~(0.09)$ 
when the hierarchy is normal (inverted) 
with $\numt{5}{21}$ POT exposure,
or 5 years of the T2K experiment,
if a 100~kton water \cerenkov detector is placed in Korea.
The $\pi^0$ backgrounds deteriorate the capability
of the mass hierarchy determination,
whereas the events from nuclear resonance productions contribute
positively to the hierarchy discrimination power.
We also find that the $\pi^0$ backgrounds seriously affect
the CP phase measurement.
Although $\dmns$ can still be constrained
with an accuracy of $\sim\pm45^\circ$ ($\pm60^\circ$)
at $1\sigma$ level for the normal (inverted) hierarchy
with the above exposure if $\srct{2}\gsim0.04$, 
CP violation can no longer be established at
$3\sigma$ level even for 
$\dmns=\pm 90^\circ$ and $\srct{2}=0.1$.
About four times higher exposure will be needed to measure $\dmns$
with $\pm30^\circ$ accuracy.
\end{abstract}
\newpage
\renewcommand{\thefootnote}{\fnsymbol{footnote}}
\setcounter{footnote}{0}
% ======================================================================== %
\section{Introduction}
\label{sec:1}
%%------------------------
 The SNO experiment found that the $\nu_e$ from the sun changes
into the other active neutrinos~\cite{SNO}.
 The atmospheric neutrino observation at SK reported that $\nu_\mu$
and $\bar{\nu}_\mu$ oscillate into
the other active neutrinos~\cite{SKatm}.
 Recently, the MiniBooNE experiment~\cite{miniboone}
reported that the LSND~\cite{LSND}
observation of rapid $\bar{\nu}_\mu \to \bar{\nu}_e$ oscillation
has not been confirmed.
Consequently, the three active neutrinos are sufficient to describe
all the observed neutrino oscillation phenomena.

%%-----------------------
Under the three generation framework,
neutrino flavor oscillation~\cite{MNS,Ponte}
is governed by 2 mass-squared differences
and 4 independent parameters in the MNS
(Maki-Nakagawa-Sakata)
matrix~\cite{MNS}, that is 3 mixing angles and 1 CP phase ($\dmns{}$).
The absolute value of the larger mass-squared difference,
$|\dm13|$
and 
one combination of the MNS matrix elements
$\satm{2}\equiv 4|U_{\mu3}|^2(1-|U_{\mu3}|^2)$,
are determined by the atmospheric
neutrino observation~\cite{SK98,SK,SKatm,atm},
which have been confirmed by the accelerator based long baseline
neutrino oscillation experiments
K2K~\cite{K2K} and MINOS~\cite{minos}.
However, the sign of $\dm13 \equiv m^2_3-m^2_1$ has not been determined.
Both the magnitude and the sign of the smaller mass-squared
difference $\dm12=m^2_2-m^2_1$,
another combination of the MNS matrix elements
$\ssun{2}\equiv 4|U_{e1}U_{e2}|^2$
are determined by the solar neutrino observations~\cite{sun,SNO}
and the KamLAND experiment~\cite{KamLAND}.
The last independent mixing angle ($\theta_{\rct}$)
has not been measured yet, and
the reactor experiments~\cite{CHOOZ} give upper bound on
the combination $\srct{2}\equiv4|U_{e3}|^2(1-|U_{e3}|^2)$.
The leptonic CP phase,
$\dmns{}=-\mbox{arg}U_{e3}$~\cite{PDB},
is unknown.

%%-----------------------
There are many experiments which 
plan to measure the unknown parameters of the three neutrino model.
In the coming reactor experiments, 
Double~CHOOZ~\cite{DoubleCHOOZ},
Daya~Bay~\cite{DayaBay},
and
RENO~\cite{reno}
plan to measure the unknown element $|U_{e3}|$
from the $\bar{\nu}_e$ survival probability.
The Tokai-to-Kamioka (T2K) neutrino oscillation experiment~\cite{T2K},
which is one of the next generation accelerator based long baseline
experiments, also plans to measure $|U_{e3}|$
by observing the $\nu_\mu \to \nu_e$ transition event,
whose rate is proportional to $|U_{e3}U_{\mu3}|^2$.

%%-----------------------
 However, the sign of $\dm13$,
or the mass hierarchy pattern,
will remain undetermined even after these experiments.
It is not only one of the most important parameters 
in particle physics
but also has serious implications in astronomy and cosmology.
For instance, 
if $\dm13$ is negative (inverted hierarchy),
the prospects of observing the neutrino-less double beta decay
are good, 
while the matrix element $|U_{e2}|$ is affected by quantum corrections
such that its high energy scale value 
depends on the Majorana phases~\cite{RGE}
in the large $\tan\beta$ supersymmetric See-Saw scenario~\cite{seesaw}.
In astronomy, the mass hierarchy pattern affects
the light elements synthesis in the supernova
through neutrino-nucleon interactions;
the yields of $^7$Li and $^{11}$Be increase 
for the normal hierarchy ($\dm13>0$) 
if $\srct{2}\gsim 10^{-3}$~\cite{SN-make}.
In cosmology,
the dark matter content of the universe
depends on the mass hierarchy.

%%-----------------------
 In the previous studies~\cite{HOS1,HOS2,HO-atm},
we explored the physics impacts of the idea~\cite{HOSetc}
of placing an additional far detector in Korea
along the T2K neutrino beam line,
which is now called as the T2KK (Tokai-to-Kamioka-and-Korea)
experiment.
 In particular, we studied semi-quantitatively
the physics impacts of placing a 100~kton water
\cerenkov detector in Korea,
about $1000$~km away from J-PARC
(Japan Proton Accelerator Research Complex)~\cite{j-parc},
during the T2K experiment period~\cite{T2K},
which plans to accumulate 
$5\times10^{21}$ POT (protons on target) in 5 years.
We find that the neutrino-mass hierarchy
pattern can be determined by comparing the $\nu_{\mu} \to \nu_e$
transition probability measured at SK ($L=295$~km)
and that at a far detector in Korea~\cite{HOS1},
if $\srct{2}\gsim0.05$ for $3\sigma$.
The CP phase can also be measured 
if $\srct{2}\gsim 0.02$ with $\pm30^\circ$ accuracy,
since the amplitude and the oscillation phase of the
$\nu_{\mu}^{} \to \nu_e^{}$ transition probability
are sensitive to $\sin \dmns$ and $\cos \dmns$,
respectively~\cite{HOS1,HOS2}.
We also find that the octant degeneracy 
between $\satm{}=0.4$ and $0.6$ for $\satm{2}=0.96$
can be resolved if $\srct{2}\gsim 0.12$~\cite{HO-atm}.
In the above studies~\cite{HOS1,HOS2,HO-atm},
a combination of $3.0^\circ$ OAB (off-axis beam) at SK
and $0.5^\circ$ OAB at $L=1000$~km in Korea
is found to be most efficient, mainly because of the
hard neutrino spectrum of the $0.5^\circ$ OAB.
In alternative studies~\cite{Kajita} of the T2KK setup,
an idea of placing two identical detectors 
at the same off-axis angle 
in Kamioka and Korea has been examined.
The idea of placing far and very far detectors along one neutrino
baseline has also been studied for the Fermi Lab. 
neutrino beam~\cite{supernova}.

%------
The T2KK experiment has a potential of becoming the most economical
experiment to determine the mass hierarchy and the CP phase,
if $\srct{2}$ is not too small.
In this paper, we re-evaluate the T2KK physics potential
by taking into account 
smearing of the reconstructed neutrino energy
due to finite resolution of electron or muon energies
and the Fermi motion of the target nucleon,
as well as those events from the nuclear resonance production
which cannot be distinguished from the quasi-elastic events by
water \cerenkov detectors.
We also study contribution from the neutral current $\pi^0$
production processes which can mimic the $\nu_e$ appearance signal.

% ----- show article this organization -----
 This article is organized as follows.
 In section~\ref{sec:2}, 
we fix our notation and give approximate analytic expressions for
the neutrino oscillation probabilities including the matter effect.
The relations between the experimental observables and 
the three neutrino model parameters are then explained by using the
analytic formulas.
 In section~\ref{sec:3},
we show how we estimate the event numbers from
the charged current (CC) and neutral current (NC) interactions
by using the event generator {\sf nuance}~\cite{nuance}.
 In section~\ref{sec:4},
we present the $\chi^2$ function which we adopt in estimating
the statistical sensitivity of the T2KK experiment
on the neutrino oscillation parameters.
 In section~\ref{sec:5},
we show our results 
on the mass hierarchy determination.
 In section~\ref{sec:6},
we show our results on the CP phase measurement.
 In section~\ref{sec:7},
we give the summary and conclusion.
In Appendix~\ref{sec:appA},
 we present
 a parameterization of the reconstructed neutrino energy distribution
as a function of the initial neutrino energy for CCQE and resonance
events.

% ======================================================================== %
\section{Notation and approximate formulas}
\label{sec:2}

In this section, 
we fix our notation and present an
analytic approximation for the neutrino oscillation
probabilities that is useful for understanding the physics potential
of the T2KK experiment qualitatively.

\subsection{Notation}
\label{sec:notation}

The neutrino flavor eigenstate 
$\left|\nu_{\alpha} \right\rangle$ ($\alpha=e,\mu,\tau$) 
is a mixture of the mass eigenstates 
$\left|\nu_{i} \right\rangle$
($i=1,2,3$) with the mass $m_i$ as
\begin{equation}
 \left|\nu_{\alpha} \right\rangle=
 \sum^{3}_{i=1} U_{\alpha i}
 \left|\nu_{i} \right\rangle\,,
\end{equation}
where $U$ is the 
unitary MNS (Maki-Nakagawa-Sakata)~\cite{MNS} matrix.
We adopt a convention where $U_{e1}$, $U_{e2}$, 
$U_{\mu3}$, $U_{\tau 3}\geq 0$
and $\dmns \equiv - \arg U_{e3}$~\cite{PDB,HO1}.
The 4 parameters, $U_{e2}$, $U_{\mu3}$, $\left|U_{e3}\right|$,
and $\dmns$, can then be chosen as the independent parameters 
of the $3\times3$ MNS matrix.
All the other elements are determined uniquely by
the unitarity conditions~\cite{HO1}.

The atmospheric neutrino observation~\cite{SK98,SK,SKatm,atm} and 
the accelerator based long baseline experiments~\cite{K2K,minos}, 
which measure the $\nu_\mu^{}$ survival probability,
are sensitive to the magnitude of the larger mass-squared
difference and 
$\satm{2}$~\cite{minos}:
\begin{subequations}
\begin{eqnarray}
\satm{2} 
&>& 0.90 ~~~{{\mbox{{($90\%$ C.L.)}}}}\,,
\label{eq:exp-satm}\\
 |\dm13| &=&\numt{(2.43\pm0.13)}{-3} {\mbox{{eV}}}^2 \,.
\label{eq:exp-dm13}
\end{eqnarray}
\label{eq:exp-atm}
\end{subequations}

\vspace*{-3ex}
The reactor experiments,
which observe the survival probability of $\bar{\nu}_e^{}$
at $L \sim O(1)$~km, are sensitive to $|\dm13|$ and $\srct{2}$.
The CHOOZ experiment~\cite{CHOOZ}
finds
\begin{subequations}
 \begin{eqnarray}
 \srct{2}
&<& 
(0.20,~0.16,~0.14)\label{eq:exp-srct1}\\
\mbox{{for~~~}}
\l|\dm13\r|
&=&\numt{(2.0,~2.5,~3.0)}{-3}\mbox{{eV}}^2\,,
\label{eq:exp-dmrct}
\end{eqnarray}
\label{eq:exp-rct}
\end{subequations}
$\!\!\!\!$
at the 90\% confidence level.

 The solar neutrino observations~\cite{sun}, 
and
 the KamLAND experiment~\cite{KamLAND},
which measure the survival probability of 
$\nu_e^{}$ and $\bar{\nu}_e^{}$, respectively,
at much longer distances
are sensitive to the smaller mass-squared difference, $\dm12$,
and $U_{e2}$.
The combined results~\cite{KamLAND} find
\begin{subequations}
\begin{eqnarray}
\ssun{2} &=& 0.87 \pm 0.04\,,
\label{eq:exp-ssun}\\
\dm12
&=&  (7.59 \pm 0.21) \times 10^{-5} {\mbox{{eV}}}^2 \,.
\label{eq:exp-dm12}
\end{eqnarray}
\label{eq:exp-sun}
\end{subequations}
$\!\!\!$The sign of $\dm12$ has been determined by the matter effect 
inside the sun~\cite{msw}.

With a good approximation~\cite{3G},
we can relate the above three mixing factors, 
eqs.~(\ref{eq:exp-satm}), (\ref{eq:exp-srct1}), (\ref{eq:exp-ssun})
with the elements of the $3\times3$ MNS matrix;
\begin{subequations} 
\begin{eqnarray}
\sin \theta_{\atm} &=& 
 U_{\mu3} = 
 \sin \theta_{23} \cos \theta_{13}\,,
 \label{eq:theta-atm}\\
\sin \theta_{\rct} &=& 
 |U_{e3}| =
 \sin \theta_{13}\,,
 \label{eq:theta-rct}\\
\sin 2\theta_{\sun} &=&
 2U_{e1}U_{e2}=
 \sin 2\theta_{12} \cos^2 \theta_{13}\,,
 \label{eq:theta-sun}
\end{eqnarray} 
\label{eq:theta}
\end{subequations}
\!\!\!where the three mixing angles
$\theta_{ij}=\theta_{ji}$ are defined in the region
$0\leq\theta_{12},\theta_{13},\theta_{23}\leq\pi/2$~\cite{PDB}.
In the following, we 
adopt $\sin \theta_{\atm}$, $\sin \theta_{\rct}$, and $\sin \theta_{\sun}$
as defined above as the independent real mixing parameters
of the $3\times3$ MNS matrix.

\subsection{Approximate formulas}
\label{sec:approximate-formulas}

The probability 
that an initial flavor eigenstate $\left|\nu_\alpha\right\rangle$ 
with energy $E_\nu$ is observed as a flavor eigenstate 
$\left|\nu_\beta\right\rangle$
after traveling a distance $L$ in the matter of density 
$\rho(x)$ $(0<x<L)$ along the baseline is
\begin{eqnarray}
 P_{\nu_\alpha \to \nu_\beta}
 &=&
\left|
\left\langle \nu_\beta \right|
\exp\left({-i\int_0^LH(x)dx}\right)
\left|\nu_{\alpha} \right\rangle
\right|^2 \,,
\label{eq:PPP}
\end{eqnarray}
where the Hamiltonian inside the matter is
\begin{eqnarray}
 H(x)&=&
\dfrac{1}{2E_\nu}U
\left(
\begin{array}{ccc}
 0 & 0 & 0 \\
 0 & \dm12 & 0 \\
 0 & 0 & \dm13
\end{array}
\right)
U^\dagger
+
\dfrac{a(x)}{2E_\nu}
\left(
\begin{array}{ccc}
 1 & 0 & 0 \\
 0 & 0 & 0 \\
 0 & 0 & 0
\end{array}
\right)\,,\nn\\
&=&\dfrac{1}{2E_\nu}
{\tilde U(x)}
\left(
\begin{array}{ccc}
 \lambda_1(x) & 0 & 0 \\
 0 & \lambda_2(x) & 0 \\
 0 & 0 & \lambda_3(x)
\end{array}
\right)
{\tilde U}^\dagger(x)\,,
\label{eq:Hm}
\end{eqnarray}
with
\begin{equation}
 a(x) \equiv 2\sqrt{2} G_F E_\nu^{} n_e(x)
  \simeq \numt{7.56}{-5} \mbox{{[eV$^2$]}} 
  \left(\dfrac{\rho(x)}{\mbox{{g/cm$^3$}}}\right)
  \left(\dfrac{E_\nu}{\mbox{{GeV}}}\right)\,.
\label{eq:matt_a}
\end{equation}
Here
$G_F$ is the Fermi constant, 
$E_\nu$ is the neutrino energy,
$n_e(x)$ is the electron number density,
and
$\rho(x)$ is the matter density along the baseline.
In the translation from $n_e(x)$ to $\rho(x)$,
we assume that the number of the neutron is same as that of proton.
To a good approximation~\cite{T2K,HOS-mat},
the matter profile along the T2K and T2KK baselines
can be replaced by a constant, $\rho(x)=\rho_0$,
and the probability eq.~(\ref{eq:PPP}) can be expressed compactly
by using the eigenvalues $(\lambda_i)$ and the unitary matrix
$\tilde U$ of eq.~(\ref{eq:Hm});
\begin{subequations}
\begin{eqnarray}
 P_{\nu_\alpha \to \nu_\beta}
&=&
{\delta}_{\alpha\beta}
-4\sum_{i>j}
 {\rm Re}({\tilde U}^{\ast}_{\alpha i}{\tilde U}_{\beta i}^{}
     {\tilde U}_{\alpha j}^{}{\tilde U}^{\ast}_{\beta j})
 \sin^2\dfrac{{\tilde \Delta}_{ij}}{2} 
+2\sum_{i>j}
 {\rm Im}({\tilde U}^{\ast}_{\alpha i}{\tilde U}_{\beta i}^{}
     {\tilde U}_{\alpha j}^{}{\tilde U}^{\ast}_{\beta j})
 \sin{\tilde \Delta}_{ij}\,,~~~~~~
\label{eq:PPP2}\\
 \tilde{\Delta}_{ij}
&\equiv&
 \dfrac{\lambda_j-\lambda_i}{2E}L\,.
\end{eqnarray}
\end{subequations}
All our numerical results are based on the above solution
eq.~(\ref{eq:PPP2}),
leaving discussions of the matter density profile along the baselines
to a separate report~\cite{HOS-mat}.
Our main results are not affected significantly 
by the matter density profile~\cite{HOS-mat}
as long as the mean matter density along the baseline
$(\rho_0)$ is chosen appropriately.

Although the expression eq.~(\ref{eq:PPP2}) is not particularly
illuminating,
we find the following approximations~\cite{HOS1,HOS2} useful
for the T2KK experiment.
Since the matter effect is small at sub GeV to a few GeV region 
for $\rho\sim3$ g/cm$^3$,
and the phase factor $\Delta_{12}$ in the vacuum,
where
\begin{equation}
 {\Delta}_{ij}
\equiv
 \dfrac{m^2_j-m^2_i}{2E}L\,,
\end{equation}
is also small near
the first oscillation maximum, $|\Delta_{13}|\sim \pi$,
the approximation of keeping the first and second order corrections
in the matter effect and $\Delta_{12}$
\cite{arafune97,HOS1,koike05,HOS2}
\begin{subequations} %probs
\begin{eqnarray}
 P_{\nu_\mu \to \nu_\mu} &=&
 1 - \satm{2}\left( 1 + A^{\mu} \right)
       \sin^2 \left( \dfrac{\Delta_{13}}{2} + B^{\mu} \right)
\,,\label{eq:Pmm}\\
P_{\nu_{\mu} \to \nu_e} &=& 
4 \satm{} \srct{}
\left\{
 \left( 1 + A^{e} \right)
  \sin^2 \left( \dfrac{\Delta_{13}}{2} \right)
+B^e \sin \Delta_{13}
\right\}
+C^e
 \,,\label{eq:Pme}
\end{eqnarray}
\label{eq:Prob2}
\end{subequations}
$\!\!$has been examined in ref.~\cite{HOS2}.
Here $A^{\mu}$ and $B^{\mu}$ 
are the corrections to the amplitude and the oscillation phase,
respectively, of the $\nu_\mu$ survival probability.
When $|A^e|$ and $|B^e|$ are small, eq.~(\ref{eq:Pme})
reduces to
\begin{equation}
P_{\nu_{\mu} \rightarrow \nu_e} \approx 
4 \satm{} \srct{}
 \left( 1 + A^{e} \right)
  \sin^2 \left( \displaystyle\frac{\Delta_{13}}{2} +B^e \right)
 +C^e 
 \,,
 \label{eq:Pme2}
\end{equation}
similar to the $\nu_{\mu}$ survival probability, eq.~(\ref{eq:Pmm}).
We therefore refer to $B^e$ in eq.~(\ref{eq:Pme}) as the oscillation
phase-shift, even thought it can be rather large ($\sim 0.4$).
%---

%---
For the $\nu_\mu$ survival probability, eq.~(\ref{eq:Pmm}),
it is sufficient to keep only the linear terms in $\Delta_{12}$ and $a$,
\begin{subequations}
\label{eq:ABmu}
\begin{eqnarray}
A^{\mu} &=& - \dfrac{aL}{\Delta_{13}E} 
\dfrac{\cos2\theta_{\atm}}{\catm{}}
\srct{} 
\label{eq:Amu}\,,\\
B^{\mu} &=& \dfrac{aL}{4E} 
\dfrac{\cos2\theta_{\atm}}{\catm{}}
\srct{} 
- \dfrac{\Delta_{12}}{2}
\left(\csun{}+ \tan^2 \theta_{\atm}\ssun{}\srct{} \right. \nn \\
&&
\left.\hspace*{20ex}
 - \tan \theta_{\atm}\sin 2\theta_{\sun}\sin\theta_{\rct}\cos\dmns
\right) 
\label{eq:Bmu}\,.
\end{eqnarray}
\end{subequations}
The above simple analytic expressions
reproduce the survival probability with 1$\%$ accuracy
throughout the parameter range explored in this analysis,
except where the probability is very small,
($P_{\nu_\mu\to\nu_\mu} \lsim 10^{-5}$).
%---
In eq.~(\ref{eq:Amu}), 
the magnitude of $A^{\mu}$ is much smaller than the unity
because of the constraints (\ref{eq:exp-satm}) and (\ref{eq:exp-srct1}),
and hence the amplitude of the $\nu_{\mu}$ survival probability is not 
affected significantly by the matter effect.
This means that
$\satm{2}$ can be fixed by the $\nu_\mu$
disappearance probability independent of the neutrino mass hierarchy
and the other unconstrained parameters.
The phase-shift term $B^{\mu}$ affects the measurement of $|\dm13|$.
However, the magnitude of this term is also much smaller than that of
the leading term, $\Delta_{13}/2$, around the oscillation maximum
$|\Delta_{13}|\sim\pi$,
because
$\cos2\theta_{\atm}=\sqrt{1-\satm{2}}<\sqrt{0.1}$ by 
eq.~(\ref{eq:exp-satm})
and 
$\Delta_{12}/\Delta_{13}<1/30$ by 
eqs.~(\ref{eq:exp-dm13}) and (\ref{eq:exp-dm12}).
The smallness of the phase shift term $B^\mu$ does not allow us to
determine the sign of $\Delta_{13}$,
or the neutrino mass hierarchy pattern,
from the measurements of the $\nu_\mu$ survival probability only.
%----

%----
For the $\nu_\mu\to\nu_e$ transition, eq.~(\ref{eq:Pme}),
we need to retain both linear and quadratic terms of 
$\Delta_{12}$ and $a$ to obtain a good approximation;
\begin{subequations}
\label{eq:ABCe}
\begin{eqnarray}
 A^e 
  &=& \dfrac{aL}{\Delta_{13}E}\cos 2\theta_{\rct}
  -\dfrac{\Delta_{12}}{2}
  \dfrac{\sin2\theta_{\sun}}{\tan\theta_{\atm} \sin\theta_{\rct}}
  \sin \dmns 
  \left( 1 + \dfrac{aL}{2\Delta_{13}E}\right)
  \nonumber\\
 &&+ \dfrac{\Delta_{12}}{4}
  \left( \Delta_{12} + \dfrac{aL}{2E} \right)
  \left(
   \dfrac{\sin2\theta_{\sun}}{\tan\theta_{\atm} \sin\theta_{\rct}}
   \cos \dmns - 2 \ssun{} \right)
  \nonumber \\
 &&-\dfrac{1}{2}\left( \dfrac{aL}{2E} \right)^2
  +\dfrac{3}{4}\left( \dfrac{aL}{\Delta_{13}E} \right)^2\,,
\label{eq:Ae}\\
 B^e
  &=& - \dfrac{aL}{4E}\cos 2\theta_{\rct}
  + \dfrac{\Delta_{12}}{4}
  \left(\dfrac{\sin2\theta_{\sun}}{\tan \theta_{\atm}\sin\theta_{\rct}}
   \cos \dmns - 2 \sin^2\theta_{\sun} \right)
  \left( 1 + \dfrac{aL}{2\Delta_{13}E}\right) 
\nonumber \\
 &&+ \dfrac{\Delta_{12}}{8} \left(\Delta_{12} + \dfrac{aL}{2E}\right)
  \dfrac{\sin2\theta_{\sun}}{\tan\theta_{\atm} \sin\theta_{\rct}} \sin \dmns 
  - \dfrac{1}{\Delta_{13}}\left(\dfrac{aL}{2E}\right)^2\,,
    \label{eq:Be}\\
 C^e
  &=& \dfrac{\Delta_{12}^2}{4} \ssun{2}\catm{}
  -\dfrac{\Delta_{12}}{2}\dfrac{aL}{2E}
\sin 2\theta_{\sun}\sin 2\theta_{\atm}\sin \theta_{\rct}\cos \dmns
\nn\\
&&
+\left(\dfrac{aL}{2E}\right)^2\srct{}\satm{}\,.
\label{eq:Ce}
\end{eqnarray}
\end{subequations}
Here,
the first and second terms in eqs.~(\ref{eq:Ae}) and (\ref{eq:Be})
are the linear terms of $\Delta_{12}$ and $a$ respectively,
while the other terms and
all the terms in eq.~(\ref{eq:Ce})
are quadratic in $\Delta_{12}$ and $a$.
These quadratic terms can dominate the oscillation probability 
when $\srct{}$ is very small.
We find that these analytic expressions,
eqs.~(\ref{eq:Pme}) and (\ref{eq:ABCe}),
are useful throughout the parameter range of this analysis,
down to $\srct{} = 0$,
except near the oscillation minimum.
%---
The amplitude of the $\nu_{\mu} \to \nu_e$ transition probability,
$1+A^e$, is sensitive to the mass hierarchy pattern,
because the first term of $A^e$ changes sign in eq.~(\ref{eq:Ae}),
with $\cos 2 \theta_{\rct} \sim 1$.
When $L/E$ is fixed at $|\Delta_{13}| \sim \pi$,
the difference between the two hierarchy cases grows with $L$,
because the matter effect grows with $E$;
see eq.~(\ref{eq:matt_a}).
The hierarchy pattern can hence be determined by comparing 
$P_{\nu_\mu\to\nu_e}$ near the oscillation maximum $|\Delta_{13}|\simeq\pi$
at two vastly different baseline lengths~\cite{HOS1,HOS2}.

Once the sign of $\Delta_{13}$ is fixed by the term linear in $a$, 
the terms linear in $\Delta_{12}$ 
allow us to constrain $\sin \dmns$ via the amplitude $A^e$, and
$\cos \dmns$ via the phase shift $B^e$.
Therefore, $\dmns$ can be measured uniquely
once the mass hierarchy pattern and
the value of $\srct{2}$, which may be measured at the next generation
reactor experiments~\cite{DoubleCHOOZ,DayaBay,reno},
are known.
%---

% ======================================================================== %
\section{Signals and Backgrounds}
\label{sec:3}
In this section, 
we show
how we estimate the event numbers from the charged current (CC)
and the neutral current (NC) interactions.
First, we explain how the signal CCQE events are reconstructed by
water \cerenkov detectors,
and study contributions from the inelastic processes
when none of the produced particles emit \cerenkov lights
and hence cannot be distinguished from the CCQE events.
Next in subsection~\ref{sec:nc-events},
we study NC production of single $\pi^0$,
which can mimic the $\nu_\mu \to \nu_e$ appearance signal
when the two photons from $\pi^0$ decay cannot be
resolved by the detector.
Finally, we show the sum of the signal and the background events.

\subsection{CC events}
\label{sec:cc-events}

In accelerator based long baseline experiments,
one can reconstruct the incoming neutrino energy $E_\nu$ by observing
the CCQE events 
($\nu_\ell n \to \ell p$ or $\bar{\nu}_\ell p \to \bar{\ell} n$)
if the charged lepton ($\ell = \mu$ or $e$) momenta are
measured and the target nucleons are at rest,
since the neutrino beam direction is known.
In practice, however, the lepton momentum measurements have errors,
the nucleons in nuclei have Fermi motion, and some
non-CCQE events cannot be distinguished from the CCQE events.
None of those uncertainties has been taken into account in the
previous studies of refs.~\cite{HOS1,HOS2,HO-atm}.
In this and the next subsections, we study them for CC and NC
processes, respectively,
for a water \cerenkov detector by using the event generator
{\sf nuance}~\cite{nuance}.

\subsubsection{Event selection}
\label{sec:event-selection}
%----------------------------------------------------------------------
\begin{figure}
\centering
 \includegraphics[scale=0.35]{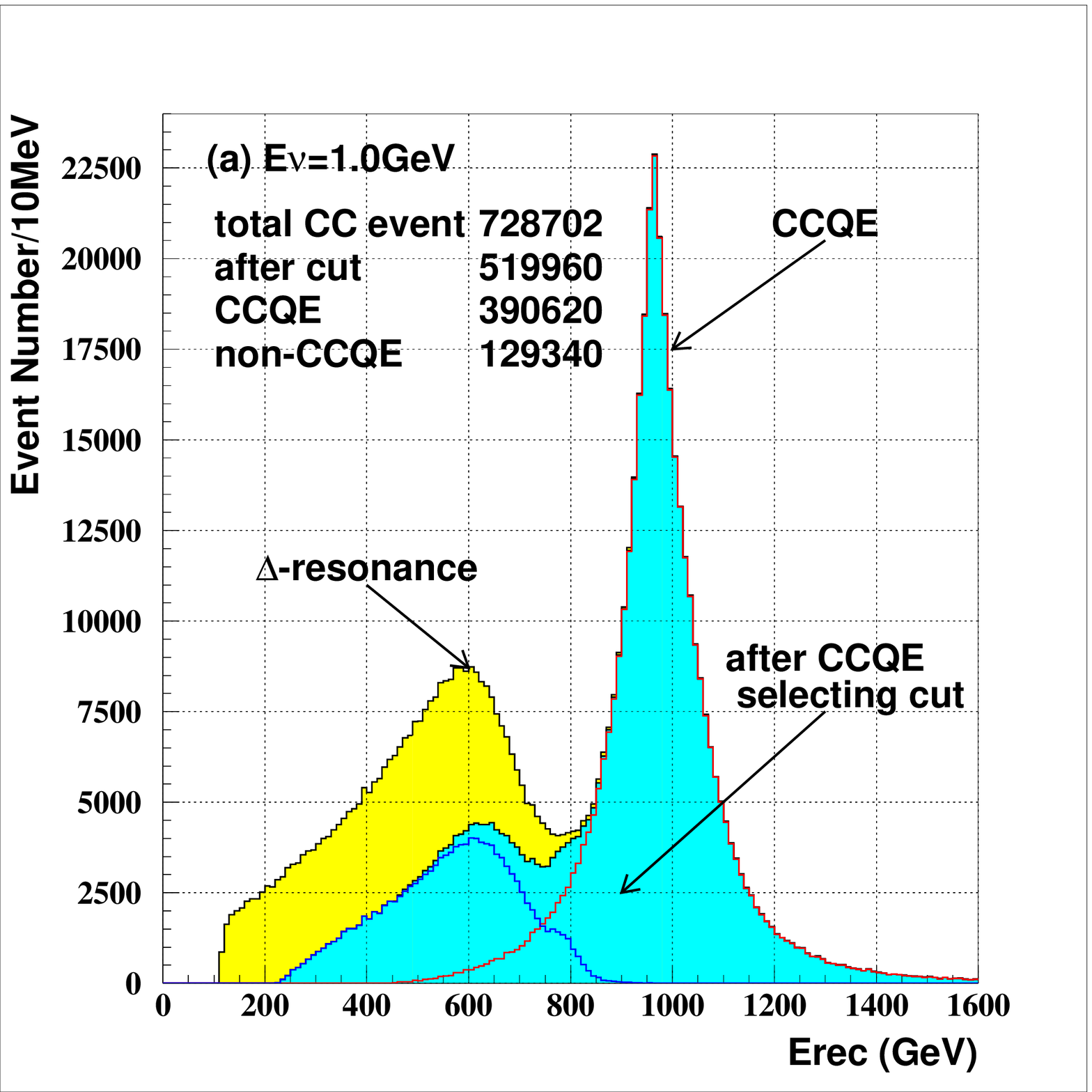}
~~~~~~~~~~
 \includegraphics[scale=0.35]{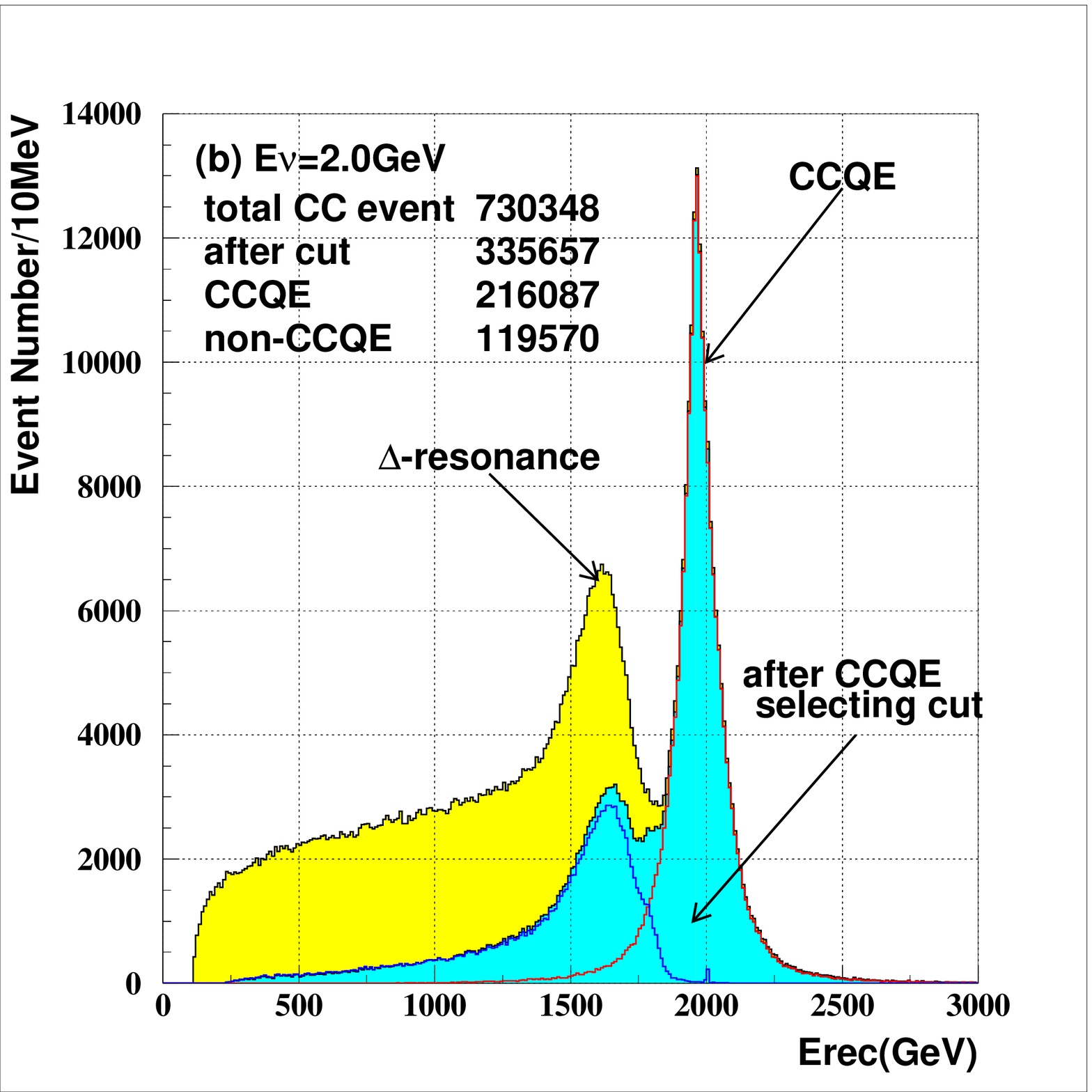}
 \caption{
Reconstructed energy distributions for $\nu_\mu$ CC events on the
water target at $E_\nu=1$~GeV (a) and $E_\nu=2$~GeV (b)
according to the event generator {\sf nuance}~\cite{nuance}
when the $\mu$ momenta are measured exactly.
Among $10^6$ generated events about 73\% are CC events
at both energies,
which consist of CCQE events, nuclear resonance production events,
and the others including deep-inelastic processes.
After the CCQE selection cuts of eq.~(\ref{eq:criteriaCC}) are applied,
the blue shaded region survive.
} 
\label{fig:event-select}
\end{figure}
%----------------------------------------------------------------------

In a CCQE event, $\nu_\ell n \to \ell p$, the neutrino energy $E_\nu$
can be reconstructed as
\begin{equation}
 \Erec = 
\dfrac{m_n E_\ell - m_\ell^2/2 - \left(m_n^2-m_p^2\right)/2}
{m_n-E_\ell+ p_\ell \cos\theta}\,,
\label{eq:erec1}
\end{equation}
in terms of the lepton energy ($E_\ell$),
total momentum ($p_\ell$),
and its polar angle $\theta$ about the neutrino beam direction,
if a target neutron is at rest.
%--
For an anti-neutrino CCQE event, $\bar{\nu}_\ell p \to \bar{\ell} n$,
$m_p$ and $m_n$ should be exchanged in eq.~(\ref{eq:erec1}).

%--
In reality, the target nucleons inside nuclei has Fermi motion
of about $100$~MeV, and the measured $e$ and $\mu$ momenta have
errors.
Therefore, $\Erec$ of eq.~(\ref{eq:erec1}) is distributed around the
true $E_\nu$, even for the CCQE processes.
%--

%--
The CCQE events are selected as 1-ring events in a water \cerenkov
detector by the following criteria
\cite{K2K, T2K}~:
\begin{subequations} %cc-criteria
\label{eq:criteriaCC}
\begin{eqnarray}
&&
\mbox{{Only one charged lepton}}~(\ell=\mu^\pm \mbox{ or } e^\pm)~
\mbox{{with}}~|p_l|>200
\mbox{{~MeV\,,}}\label{eq:CC1} \\
&&
\mbox{{No high energy}}~\pi^\pm~(|p_{\pi^\pm}|>200\mbox{{~MeV}})\,,
\label{eq:CC2}\\
&&
\mbox{{No high energy}}~\gamma~(|p_{\gamma}|>30\mbox{{~MeV)}}\,,
\label{eq:CC3}\\
&&
\mbox{{No}}~\pi^0,~K_{S},~K_{L},~\mbox{{and}}~K^\pm\,.
\label{eq:CC4}
\end{eqnarray}  
\end{subequations}
The lower limit of the total momentum in the first criterion
in eq.~(\ref{eq:CC1}) is from the threshold of the
water \cerenkov detector for $\ell=\mu$~\cite{SK}.
$\pi^\pm$ with $|p|>200$~MeV or
$\gamma$ with $|p|>30$~MeV
gives rise to an additional ring.
Also,
$\pi^0$, $K_{L}$, $K_{S}$, and $K^\pm$
always decay inside the detector,
making additional rings.

Figure \ref{fig:event-select}
shows the $\Erec$ distribution of the $\nu_\mu$ CC events
at $E_\nu=1$~GeV (a) and $E_\nu=2$~GeV (b) on the water target,
according to the event generator {\sf nuance}~\cite{nuance}.
Among the $10^6$ events at each energy, about 73\% are CC
events (the rests are NC events)
which consist of CCQE events, nuclear resonance production, and 
the others including deep inelastic events.
After the CCQE selection cuts of eq.~(\ref{eq:criteriaCC}) are applied,
the blue shaded region survives,
which consists of the CCQE events
and the other events where the produced $\pi^\pm$ are soft.
We call the non-CCQE events which survives the selection cuts of
eq.~(\ref{eq:criteriaCC}) ``resonance events'',
since most of them come from single soft $\pi^\pm$ emission from
the $\Delta$ resonance.
The CCQE events and the resonance events
are observed as two peaks in the reconstructed energy which are
separated by about 380~MeV at $E_\nu\simeq 1$~GeV,
rather independent of the initial $\nu_\mu$ energy.
This is because the origin of the distance between the two peaks
mainly comes from the mass difference between the nucleon and the 
$\Delta$ resonance,
which scales as 
\begin{equation}
\dfrac{(m_\Delta^2-m_p^2)}{2m_n}\simeq 340\mbox{{~MeV}}
\label{eq:340}
\end{equation}
in eq.~(\ref{eq:erec1}).
Because the peak value of the factor, $E_\ell - p_\ell \cos\theta$,
in the denominator of eq.~(\ref{eq:340}) decreases from about $100$~MeV
at $E_\nu = 1$~GeV to about $50$~MeV at $E_\nu = 2$~GeV,
the difference in the peak locations decreases slightly from
about $380$~MeV at $E_\nu=1$~GeV in Fig.~1(a)
to about $360$~MeV at $E_\nu=2$~GeV in Fig.~1(b).
The half width of the CCQE peak is about $60$~MeV,
almost independent of $E_\nu$,
because it comes from the Fermi motion of the target nucleons
inside nuclei.

\subsubsection{Lepton momentum resolutions}
\label{sec:lept-moment-resol}
After selecting the CCQE-like events,
we examine the detector resolution
which further smears the $\Erec$ distribution.
We use the momentum and angular resolutions of 
the muon and electron at SK~\cite{SK},
which are shown in Table~\ref{tab:resolution}.
For the momenta around 1~GeV, the momentum resolutions are 
about a few \% and the angular resolutions
are about a few degrees for both $\mu$ and $e$.

%--
\begin{table}
\centering
  \begin{tabular}{|l|c|c|}
   \hline
   & ${\delta p}/{p}~~(\%)^{}_{}$& $\delta \theta$ (degree)\\
   \hline
   \rule[-13pt]{0pt}{31pt}
   $\mu$
   & $\l( 1.7+{0.7}/{\sqrt{p\mbox{(GeV)}}} \r)$
       & $1.8^\circ$ \\
   \hline
   \rule[-13pt]{0pt}{31pt}
   $e$
   & $\l( 0.6+{2.6}/{\sqrt{p\mbox{(GeV)}}} \r)$
       & $3.0^\circ$ \\
   \hline
  \end{tabular}
 \caption{The momentum and angular resolution 
of $\mu$- and $e$-momenta at SK~\cite{SK}.}
 \label{tab:resolution}
\end{table}
%--

\begin{figure}
\centering
\includegraphics[scale=0.85]{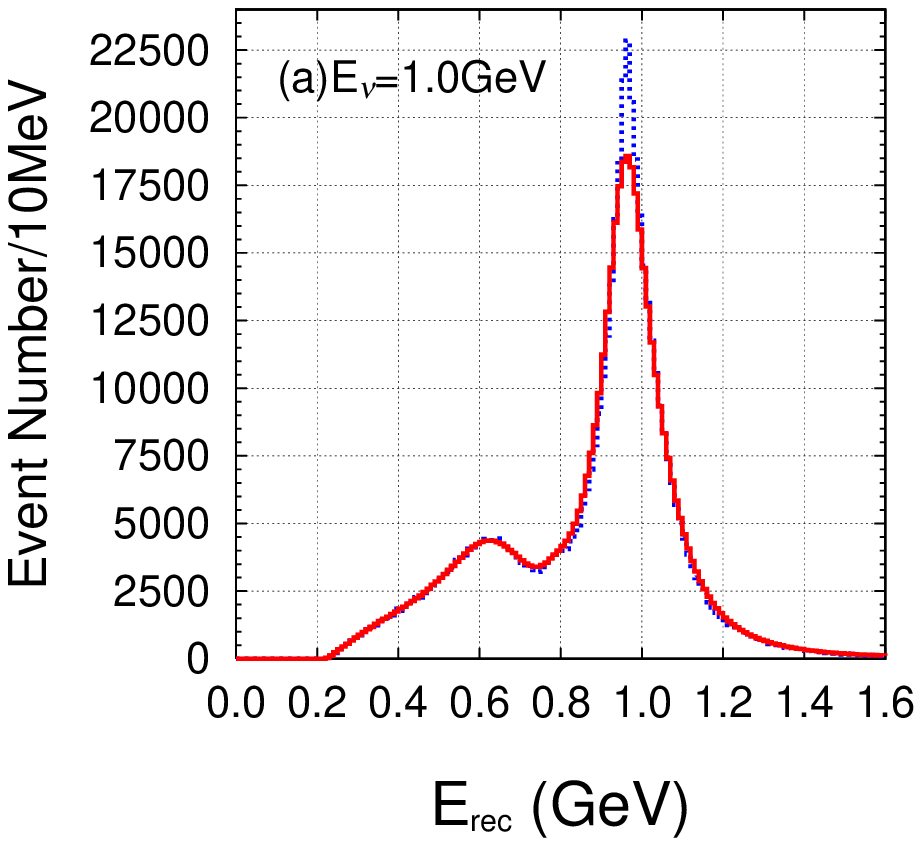}
\hspace*{-3ex}
\includegraphics[scale=0.85]{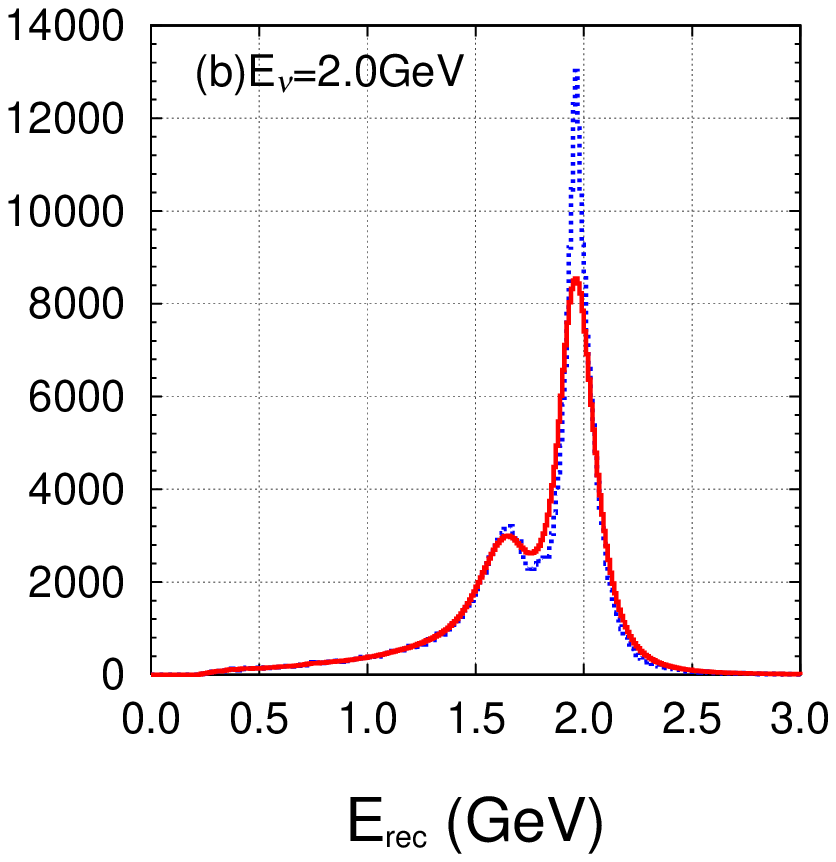} 
\caption{
$\Erec$ distribution of the CC events on the water target
for monochromatic energy $\nu_\mu$ at $E_\nu=1$~GeV
(a) and at $E_\nu=2$~GeV (b),
after the CCQE selecting cuts, generated by 
{\sf nuance}~\cite{nuance}.
The dotted curves show the distributions when the $\mu^\pm$
momenta are measured exactly,
the boundaries of the blue region in Fig.~\ref{fig:event-select},
whereas the solid lines show the distributions after the finite
momentum resolution of Table~\ref{tab:resolution} is
taken into account. 
}
\label{fig:reso:CCQE}
\end{figure}

 In Fig.~\ref{fig:reso:CCQE},
we show by solid curves the $\Erec$ distributions after taking account
of the $\mu^\pm$ momentum resolutions of Table~\ref{tab:resolution},
while the dotted lines show the distributions
when the $\mu^\pm$ momenta are measured exactly,
which are the boundaries of the blue shaded region 
in Fig.~\ref{fig:event-select}.

%---
%---

The total width of the CCQE peak is now the 
sum of the effects from 
the Fermi motion ($\sigma_{\rm Fermi}$),
the momentum resolution ($\sigma_{\delta p/p}$),
and the angular resolution ($\sigma_{\delta\theta}$);
it grows with $E_\nu$, 
because $\sigma_{\delta p/p}$ grows with the lepton momentum.
For instance, the half width is about 60~MeV for $E_\nu=1$~GeV
and 70~MeV for $E_\nu=2$~GeV.
As a consequence of the energy dependence for the total width,
the peak height of the CCQE events becomes lower,
by about $80\%$ for $E_\nu=1.0$~GeV and $67\%$ for $2.0$~GeV.

The $\Erec$ distribution for the $\nu_e$ CCQE events are very similar,
and we do not show them separately.
Small differences, due to poorer momentum resolution of electrons in
Table~\ref{tab:resolution}, are reflected in our parameterizations in
the next subsection.

\subsubsection{Parameterization for the CCQE events}
\label{sec:param-ccqe-events}

In this section, we present our parameterization of the $\Erec$
distribution of the CCQE events for a given initial $\nu_\mu$ or
$\nu_e$ energy $E_\nu$,
after taking account of the $\mu$- and $e$-momentum resolutions of
Table~\ref{tab:resolution}.

The $\Erec$ distribution from the CCQE events
can be reproduced by three Gaussians,
\begin{eqnarray}
 f^{\rm CCQE}_{\alpha}(\Erec;E_\nu) =
 \dfrac{1}{A^\alpha(E_\nu)}
\sum_{n=1}^{3}
r^\alpha_n(E_\nu) 
\exp
\left(-\dfrac{(\Erec-E_\nu+{\delta}E^\alpha_n(E_\nu))^2}
{2(\sigma^\alpha_n(E_\nu))^2}
\right)
\,,
\label{eq:fit_ccqe0}
\end{eqnarray}
where the index $\alpha$ is for $\mu$ or $e$,
with $r^\alpha_1 (E_\nu) = 1$.
The factor $A^\alpha(E_\nu)$ ensures the normalization 
\begin{equation}
 \int f^{\rm CCQE}_{\alpha}(\Erec;E_\nu) d\Erec = 1\,.
\label{eq:fit_normal_ccqe0}
\end{equation}
The variance $\sigma^\alpha_n$, 
the energy shift ${\delta}E^\alpha_n$ ($n=1,2,3$), 
and 
the coefficients, $r^\alpha_2$ and $r^\alpha_3$,
are parameterized as functions of 
the incoming neutrino energy $E_\nu$.
These parameters depend on the neutrino species,
$\nu_\mu$ or $\nu_e$,
because of the mass difference in eq.~(\ref{eq:erec1}),
the difference in the momentum resolutions in
Table~\ref{tab:resolution},
and also because of small differences in the CC cross sections
at low energies~\cite{nuance}.
Our parameterization\footnote{{%
A computer code (C/C++) for the parameterization
are available from the authors,
or directly from the web-site~\cite{mail2NO}.}}
is given in \ref{sec:appA1},
eqs.~(\ref{eq:sig_ccqe_mu})-(\ref{eq:r_ccqe_e})
which is valid in the region
$0.3$~GeV $\leq E_\nu \leq$ $6.0$~GeV and 
$0.4$~GeV $\leq \Erec \leq$ $5.0$~GeV
for both $\nu_\mu$ and $\nu_e$.
For the sake of keeping the consistency with 
the previous studies in ref.~\cite{HOS1,HOS2,HO-atm},
those events with $\Erec<0.4$~GeV are not used in the
present analyses.
%---

\begin{figure}
\centering
\includegraphics[scale=0.65]{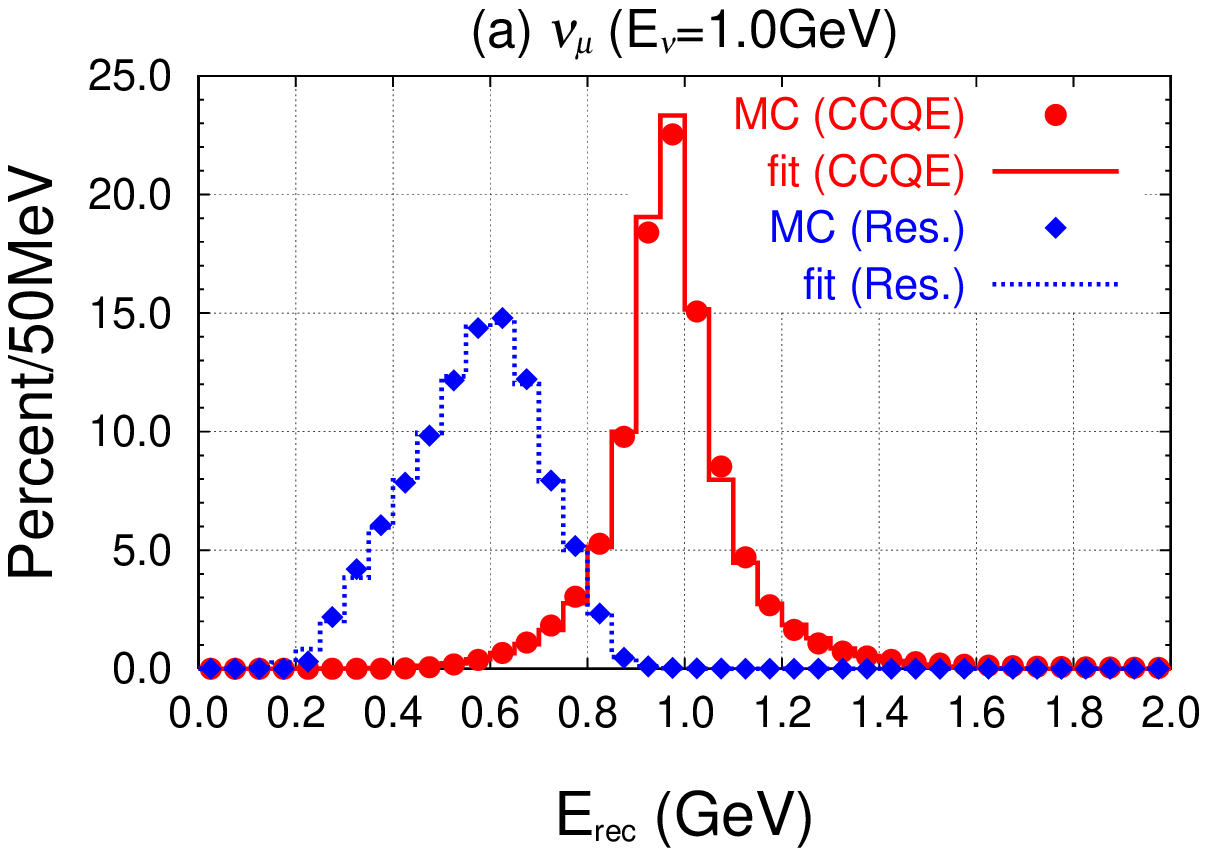}
\hspace*{-5ex}
\includegraphics[scale=0.65]{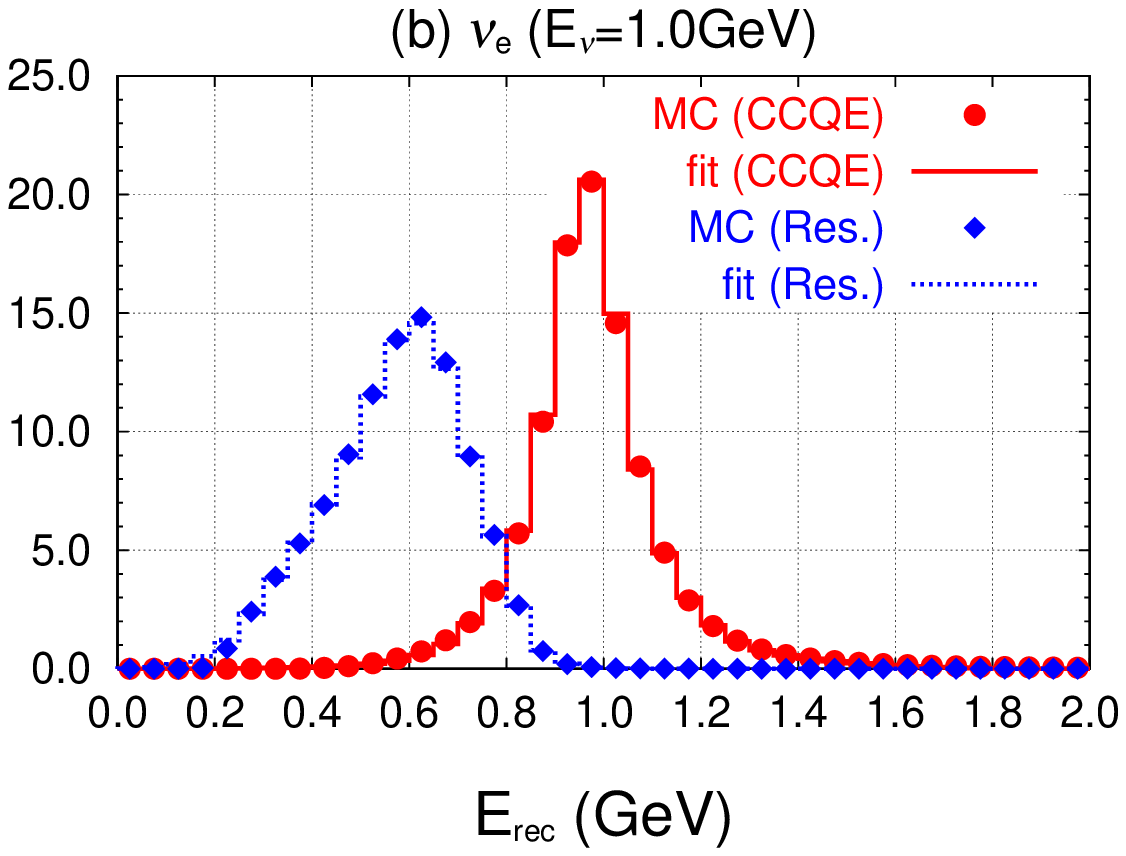}

\bigskip

\includegraphics[scale=0.65]{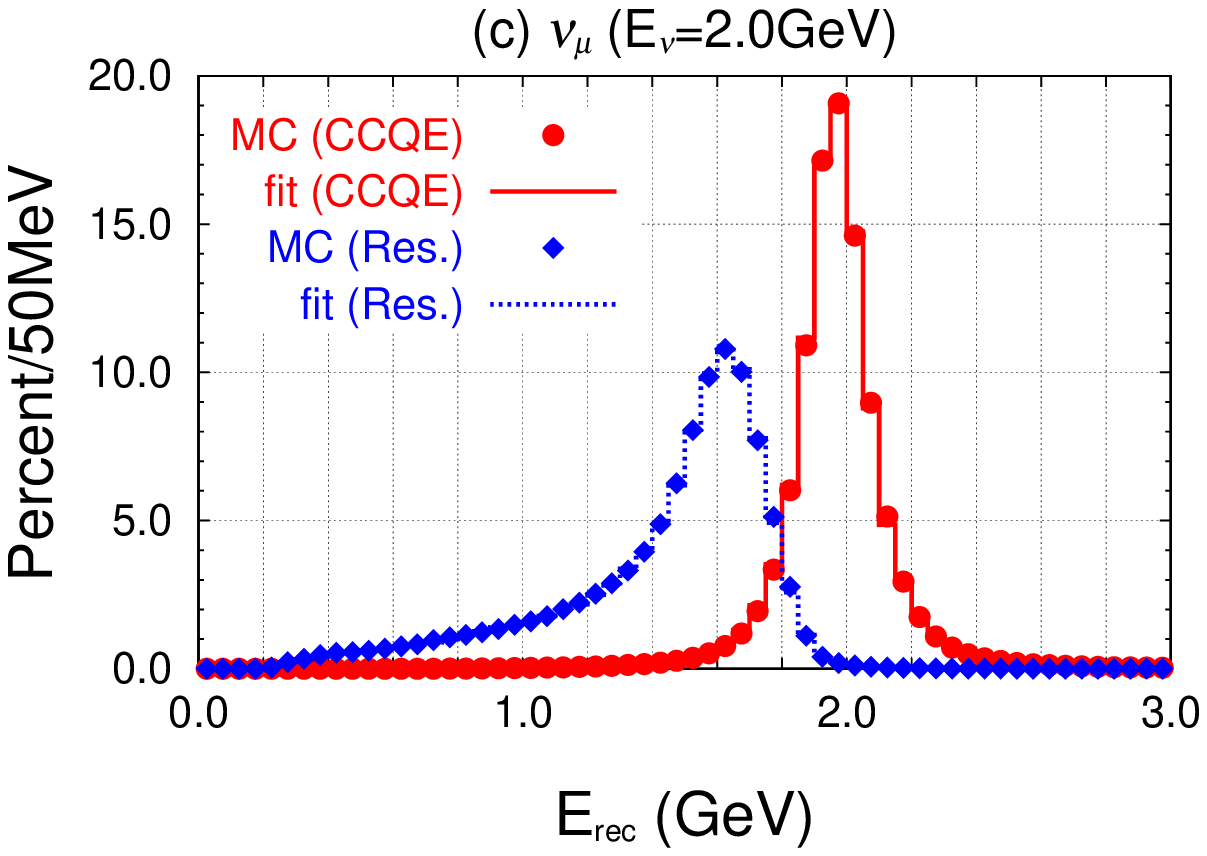}
\hspace*{-5ex}
\includegraphics[scale=0.65]{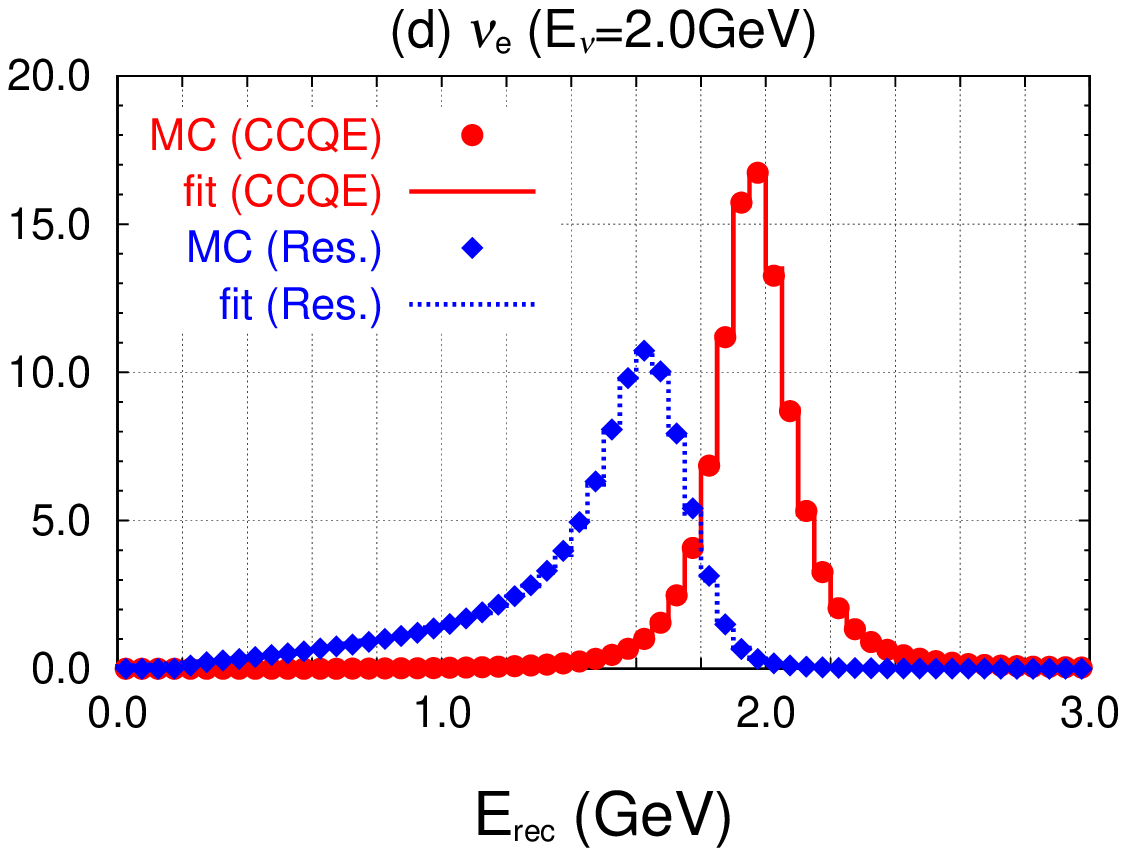}
 \caption{
Normalized $\Erec$ distribution of the CCQE events (solid circles)
and the resonance events (solid diamonds) after the CCQE selection
cuts of eq.~(\ref{eq:criteriaCC})
and the momentum resolutions for $\mu$ and $e$ in
Table~\ref{tab:resolution} are applied.
The events are generated by {\sf nuance}~\cite{nuance}.
The solid line shows our parameterization for the CCQE events and
the dotted line is for the resonance events:
(a) $\nu_\mu$ and (b) $\nu_e$ at $E_\nu=1$~GeV;
(c) $\nu_\mu$ and (d) $\nu_e$ at $E_\nu=2$~GeV.
}
\label{fig:fitfc:1GeV}
\end{figure}

In Fig.~\ref{fig:fitfc:1GeV},
we show the $\Erec$ distribution of the CCQE events.
The solid circles show the distributions generated 
by {\sf nuance}~\cite{nuance},
and the histograms show our smearing functions of
eq.~(\ref{eq:fit_ccqe0}).
Figures.~\ref{fig:fitfc:1GeV}(a) and (b)
are for $\nu_\mu$ and $\nu_e$, respectively,
at $E_\nu=1$~GeV,
and
(c) and (d) are for those at $E_\nu=2$~GeV.
The area under each distribution is normalized to unity.

%--------------------

\subsubsection{Nuclear resonance contributions}
\label{sec:nucl-reson-contr}

The $\Erec$ distribution of the non-CCQE events which
pass the CCQE selection cuts of eq.~(\ref{eq:criteriaCC})
can also be parameterized.
Most of them come from the $\Delta$ resonance production, and
the resonance peak in the $\Erec$ distribution is observed
in Figs.~\ref{fig:event-select} and \ref{fig:reso:CCQE}.
For $E_\nu \leq 1.2$~GeV, 3 Gaussians suffice to reproduce the
$\Erec$ distributions generated by {\sf nuance}~\cite{nuance};
\begin{eqnarray}
 f^{\rm res}_{\alpha}(\Erec;E_\nu\leq1.2\mbox{GeV}) = 
\dfrac{1}{\hat{A}^\alpha(E_\nu)}
\sum_{n=1}^{3}
\hat{r}^\alpha_n(E_\nu) 
\exp
\left(-\dfrac{(\Erec-E_\nu+\delta\hat{E}_n^\alpha(E_\nu))^2}
     {2(\hat{\sigma}_n^\alpha(E_\nu))^2}
\right)
\,,
\label{eq:fit_res3}
\end{eqnarray}
while at high energies ($E_\nu>1.2$~GeV),
we need 4 Gaussians, because the number of contributing 
resonances grows with $E_\nu$;
\begin{eqnarray}
f^{\rm res}_{\alpha}(\Erec;E_\nu>1.2\mbox{GeV})
= \dfrac{1}{\tilde{A}^\alpha(E_\nu)}
\sum_{n=1}^{4}
\tilde{r}^\alpha_n(E_\nu) 
\exp
\left(-\dfrac{(\Erec-E_\nu+{\delta}\tilde{E}^\alpha_n(E_\nu))^2}
 {2(\tilde{\sigma}_n^\alpha(E_\nu))^2}
\right)
\,.
\label{eq:fit_res4}
\end{eqnarray}
Around $E_\nu\sim1.2$~GeV, both parameterizations are valid.
Here again $\alpha$ is $\mu$ or $e$,
$\hat{r}^{\mu,e}_1(E_\nu)=\tilde{r}^{\mu,e}_1(E_\nu)=1$,
and the factors $\hat{A}(E_\nu)$ and $\tilde{A}(E_\nu)$
assure that the smearing functions are normalized to 1
as in eq.~(\ref{eq:fit_normal_ccqe0}).
The variances $\hat{\sigma}^\alpha_n$ and 
$\tilde{\sigma}^\alpha_n$,
the energy shifts ${\delta}\hat{E}^\alpha_n$,
${\delta}\tilde{E}^\alpha_n$,
and
the relative normalization factors
$\hat{r}^\alpha_n$ and $\tilde{r}^\alpha_n$ ($n\neq 1$)
are all parameterized as functions of the incoming energy $E_\nu$,
which are given in \ref{sec:appA2}.
The shape of the $\Erec$ distribution for the ``resonance'' events
are also shown in Fig.~\ref{fig:fitfc:1GeV}.
The solid diamonds show the distribution of non-CCQE events generated
by {\sf nuance}~\cite{nuance}
after the CCQE selection cuts of eq.~(\ref{eq:criteriaCC}) and the
momentum resolutions of Table~\ref{tab:resolution} are applied.
The dotted histograms show our smearing functions,
eqs.~(\ref{eq:fit_res3}) and (\ref{eq:fit_res4}).

%===========================================
\subsection{NC events}
\label{sec:nc-events}

The key observation of ref.~\cite{HOS1,HOS2} for 
the T2KK proposal is that
it is advantageous to observe the first oscillation maximum 
($|\Delta_{13}|\sim \pi$) at two vastly different baseline lengths,
$L=295$~Km at SK and $L\simeq1000$~km in Korea.
Higher energy neutrino beam, or small off-axis angle, is 
hence desired for the far detector in Korea.
However, the use of high energy (broad band) beam gives rise to a serious
background for the $\nu_\mu\to\nu_e$ oscillation signal.
The single $\pi^0$ production via the neutral current (NC),
whose cross section grows with $E_\nu$,
cannot always be distinguished from the $\nu_\mu\to\nu_e$ signal
in a water \cerenkov detector.
In this subsection, we study the NC $\pi^0$ production background in
detail and estimate its $\Erec$ distribution by using the momentum 
distribution of misidentified $\pi^0$'s.

\subsubsection{Event selection}
\label{sec:event-selection-1}

\begin{figure}
\centering
\includegraphics[scale=0.8]{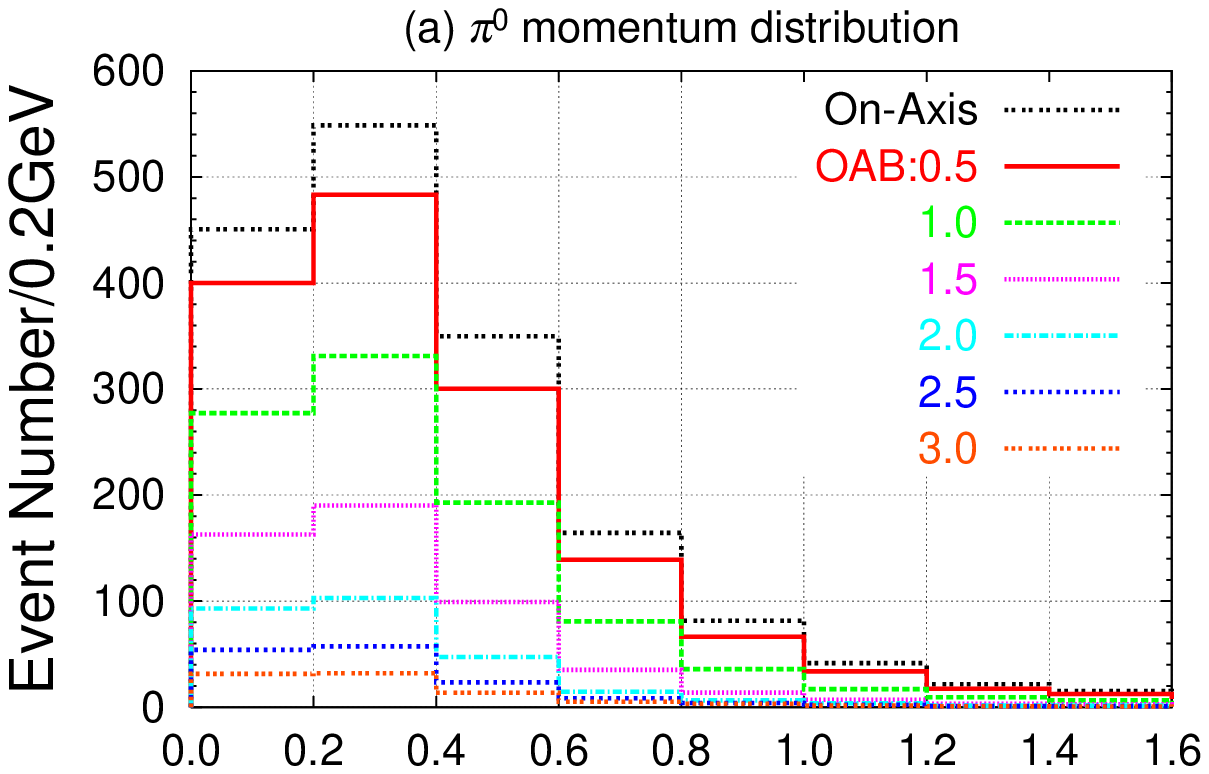}

\vspace*{-3ex}

\includegraphics[scale=0.8]{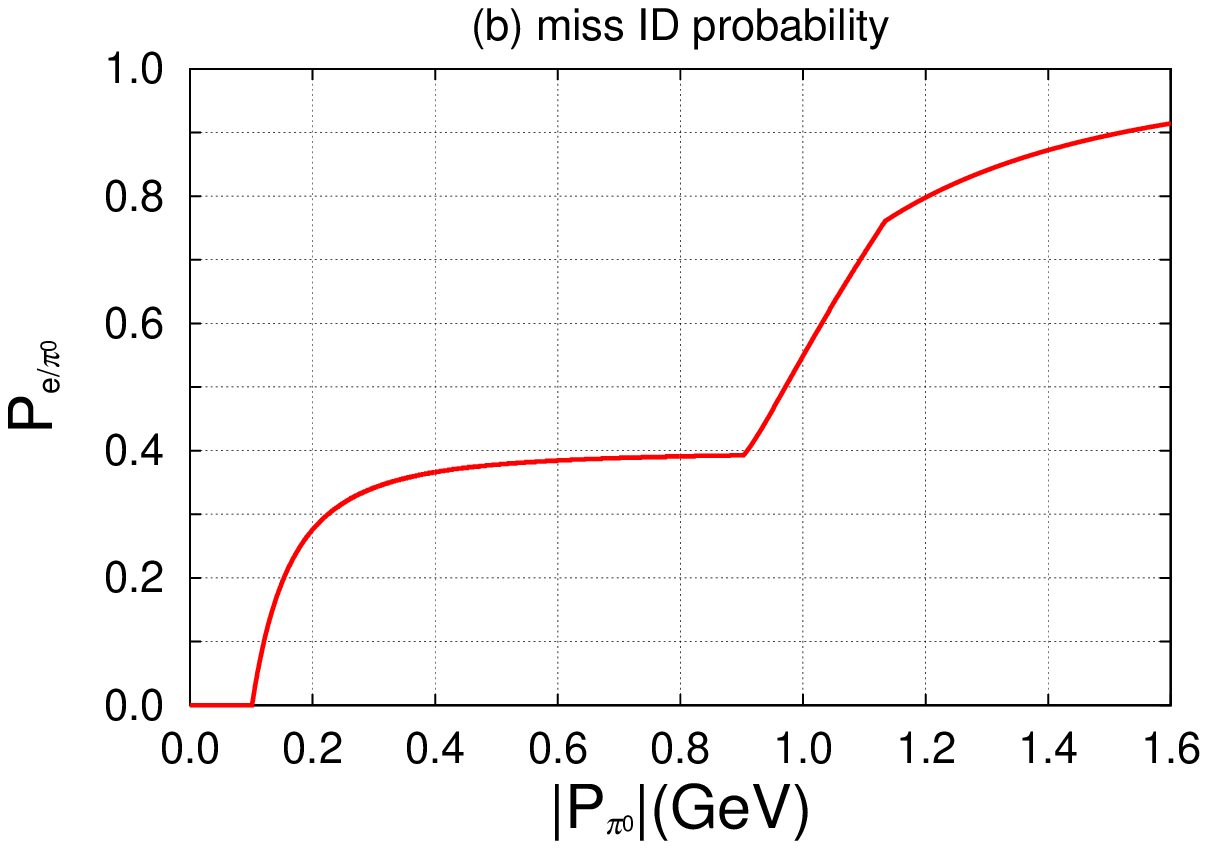}
  \caption{
(a): The $\pi^0$ momentum distribution of the single $\pi^0$
NC events selected by the criteria, eq.~(\ref{eq:criteriaNC}),
at various off-axis angles.
The event numbers are obtained for a 100~kton water target at
$L=1000$~km
with $\numt{5}{21}$POT, according to {\sf nuance}~\cite{nuance}.
(b): Probability that a $\pi^0$ cannot be distinguished from
$e^\pm$, according to eq.~(\ref{eq:def_P}).
The common horizontal axis measures the $\pi^0$ momentum.
}
\label{fig:pi0pt}
\end{figure}

We simulate the NC $\pi^0$ production background as follows.
By using the neutrino flux\footnote{{%
All the on- and off-axis neutrino flux 
distributions of the T2K beam used in this report
are available from the authors, or
directly from the web-site~\cite{mail2NO}.}}
of the T2K beam at various off-axis angles
between $0.0^\circ$(on-axis) and $3.0^\circ$,
and by using the total cross section
$\sigma_{\rm tot}^\alpha$ 
($\alpha=\nu_\mu,\bar{\nu}_\mu,\nu_e,$ and $\bar{\nu}_e$)
off the water target~\cite{nuance},
both CC and NC events are generated by {\sf nuance}~\cite{nuance}
for a water \cerenkov detector of 100~kton fiducial volume
at $L=1000$~km, with $\numt{5}{21}$ POT.
All the generated events are then confronted against
the following selection criteria : 
\begin{subequations} %nc-criteria
\begin{eqnarray}
&&
\mbox{{No charged leptons.}} \label{eq:NC1} \\
&&
\mbox{{Only one}}~\pi^0\,. \label{eq:NC2} \\
&&
\mbox{{No high energy}}~\pi^\pm~(|p_{\pi^\pm}|<200\mbox{{~MeV.}})
\label{eq:NC3}\\
&&
\mbox{{No high energy}}~\gamma~(|p_{\gamma}|<30\mbox{{~MeV.}})
\label{eq:NC4}\\
&&
\mbox{{No}}~K_L,~K_S,~K^\pm.
\label{eq:NC5}
\end{eqnarray}  
\label{eq:criteriaNC}
\end{subequations}
$\!\!\!\!$
The first condition, eq.~(\ref{eq:NC1}), selects
NC events, and the others eliminate multi-ring events.
The $\pi^0$ momentum distribution after the above cuts
is shown in 
Fig.~\ref{fig:pi0pt}(a) for various off-axis beams.
We find that the number of single $\pi^0$ events grows with
decreasing off-axis angle, especially for the angles below $2.0^\circ$
which have been envisaged in ref.~\cite{HOS1,HOS2,HO-atm}
as an optimal choice for the far detector in Korea.
%----
%----

\subsubsection{$\pi^0$ - $e^\pm$ misidentification probability}
\label{sec:pi0-epm-misid}

Figure.~\ref{fig:pi0pt}(a) shows that there are many single-$\pi^0$ 
events from the NC interactions,
especially for smaller off-axis angles.
Some of them become backgrounds of the $\nu_\mu\to\nu_e$ 
oscillation signal,
because the two photons from $\pi^0$ 
are not always resolved by a water \cerenkov detector.
When one of the two photons is much softer than the other,
the soft photon dose not give a clear ring,
resulting in a single-ring ($e$-like) event.
In addition, when the photons have a small opening angle 
the overlapping rings cannot always be resolved.

We therefore parameterize the probability of misidentifying
$\pi^0$ as an $e^\pm$ in terms of the energy ratio and the
opening angle of the two photons in the laboratory frame.
The energy fraction of the softer photon in the laboratory frame
\begin{equation}
x =\dfrac{E_2}{E_1+E_2}
~~~(E_2<E_1)
\label{eq:defR}
\end{equation}
can be expressed as
\begin{equation}
x = \dfrac{1}{2}\l(1-\beta \cos \hat{\theta}\r)\,,~~~~~~
\label{eq:defx2}
\end{equation}
in terms of the smaller polar angle
($\cos\hat{\theta}>0$)
of the photon momentum in the $\pi^0$ rest frame
about the polar axis along 
the $\pi^0$ velocity ($\beta$) in the laboratory frame.
The opening angle between the two photons in the laboratory frame
is then
\begin{eqnarray}
 \cos \theta_{\gamma\gamma} 
=1-\dfrac{1-\beta^2}{2x\left(1-x\right)}\,.
\label{eq:oa}
\end{eqnarray}
It is clear from eqs.~(\ref{eq:defx2}) and (\ref{eq:oa}) that
when the $\pi^0$ momentum is relativistic ($\beta\to 1$) 
either one of the photons becomes soft $(x\ll1)$
around $\cos\hat{\theta}\sim1$,
or the two photons become collinear, $\cos \theta_{\gamma\gamma}\sim 1$.

By using the energy fraction $x$ and
$\cos \theta_{\gamma\gamma}$, the $\pi^0$-$e^\pm$
misidentification probability can be parameterized as 
\begin{equation}
 P_{e/\pi^0} (|p_{\pi^0}|) =
\int_0^1 
\left[
\Theta(x^0-x)+ 
\Theta(x-x^0)
\Theta(\cos\theta_{\gamma\gamma}-\cos\theta_{\gamma\gamma}^0)
f(x,\cos\theta_{\gamma\gamma})
\right]
 d\cos \hat{\theta}\,,
\label{eq:def_P}
\end{equation}
where
$\Theta(x)$ is the step function.
The first step function in the r.h.s. tells that the
$\pi^0$ is misidentified as an $e^\pm$ 
when the energy fraction $x$
of the soft photon is smaller than $x^0$.
When both photons are hard $(x^0<x<0.5)$, it is still misidentified as
an $e^\pm$ when $\cos \theta_{\gamma\gamma} > \cos \theta_{\gamma\gamma}^0$.
We introduce a fudge factor
\begin{eqnarray}
f(x,\cos\theta_{\gamma\gamma})=
1.0-\l(\dfrac{x-x^0}{0.5-x^0}\r)^{1/2} 
    \l(\dfrac{1.0-\cos\theta_{\gamma\gamma}}
    {1.0-\cos\theta_{\gamma\gamma}^0}\r)^{3/2}\,,
\label{eq:def_Ppie2}
\end{eqnarray}
in order to take account of detector performance.
We show in Fig.~\ref{fig:pi0pt}(b) the $\pi^0$-$e^\pm$ misidentification
probability, $P_{e/\pi^0} (|p_{\pi^0}|)$, of eq.~(\ref{eq:def_P})
for $x^0=0.2$ and $\theta^0=17^\circ$, 
which reproduces qualitatively the typical performance of
water \cerenkov detectors.
The leadoff energy, $|p_{\pi^0}|=0.1$~GeV, and 
the height of the plateau, $P_{e/\pi^0}=0.4$,
are dictated by the first step function in eq.~(\ref{eq:def_P}),
which tells that the two photons are not resolved 
when the softer photon has an energy fraction less than 0.2.
The second term in eq.~(\ref{eq:def_P}) determines the
kink structure around $|p_{\pi^0}|=0.9$ and $11.0$~GeV,
as well the asymptotic behavior at high $\pi^0$ momentum.

The number of the $e$-like events from the $\pi^0$ background
can now be calculated as the product of the $\pi^0$ event number
in Fig.~\ref{fig:pi0pt}(a)
and the probability $P_{e/\pi^0} (|p_{\pi^0}|)$
in Fig.~\ref{fig:pi0pt}(b).
The reconstructed energy $\Erec$ of each $\pi^0$ background event is
calculated from the $\pi^0$ energy and the scattering angle by
assuming the electron mass.

\subsection{The event numbers}
\label{sec:event-numbers}

We calculate the numbers of $\nu_\mu$ and $\nu_e$ CC events
from the primary and the secondary beam
in the $i$-th energy bin, 
$\Erec^i < E <\Erec^{i+1}$, as
\begin{equation}
N_{\beta,D}^{i,X} (\nu_\alpha)=
 M N_A
 \int_{\Erec^i}^{\Erec^{i+1}}
\!\!\!\!\!
 d\Erec
 \int_{0}^{\infty}
 dE_\nu
\left[
 \Phi_{\nu_\alpha}^{D}(E_\nu)~
 P_{\nu_\alpha\to\nu_\beta}^{D}(E_\nu)~ 
 \hat{\sigma}_\beta^{X}(E_\nu)~ 
 f^X_\beta(\Erec;E_\nu)
\right]\,,
\label{eq:N}
\end{equation}
where $\Erec^i = 0.2{\rm GeV} \times i$.
Here
$M$ is the detector mass~(g),
$N_{A} = 6.017\times10^{23}$~(mol$^{-1}$) is the Avogadro number,
$\Phi_{\nu_\alpha}$ is the $\nu_{\alpha}$ 
flux\footnote{{%
The flux distribution used in this report are available
from the authors, or directly from the web-site~\cite{mail2NO}.}}
($\nu_{\alpha}=\nu_\mu,\bar{\nu}_\mu,\nu_e,\bar{\nu}_e$)
of the T2K $\nu_\mu$-beam~\cite{ICHIKAWA},
which is dominated by $\nu_\mu$ 
but has secondary $\bar{\nu}_\mu$, $\nu_e$, $\bar{\nu}_e$ components.
$P_{\nu_{\alpha}\to \nu_{\beta}}$
denotes the neutrino oscillation probability 
for $\nu_\mu,\nu_e \to \nu_\mu,\nu_e$ or
$\bar{\nu}_\mu,\bar{\nu}_e \to \bar{\nu}_\mu,\bar{\nu}_e$,
including the matter effect.
$\hat{\sigma}_\beta^{X}(E_\nu)~ $ is the 
cross section of the $\nu_\beta$ CC events for the
CCQE process ($X=$ CCQE) and the non-CCQE processes ($X=$ Res)
per nucleon in water.
The last term of eq.~(\ref{eq:N}), $f^X_\beta(\Erec;E_\nu)$
is the smearing function of eq.~(\ref{eq:fit_ccqe0}) for
the CCQE events, and
that of eqs.~(\ref{eq:fit_res3}) and (\ref{eq:fit_res4}) for
the ``resonance'' events.
The index $D$ tells the detector location;
the baseline length for $D=$ SK is 295~km 
and 
that for the far detector
$D=$ Kr is chosen between $L=1000$~km and $1200$~km.

The effective CCQE cross section per nucleon is slightly smaller
than the naive cross section at high energies;
\begin{equation}
{\hat{\sigma}_\beta^{\rm CCQE}(E_\nu)}
=
{{\sigma}_\beta^{\rm CCQE}(E_\nu)} \times
\left\{
\begin{array}{ll}
1.0\,,
&(\mbox{for } E_\nu\mbox{[GeV]}<0.9 )\, \\
1.0-0.054\left(\dfrac{E_\nu-0.90}{E_\nu-0.26}\right)\,,
&(\mbox{for } E_\nu\mbox{[GeV]}>0.9)\,
\end{array}
\right.
\label{eq:Xsec_CCQE}
\end{equation}
because of occasional emission of $\pi^0$ or $\gamma$ from the oxygen
nuclei.
As for the naive CCQE cross section per nucleon,
${\sigma}_\beta^{\rm CCQE}(E_\nu)$
for $\nu_\beta$ 
$(\nu_\beta=\nu_\mu,\bar{\nu}_\mu,\nu_e,\bar{\nu}_e)$
in water, we use the estimates of ref.~\cite{Xsec}
throughout the present analysis.
The reduction factor in eq.~(\ref{eq:Xsec_CCQE}) is our
parameterization of the outputs of {\sf nuance}~\cite{nuance}.

The effective resonance event cross section
$\hat{\sigma}_\beta^{\rm Res} (E_\nu)$
is the total cross section of all the non-CCQE
CC events that satisfy the CCQE selection criteria of
eq.~(\ref{eq:criteriaCC}).
They are slightly different between $\nu_\mu$ and $\nu_e$ CC events,
and we find that the following parameterizations 
\begin{subequations} %cross-sec-res
\begin{eqnarray}
\hat{\sigma}_e^{\rm Res} (E_\nu)\!\!\!
&=&\!\!\!
 {\sigma}_e^{\rm CCQE} (E_\nu)
\left(
0.789
+0.00738\log E_\nu
-\dfrac{0.455}{E_\nu}
\right)\,,
~(\mbox{ for } E_\nu\mbox{[GeV]}>0.51)\,,
\label{eq:Xsec_ResE}\\
\hat{\sigma}_\mu^{\rm Res} (E_\nu)\!\!\!
&=&\!\!\!
 {\sigma}_\mu^{\rm CCQE} (E_\nu)
\left(
0.810
+0.00738\log E_\nu
-\dfrac{0.436}{E_\nu}
\right)\,,
~(\mbox{ for } E_\nu\mbox{[GeV]}>0.54)\,,~~~~~
\label{eq:Xsec_ResM}
\end{eqnarray} 
\label{eq:Xsec_Res}
\end{subequations}
$\!\!\!\!$
reproduce well the results of {\sf nuance}~\cite{nuance}.
The gradual increase of the non-CCQE rates with $E_\nu$ reflects the
growth of the number of contributing resonances and deep-inelastic
events at high energies.

Both the fudge factors in eqs.~(\ref{eq:Xsec_CCQE}) and
(\ref{eq:Xsec_Res})
and the smearing functions 
eqs.~(\ref{eq:fit_ccqe0}),
(\ref{eq:fit_res3}),
and (\ref{eq:fit_res4}),
are obtained for $\nu_\mu$ and $\nu_e$
CC events.
They can be slightly different for $\bar{\nu}_\mu$ and $\bar{\nu}_e$
CC events because of isospin breaking 
($m_p\neq m_n$, $m_{\Delta^+} \neq m_{\Delta^0}$, \etc)
and the presence of isolated protons in a water molecule.
However, because the secondary anti-neutrino fluxes are small,
we use the same fudge factors and the smearing functions for
anti-neutrinos,
simply by replacing the CCQE cross sections by those of anti-neutrinos.

The total number of the signal CC events in each bin is now expressed
as
\begin{equation}
 N_{\alpha,D}^{i,{\rm CC}} = 
\varepsilon_{\alpha}
\sum_{X={\rm CCQE},{\rm Res}}
\left[
N_{\alpha,D}^{i,X}(\nu_\mu)+
N_{\alpha,D}^{i,X}(\nu_e) +
N_{\bar{\alpha},D}^{i,X}(\bar{\nu}_\mu)+
N_{\bar{\alpha},D}^{i,X}(\bar{\nu}_e)
\right]
\,,
\label{eq:totalN}
\end{equation}
for $\alpha=\mu$ and $e$, if there are no background.
Here $\effm$ and $\effe$ are the detection efficiencies 
for observing the $\mu^\pm$ or $e^\pm$ signal, respectively.
In actual experiments, there is a small probability of a percent level
that
a $\mu^\pm$ is misidentified as an $e^\pm$ signal,
$P_{e/\mu}$, 
and also the reciprocal probability, $P_{\mu/e}$, of taking
$e^\pm$ as $\mu^\pm$.
In addition,
significant fraction of single $\pi^0$ production events via NC
cannot be distinguished from the $e^\pm$ CCQE signal as explained in
the previous subsection.
After adding those backgrounds the total number of a observed events
can be expressed as
\begin{subequations} %final event numbers
\begin{eqnarray}
 N_{\mu,D}^{i} &=& 
  (1-P_{e/\mu})N_{\mu,D}^{i,{\rm CC}}
+ P_{\mu/e}\cdot N_{e,D}^{i,{\rm CC}}\,,
\label{eq:Nm}\\
 N_{e,D}^{i} &=& 
 P_{e/\mu} \cdot N_{\mu,D}^{i,{\rm CC}}
+(1-P_{\mu/e})N_{e,D}^{i,{\rm CC}}
+ N_{\pi^0,D}^{i,{\rm NC}}\,, 
\label{eq:Ne}
\end{eqnarray}
\label{eq:FinalN}
\end{subequations}
$\!\!\!$
where $N_{\pi^0,D}^{i,{\rm NC}}$ is the event numbers from
the NC $\pi^0$ background in the $i$-th bin.

%---
\begin{figure}[t]
 \centering
\includegraphics[scale=0.62]{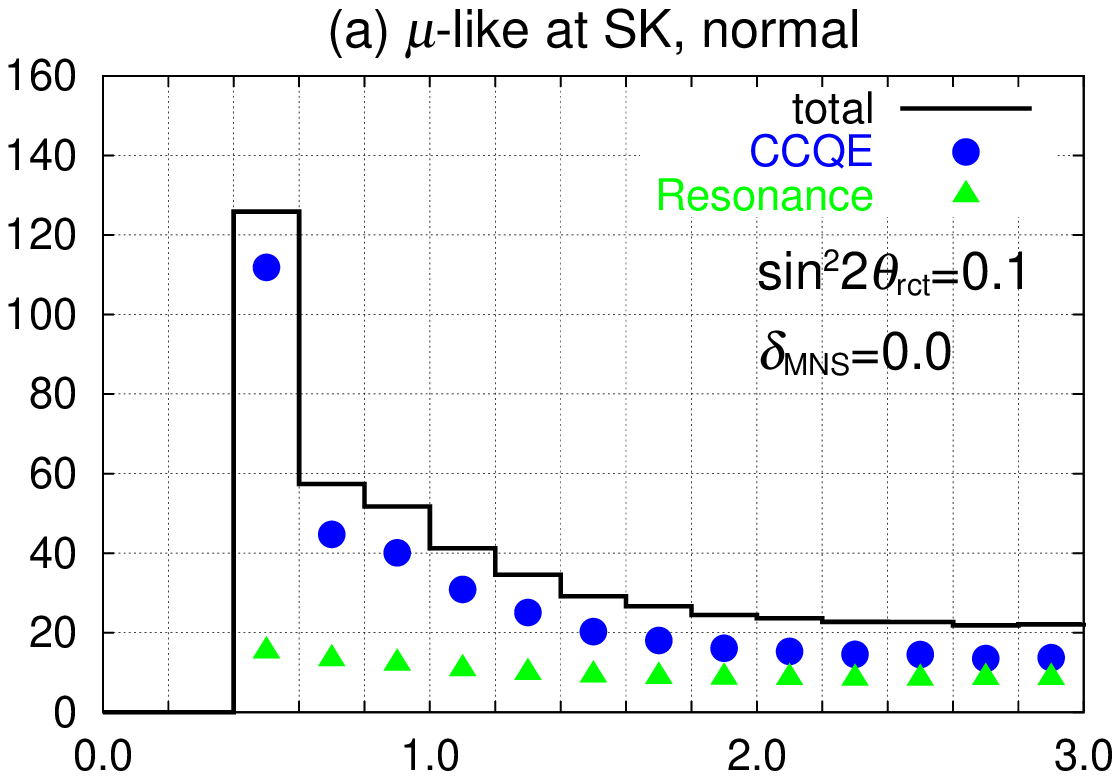}
\hspace*{-3ex}
\includegraphics[scale=0.62]{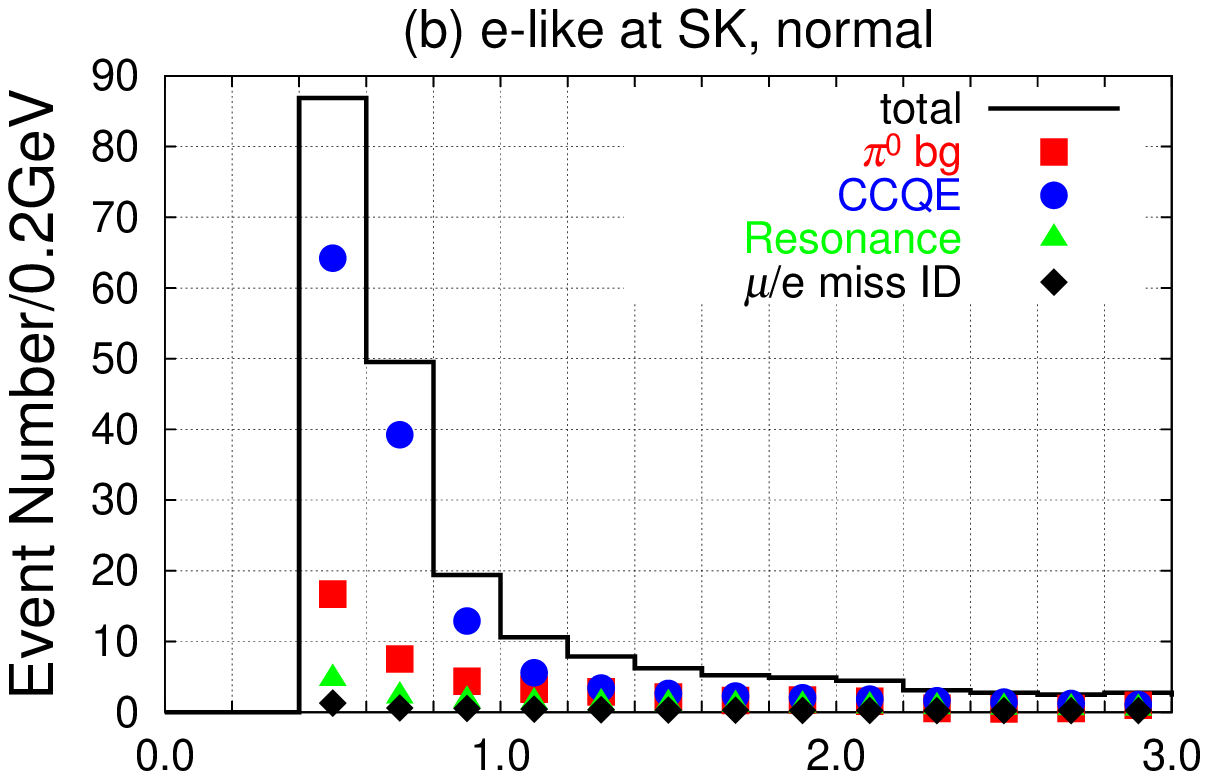}

\includegraphics[scale=0.62]{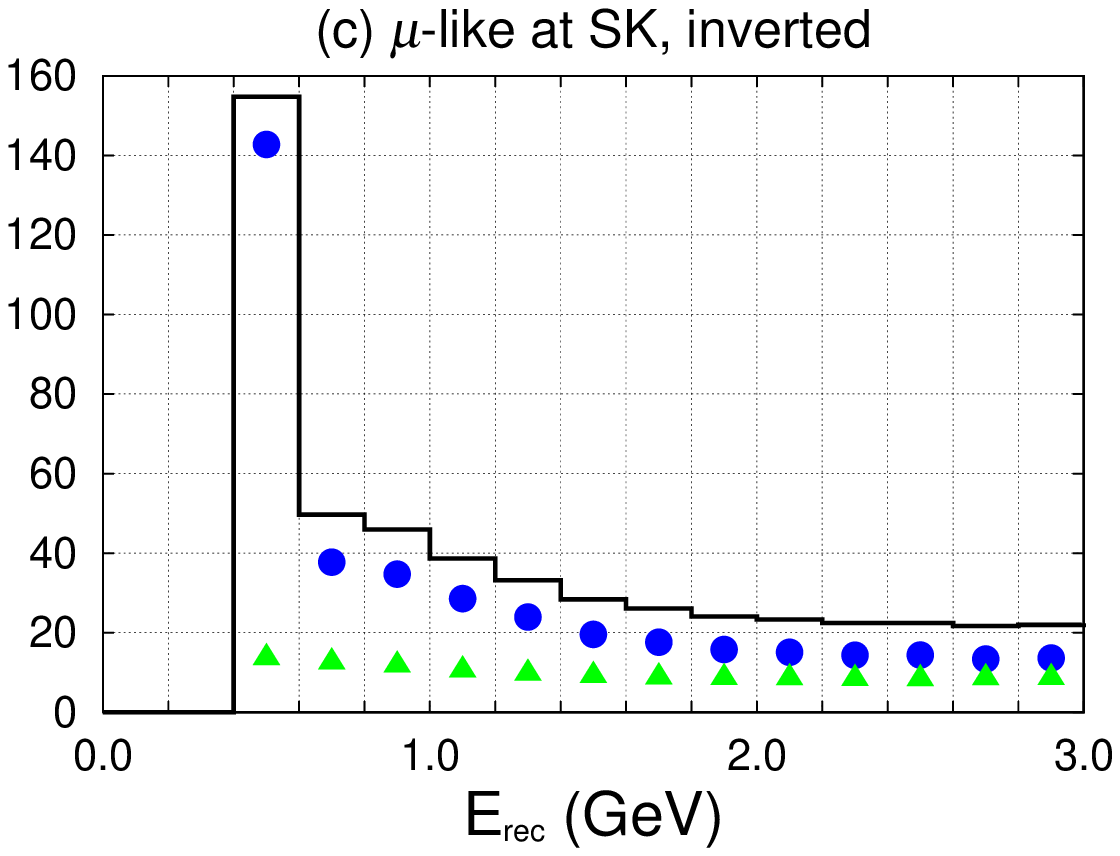}
\hspace*{-3ex}
\includegraphics[scale=0.62]{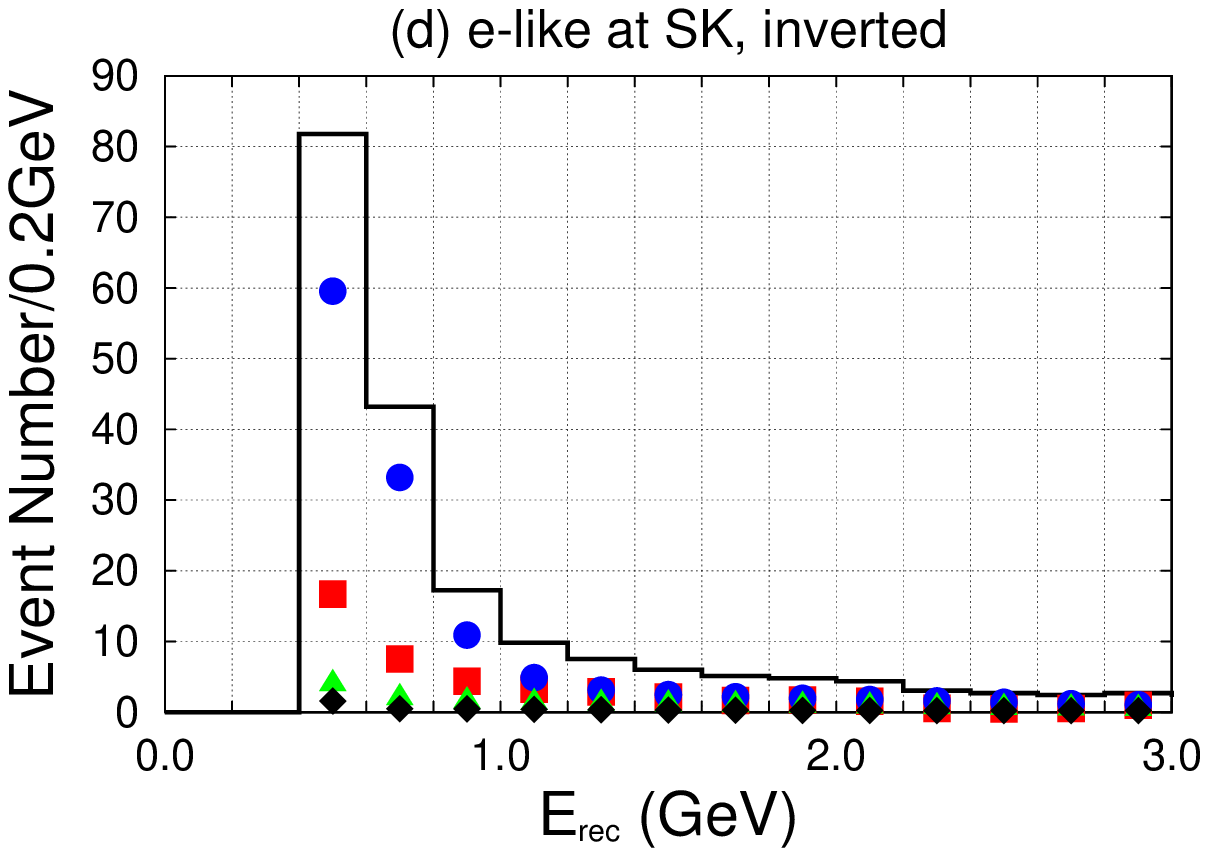}
  \caption{Typical numbers of the 
$\mu$-like events, (a) and (c),
and
$e$-like events, (b) and (d),
for the $3.0^\circ$ OAB at SK with $\numt{5}{21}$POT.
(a) and (b) are for the normal hierarchy,
and
(c) and (d) are for the inverted hierarchy.
The histograms gives the total event numbers,
the circles and the triangles
are the CCQE 
and the ``resonance'' event numbers, respectively.
The squares and diamonds in (a) and (c)
stand for the background event numbers from the misidentified
$\pi^0$ and $\mu^\pm$, respectively.
The inputs are listed in 
eqs.~(\ref{eq:input_physet})-(\ref{eq:eff_and_missID}).
We show only those events with $\Erec>0.4$~GeV used in our analysis.
}
\label{fig:eventSK}
\end{figure}

In Fig.~\ref{fig:eventSK}(a) and (b),
typical $e$- and $\mu$-like event numbers 
with $\numt{5}{21}$POT for the
$3.0^\circ$ OAB at SK is shown,
when the normal hierarchy is assumed.
Figures~\ref{fig:eventSK}(c) and (d)
are
for the inverted hierarchy.
The histogram gives the total event numbers,
and the circles and the triangles give
the CCQE and non-CCQE ``resonance'' event numbers, respectively.
The squares and the diamonds in (b) and (d)
show the background event numbers from the misidentified
$\pi^0$ and $\mu^\pm$, respectively.
Events with $\Erec<0.4$~GeV $(i=0$ and $1)$
are not shown because we do not use them
in our analysis.

The input values of the neutrino mass and mixing parameters
adopted for Fig.~\ref{fig:eventSK} are
\begin{subequations} %Phys Para
\begin{eqnarray}
\left|\dm13\right| &=&
 \numt{2.5}{-3} \mbox{{eV$^2$}}\,,\hspace{5ex}
\satm{} = 0.5\,, 
\\
\dm12 &=&
 \numt{8.2}{-5} \mbox{{eV$^2$}}\,, \hspace{5ex}
\ssun{2} = 0.83\,,  
\\
\srct{2} &=& 0.10\,, \hspace{5ex}
\dmns = 0^\circ\,.
\end{eqnarray}
\label{eq:input_physet}
\end{subequations}
Although the central values of the most recent measurements in
eqs.~(\ref{eq:exp-atm}) and (\ref{eq:exp-sun})
are slightly different,
we use the above values in order to compare our results quantitatively
with those of the previous studies in ref.~\cite{HOS1,HOS2}.

The matter density along the baseline between J-PARC and SK,
and that between J-PARC and the far detector in Korea are 
taken as
\begin{subequations} %matter density
\begin{eqnarray}
\rho_{\rm SK} &=&2.6 \mbox{{g/cm$^3$ for SK}}\,,
\label{eq:input_mden_SK} \\
\rho_{\rm Kr} &=&3.0 \mbox{{g/cm$^3$ for Korea}}\,.
\label{eq:input_mden_Kr}
\end{eqnarray} 
\label{eq:input_mden}
\end{subequations}
%---
$\!\!\!\!$
These average matter densities along the baseline
are obtained~\cite{HOS-mat}
from the recent geophysical measurements~\cite{Jmat,Kmat}
which have typical errors of about $6\%$.
%---
The value for the T2K baseline eq.~(\ref{eq:input_mden_SK}) is
slightly lower than $2.8$g/cm$^3$ quoted in ref.~\cite{T2K},
because of the ``Fossa Magna'' along the baseline,
in which the average density is as low as $2.5$g/cm$^3$.
The average matter density along the baseline
for the far detector in Korea
depends slightly on the baseline length between $L=1000$~km and
$1200$~km,
because it goes through the upper mantle.
Those details as well as the impacts of the matter profile along the
baseline will be reported elsewhere~\cite{HOS-mat}.
%---

%---
Finally, 
the efficiencies for detecting $\mu^\pm$ and $e^\pm$ sinal events 
in eq.~(\ref{eq:totalN})
and the probability of misidentifying
$\mu^\pm$ as $e^\pm$ ($P_{e/\mu}$) and that of misidentifying
$e^\pm$ as $\mu^\pm$ ($P_{\mu/e}$) in eq.~(\ref{eq:FinalN})
are respectively,
\begin{subequations} %effi
\begin{eqnarray}
 \effm &=& 100\%\,, \hspace{5ex}
 \effe = 90\%\,, 
 \label{eq:input_eff}   \\
  P_{e/\mu} &=& 1\%\,, \hspace{7ex}
  P_{\mu/e} = 0\%\,.
\label{eq:input_missID}
\end{eqnarray}
\label{eq:eff_and_missID}
\end{subequations}
$\!\!\!\!$
%---
Hereafter
we set $P_{\mu/e} = 0$ for simplicity, because 
$P_{\mu/e} \sim 1\%$ does not affect our results significantly
due to the smallness of the expected number of $e^\pm$ events.

%-----
The $\nu_\mu$ survival probability is less than $40\%$ in 
the region $0.4$~GeV $<E_\nu<1.0$~GeV,
because of the oscillation dip for $P_{\nu_\mu \to \nu_\mu}$
at $E_\nu\simeq0.6$~GeV.
Nevertheless,
we expect many CCQE events with $\Erec<1.0$~GeV
in Fig.~\ref{fig:eventSK}(a) and (c)
due to the high intensity of the $\nu_\mu$ flux
at $3.0^\circ$ off-axis angle,
which has a peak at $E_\nu \simeq 0.5$~GeV.
It catches our eyes that
the $\mu$-like event rate in the first bin 
(0.4~GeV$\leq\Erec\leq$0.6~GeV)
is significantly larger for the inverted hierarchy
than for the normal hierarchy.
This is because the oscillation phase shift, 
the factor $B^\mu$ in eqs.~(\ref{eq:Pmm}) and (\ref{eq:Bmu}),
is negative for the parameters of
eq.~(\ref{eq:input_physet})
so that the location of the dip occurs at slightly higher
$E_\nu$ for the inverted hierarchy.
%---
Such small difference in the dip location between 
the two hierarchies, however, can be compensated by a
small shift in $|\dm13|$ of several percent order.
This in turn tells that $|\dm13|$ cannot be measured beyond the
accuracy of several percent unless the mass hierarchy pattern is
determined;
see discussions in section~\ref{sec:dm13} for more details.

%------
Typical $e$-like events at SK are shown in 
Figs.~\ref{fig:eventSK}(b) and (d).
The CCQE events dominate the $e$-like events 
for both mass hierarchies.
Because there is little high energy tail for the $3.0^\circ$ OAB and
the probability of misidentifying $\pi^0$ as $e^\pm$
is not large at $E_\nu<1.0$~GeV,
as can be seen from Figs.~\ref{fig:pi0pt}(a) and (b), respectively,
the $\pi^0$ background events 
given by the squares do not dominate over the CCQE signal events.
Nevertheless, they consist of about $20\%$ of the total number of
$e$-like events in the first three bins of $E_\nu<1.0$~GeV.
Quantitative estimate of the $\pi^0$ background should hence
be essential to measure the $\nu_\mu\to\nu_e$ transition probability
with confidence.

\begin{figure}[t]
 \centering
 \includegraphics[scale=0.62]{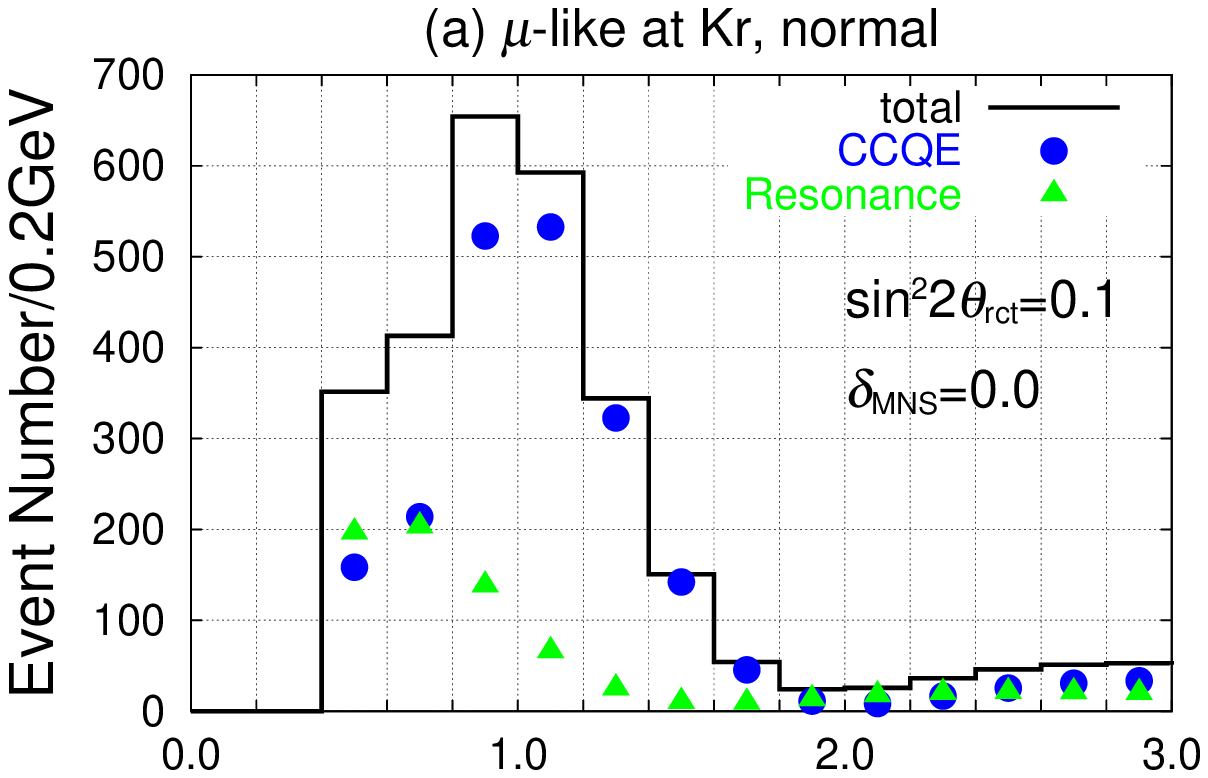}
 \hspace*{-3ex}
 \includegraphics[scale=0.62]{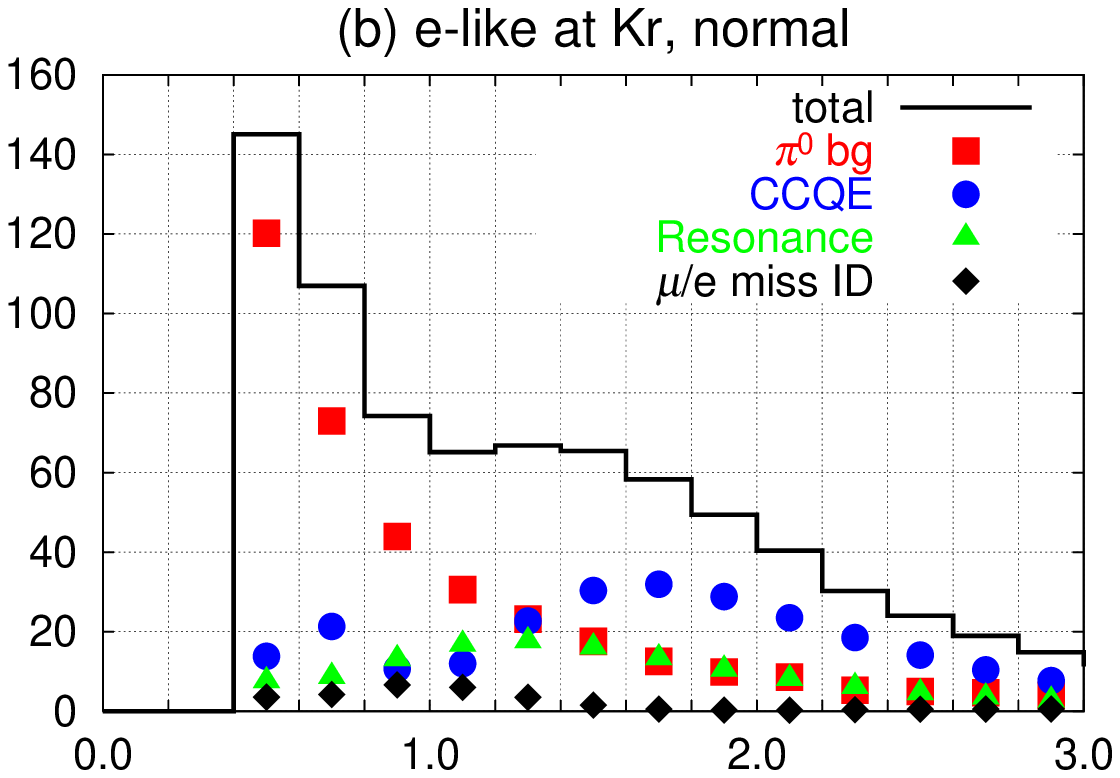}
 
 \includegraphics[scale=0.62]{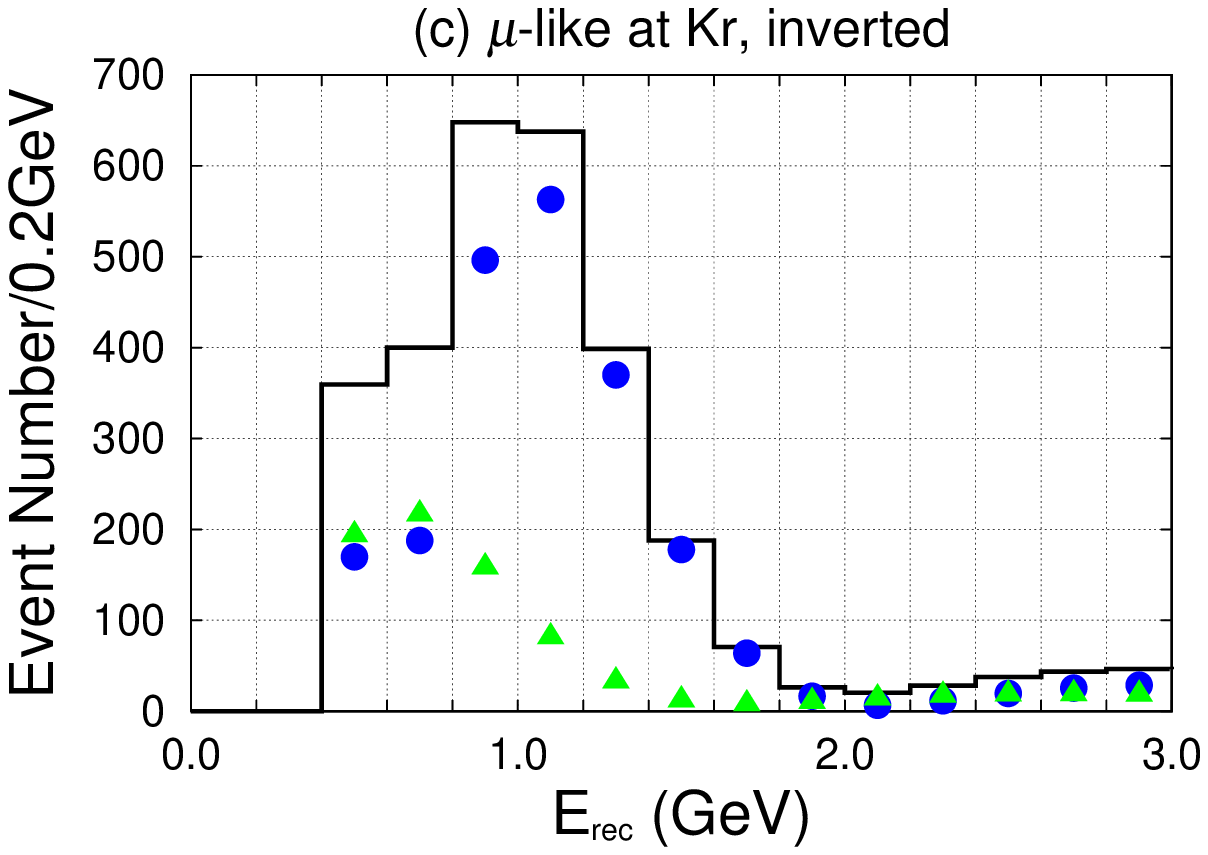}
 \hspace*{-3ex}
 \includegraphics[scale=0.62]{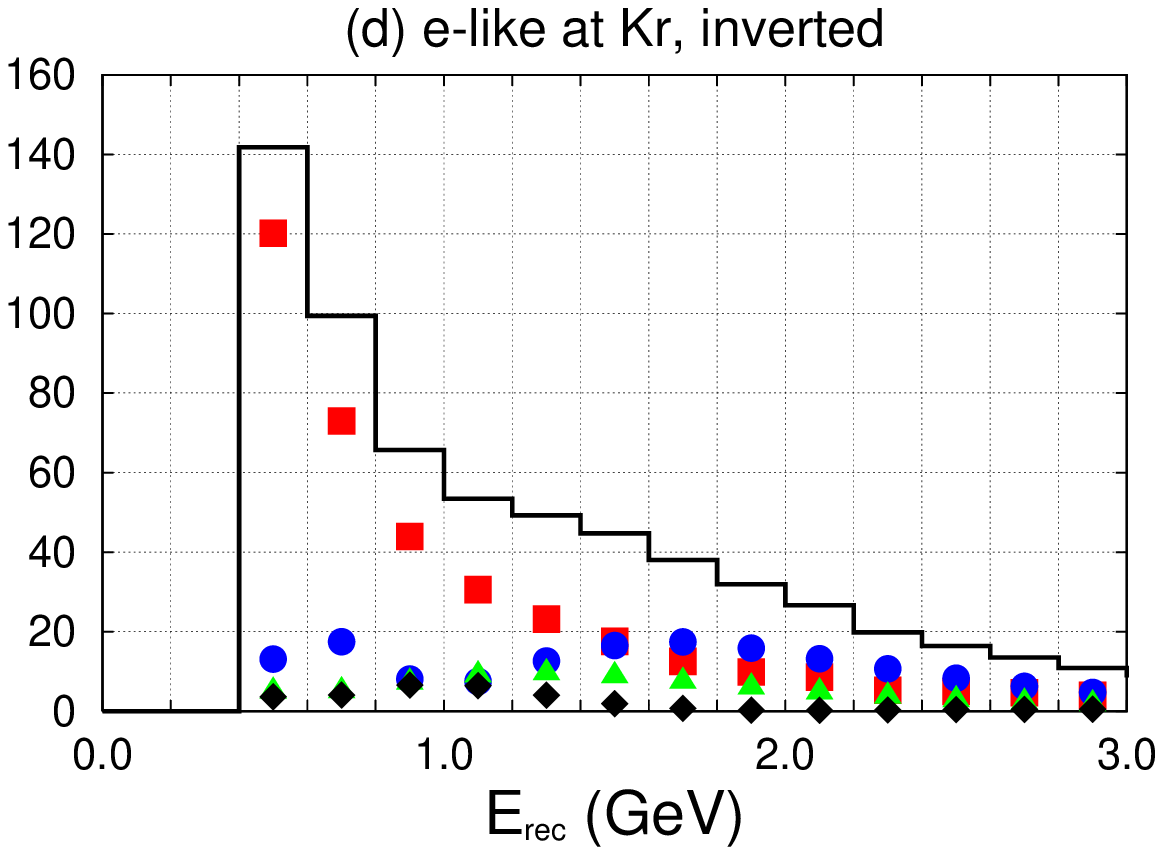}
  \caption{The same as Fig.~\ref{fig:eventSK},
but for the $0.5^\circ$ OAB at $L=1000$~km
with a 100~kton water \cerenkov
detector.}
\label{fig:eventKr}
\end{figure}

In Fig.~\ref{fig:eventKr}, we show the $\Erec$ distributions of the
$\mu$-like and $e$-like events expected for a 100~kton far detector
at $L=1000$~km and with the $0.5^\circ$ OAB,
for exactly the same model parameters of eq.~(\ref{eq:input_physet})
and the systematics of eq.~(\ref{eq:eff_and_missID}),
but with the average matter density of eq.~(\ref{eq:input_mden_Kr}).

The $\Erec$ distributions of the $\mu$-like events are shown for the
normal and inverted hierarchy in Figs.~\ref{fig:eventKr}(a) and (b),
respectively,
where little dependence on the mass hierarchy pattern can be observed.
The $\nu_\mu \to \nu_\mu$ oscillation dip at $E_\nu \sim 2.0$~GeV
is clearly seen in both cases,
despite the contribution from the non-CCQE ``resonance'' events
shown by the triangles,
which has a dip at lower $\Erec$.

What is most surprising in Fig.~\ref{fig:eventKr} is the overwhelmingly
large contribution of the $\pi^0$ background events,
shown by the squares,
in the $e$-like event distributions,
both in (b) and (d), respectively, for the normal and the inverted
hierarchies.
They dominate the CCQE signal at low $\Erec$, $\Erec < 1.4$~GeV for
the normal hierarchy and $\Erec<1.6$~GeV for the inverted hierarchy.
This is essentially because of the hard energy (broad band) spectrum
of the $0.5^\circ$ OAB,
which gives rise to copious production of single $\pi^0$ events
via the NC.
Nevertheless, 
the CCQE event numbers supersede the $\pi^0$ background
at high $\Erec$, $\Erec > 1.4$~GeV for the normal hierarchy,
and $\Erec$, $\Erec>1.6$~GeV for the inverted hierarchy.
The significant difference in the $\Erec$ distributions of the
$e$-like events expected at a far detector,
between Figs.~\ref{fig:eventKr}(b) and (d),
in contrast to the similarity of the corresponding distributions at
SK,
between Figs.~\ref{fig:eventSK}(b) and (d),
may allow us to determine the neutrino mass hierarchy even in the
presence of the $\pi^0$ background,
since the $\pi^0$ background due to the NC events
do not depend on the mass hierarchy.
The non-CCQE ``resonance'' events, shown by the triangle,
behave similarly to the CCQE signal events;
the number of events is enhanced for the normal hierarchy and
suppressed for the inverted hierarchy.
Therefore, we expect that the contribution from the ``resonance''
events will enhance the sensitivity of the T2KK experiment to the mass
hierarchy.

% ======================================================================== %
\section{Analysis Method}
\label{sec:4}
In order to quantify the physics potential of 
the T2KK neutrino oscillation experiment,
we introduce a $\chi^2$ function
\begin{equation}
\chi^2 \equiv
 \chi^2_{\rm SK}
+ \chi^2_{\rm Kr}
+ \chi^2_{\rm sys}
+ \chi^2_{\rm para}\,,
\label{eq:def_chi^2}
\end{equation}
which measures the sensitivity of the expected measurements
on the physics parameters such as the neutrino mass hierarchy,
$\srct{2}$ and $\dmns$,
in the presence of statistical errors as well as
various systematic errors including the uncertainties in the other
parameters of the three neutrino model.
%---

%---
The first two terms in eq.~(\ref{eq:def_chi^2}), 
$\chi^2_{\rm SK}$ and $\chi^2_{\rm Kr}$, respectively,
measure the constraints from the measurements
at SK and a far detector in Korea;
\begin{eqnarray}
 \chi^2_{D}
 = \sum_{i} \left\{
\left(
\dfrac
{(N_{\mu,D}^{i})^{\rm fit} - (N_{\mu,D}^{i})^{\rm input}}
{\sqrt{(N^i_{\mu,D})^{\rm input}}}
\right)^2
+
\left(
\dfrac
{(N_{e,D}^{i})^{\rm fit} - (N_{e,D}^{i})^{\rm input}}
{\sqrt{(N^i_{e,D})^{\rm input}}}
\right)^2
\right\}\,,(D={\rm SK, Kr})\,.
\label{eq:chi_N}
\end{eqnarray}
Here
$(N_{\mu,D}^{i})^{\rm input}$ and 
$(N_{e,D}^{i})^{\rm input}$
denotes the $\mu$-like and $e$-like event numbers,
respectively, at SK ($D=$ SK)
and at a far detector in Korea ($D=$ Kr),
in the $i$-th bin of $\Erec$
calculated as in eq.~(\ref{eq:N})-(\ref{eq:FinalN}),
and its square root gives the statistical error.
The summation is over all bins
from $0.4$~GeV to $5.0$~GeV at both detectors for $N_{\mu}$,
$0.4$~GeV to $1.2$~GeV at SK,
and
$0.4$~GeV to $2.8$~GeV at Korea for $N_e$.
In order to compare our results quantitatively with those of the
previous studies in ref.~\cite{HOS1,HOS2,HO-atm},
we use
the same input values of the neutrino model parameters, 
as in eq.~(\ref{eq:input_physet}),
when calculating the expected number of events in each bin.

The event numbers for the fit,
$(N_{\mu,D}^{i})^{\rm fit}$
and
$(N_{e,D}^{i})^{\rm fit}$
are calculated as
\begin{subequations} %Nfit
 \begin{eqnarray}
\left(N_{\mu,D}^{i}\right)^{\rm fit}\!\!\!&=&\!\!\! 
f_{\rm V}^{D}
\left[
\left(
1-P_{e/\mu}
\right)
  \sum_{X,\alpha,\beta}
  \effm 
  f_{\nu_\alpha}^{D}
  f_\beta^{X}
  N_{\mu,D}^{i,X}(\nu_\alpha)
\right]\,,\label{eq:Nfitm}
\\
\left(N_{e,D}^{i}\right)^{\rm fit}\!\!\!&=&\!\!\!
f_{\rm V}^{D}
\left[
  \sum_{X,\alpha,\beta}
  \left\{
   \effe 
   f_{\nu_\alpha}^{D}
   f_\beta^{X}
   N_{e,D}^{i,X}(\nu_\alpha)
   +
   P_{e/\mu}
   f_{\nu_\alpha}^{D}
   f_\beta^{X}
   N_{\mu,D}^{i,X}(\nu_\alpha)
	\right\}
  +
  f_{\nu_\mu}^{D}
  f_{\pi^0}^{}
  N^{i,NC}_{\pi^0,D}
\right]\,,~~~~~~~~
\label{eq:Nfite}
 \end{eqnarray}
\label{eq:Nfit}
\end{subequations}
$\!\!\!\!$
where 
the initial neutrino flavor,
$\nu_\mu$, $\bar{\nu}_\mu$, $\nu_e$, $\bar{\nu}_e$,
are denoted as $\nu_\alpha$ with
$\alpha=\mu$, $\bar{\mu}$, $e$, $\bar{e}$, respectively,
and 
the superscript $X$ denotes the event type, 
$X=$ CCQE for the signal, or
$X=$ Res for the non-CCQE ``resonance'' events
that pass the CCQE selection criteria of eq.~(\ref{eq:criteriaCC}).
The subscript $\beta$ distinguishes neutrinos
($\beta=\nu$ for $\nu_\mu$ or $\nu_e$)
and anti-neutrino ($\beta = \bar{\nu}$ for $\nu_\mu$ or $\nu_e$),
while $D=$ SK or $D=$ Kr as in eq.~(\ref{eq:chi_N}).
We introduce 17 normalization factors whose deviation from
unity measures systematic uncertainties,
15 of which appear explicitly in eq.~(\ref{eq:Nfit});
$f_V^D$ for the fiducial volume and
$f_{\nu_\alpha}^D$ for the initial neutrino flux 
at $D=$ SK and $D=$ Kr,
$f_\beta^X$ for the CC cross section of $X=$ CCQE or $X=$ Res
with neutrino ($\beta=\nu$) or anti-neutrino ($\beta=\bar{\nu}$),
and
$f_{\pi^0}$ for the NC cross section of producing
the single $\pi^0$ background.
In addition the factor $f_\rho^D$ takes account of the uncertainty in
the average matter density along the baseline between J-PARC
and SK ($D=$ SK) or Korea ($D=$ Kr),
which appear in the computation of the oscillation probability
$P_{\nu_\alpha \to \nu_\beta}$ by modifying the
matter density as 
\begin{eqnarray}
\rho_{D}^{\rm fit} &=& f^{D}_{\rho}\,\rho^{\rm input}_D \,,
\hspace{5ex}
(D = {\rm SK,~Kr}) \,.
\label{eq:sys_matt}
\end{eqnarray}

By using the above 17 normalization factors, 
the detection efficiencies ($\effe$ and $\effm$)
and 
the $\mu$-to-$e$ misidentification probability ($P_{e/\mu}$),
we estimate the systematic effects as follows;
\begin{eqnarray}
 \chi^2_{\rm sys} &=& 
\sum_{ D = {\rm SK,~Kr}}
\left\{
\left(
\dfrac{f^{D}_{\rm V}-1}{0.03}
\right)^2
+
\left(
\dfrac{f^{D}_{\rho}-1}{0.06}
\right)^2
+
\sum_{\alpha = e,\bar{e},\mu,\bar{\mu}}
\left(
\dfrac{f_{\nu_{\alpha}}^D-1}{0.03}
\right)^2
\right\} 
\nn\\
&&
+
\sum_{\beta = \nu, \bar{\nu}}
\left\{
\left(
\dfrac{f^{\rm CCQE}_{\beta}-1}{0.03}
\right)^2
+
\left(
\dfrac{f^{\rm Res}_{\beta}-1}{0.20}
\right)^2
\right\}
+
\left(
\dfrac{f_{\pi^0}-1}{0.50}
\right)^2 
\nn\\
&&
+
\left(
\dfrac{\effe-0.9}{0.05}
\right)^2 
+
\left(
\dfrac{\effm-1}{0.01}
\right)^2
+
\left(
\dfrac{P_{e/\mu}-0.01}{0.01}
\right)^2
\,.~~
\label{eq:sys_chi2}
\end{eqnarray}
All the errors in the first row of eq.~(\ref{eq:sys_chi2})
depend on the detector and its location,
$D=$ SK and $D=$ Kr.
The first term is the uncertainty of the fiducial volume,
for which we assign $3\%$ error independently for SK
($f_{\rm V}^{\rm SK}$) and 
a far detector in Korea ($f_{\rm V}^{\rm Kr}$).
The second one is for the matter density uncertainties along the
T2K ($f^{\rm SK}_{\rho}$) and
the Tokai-to-Korea ($f^{\rm Kr}_{\rho}$) baseline.
The dominant source of the error in the matter density
arises when the sound velocity data
are translated into the matter density~\cite{HOS-mat,dvconv},
and we assign 6$\%$ error independently for each baseline.
The last term of the first row is 
for the overall normalization of each neutrino flux,
which are taken independently for each 
neutrino species and the detector location.
This is a conservative estimate,
since it is likely that all the flux normalization
errors are positively correlated.
The second row gives the uncertainty in
the cross sections.
Because the CCQE cross section for $\nu_e$ and $\nu_\mu$ are expected
to be very similar theoretically, 
we assign a common overall error of $3\%$ for $\nu_e$ and $\nu_{\mu}$ 
($f_{\nu}^{\rm CCQE}$)
and an independent $3\%$ error for $\bar{\nu}_e$ and $\bar{\nu}_\mu$
($f_{\bar{\nu}}^{\rm CCQE}$).
For non-CCQE ``resonance'' events ($f_{\beta}^{\rm Res}$),
we assume $20\%$ error for $\beta=\nu$ and $\beta=\bar{\nu}$
independently,
since it depends not only on the single $\pi$ production cross section
but also on the momentum distribution and the detector performance.
We allow $50\%$ error for the NC cross section
of producing single $\pi^0$ background ($f_{\pi^0}$),
since it takes account of the uncertainty in the $\pi^0$-to-$e$
misidentification probability ($P_{\pi^0/e}$).
The systematic errors in the last row of eq.~(\ref{eq:sys_chi2}) 
account for the performance of a water \cerenkov detector.
The first and the second terms denote
the uncertainty of the detection efficiency
for $e$- and $\mu$-like events, respectively.
In this analysis, we adopt $\delta \effe=5\%$ and $\delta \effm=1\%$,
which are taken common for SK and a far detector in Korea.
The last one is the probability of misidentifying a $\mu$-event as an
$e$-event, for which a common error of 1$\%$ is assumed.
In total, we adopt 20 parameters in simulating the systematic errors.

Finally, $\chi^2_{\rm para}$ accounts for external constraints 
on the model parameters:
\begin{eqnarray}
\label{eq:para_chi2}
\chi^2_{\rm para}
&=&
\left(
\dfrac{\left( \dm12 \right)^{\rm fit}-
\numt{8.2}{-5}\mbox{{eV}$^2$}}
{\numt{0.6}{-5}}
\right)^2
+
\l(\dfrac{\ssun{2}^{\rm fit}-0.83}{0.07}\r)^2 
\nn\\
&&+
\l(\dfrac{\srct{2}^{\rm fit}- \srct{2}^{\rm input}}{0.01}\r)^2\,.
\label{eq:chi-para}
\end{eqnarray}
Although the errors of the smaller mass-squared difference
and the solar mixing angle
in eq.~(\ref{eq:chi-para}) are somewhat larger than
their most recent values in eq.~(\ref{eq:exp-sun}),
we stick to the above estimates in order to compare our results
quantitatively with those of the previous studies in
ref.~\cite{HOS1,HOS2,HO-atm}.
In the last term,
we assume that the planned future reactor experiments
\cite{DoubleCHOOZ,DayaBay,reno}
will measure $\srct{2}$ with the uncertainty of 0.01.

% ======================================================================== %
\section{Mass hierarchy}
\label{sec:5}

In this section, we study the sensitivity of the T2KK experiment 
on the neutrino mass hierarchy.
First, we look for the best combination of the off-axis angle at SK
and 
the location of a far detector in Korea,
which can be parameterized in terms of the baseline length $L$
and the off axis angle from the beam center.
Second, we examine carefully the impacts of the systematic errors,
including the contribution from the uncertainty 
in the $\pi^0$ background.
In subsection~\ref{sec:dependence-oab-sk},
we show the sensitivity of the T2KK experiment 
on the neutrino mass hierarchy,
as contour plots on the plane of $\srct{2}$ and $\dmns$.
In last subsection,
we show the impacts of the mass hierarchy uncertainty
on the measurement of $|\dm13|$.

\subsection{The best combination}
\label{sec:best-combination}

Here we repeat the analysis of ref.~\cite{HOS1,HOS2} in which
the combination of the off-axis angle at SK and the location of a far
detector in Korea that maximizes the sensitivity to the neutrino mass
hierarchy has been looked for, by assuming a water \cerenkov detector of
100~kton fiducial volume at a distance between $L=1000$~km and
$1200$~km from J-PARC.
It should be noted here that because 
the detector should be placed on the earth surface, the allowed range of
the off-axis angle at a far detector depends on the off-axis angle
at SK.
For instance, the off-axis angle observable in Korea is larger than
$0.5^\circ$ for the $3.0^\circ$ OAB at SK,
while it is larger than $1.0^\circ$ for the $2.5^\circ$ OAB at SK.

%-----------------------------------------------------------------
% Position select 3.0@SK
%-----------------------
\begin{figure}
 \centering
(a) normal hierarchy (OAB:3.0@SK)

 \includegraphics[width=0.24\textwidth]{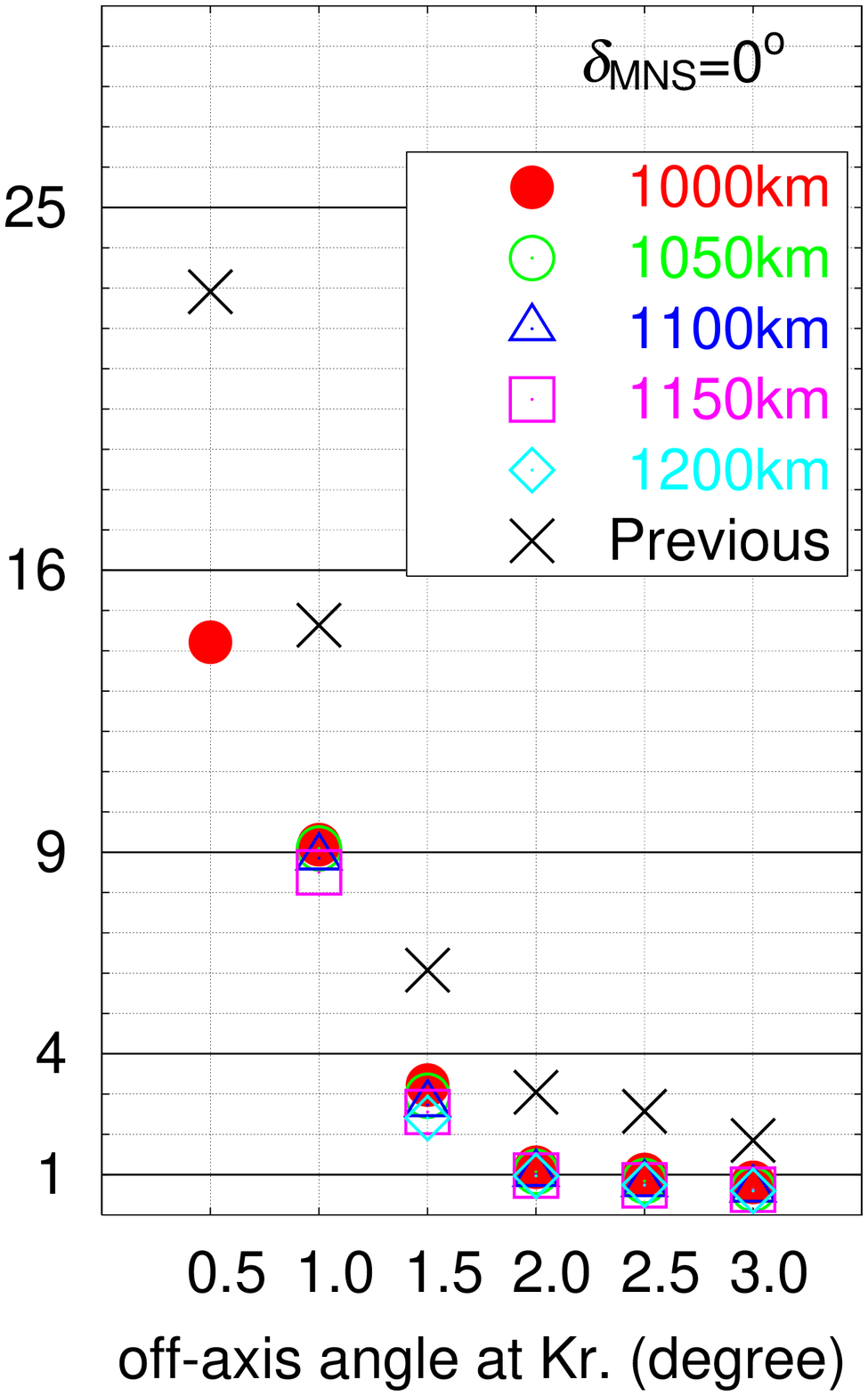}
 \includegraphics[width=0.24\textwidth]{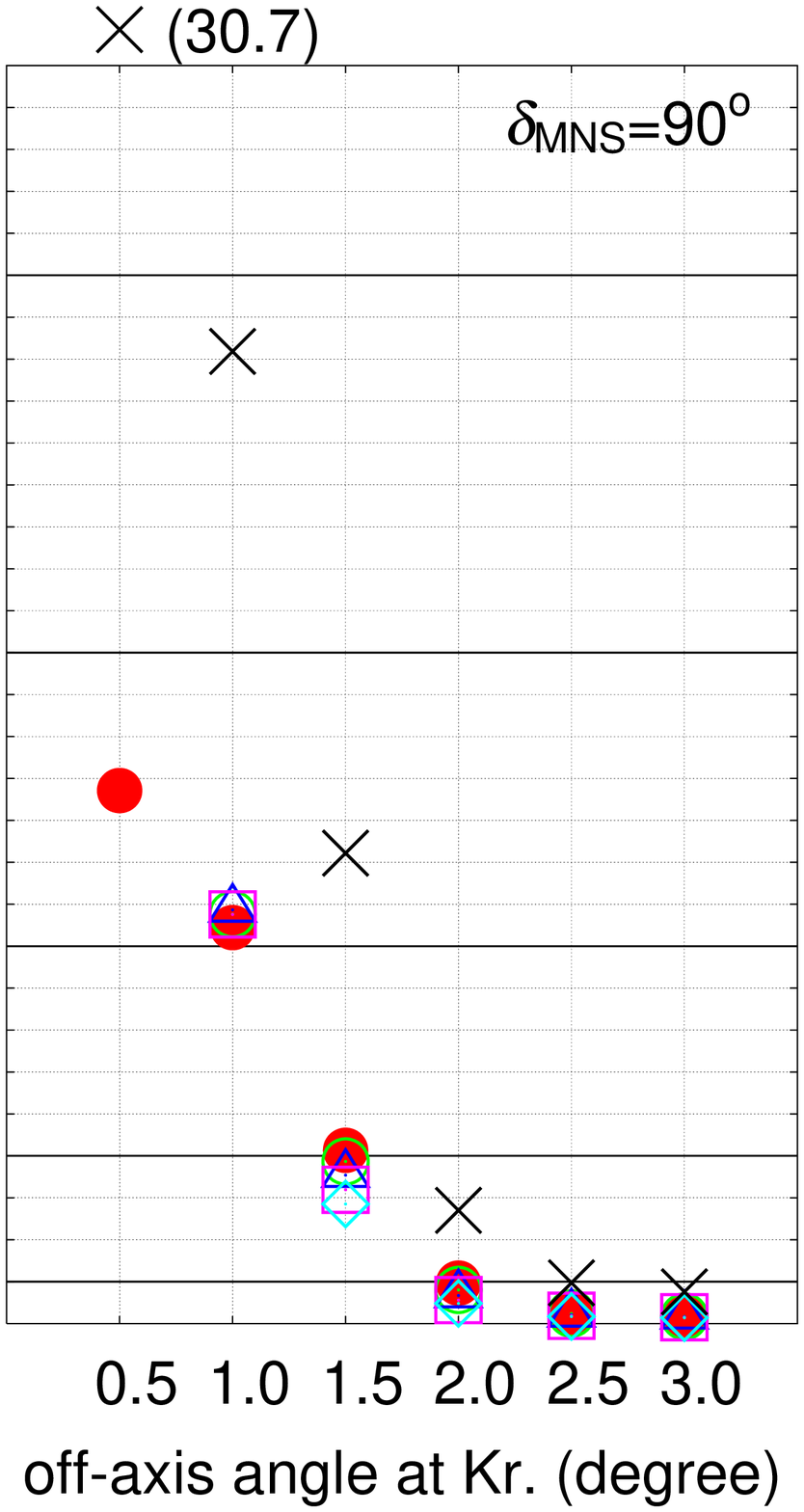}
 \includegraphics[width=0.24\textwidth]{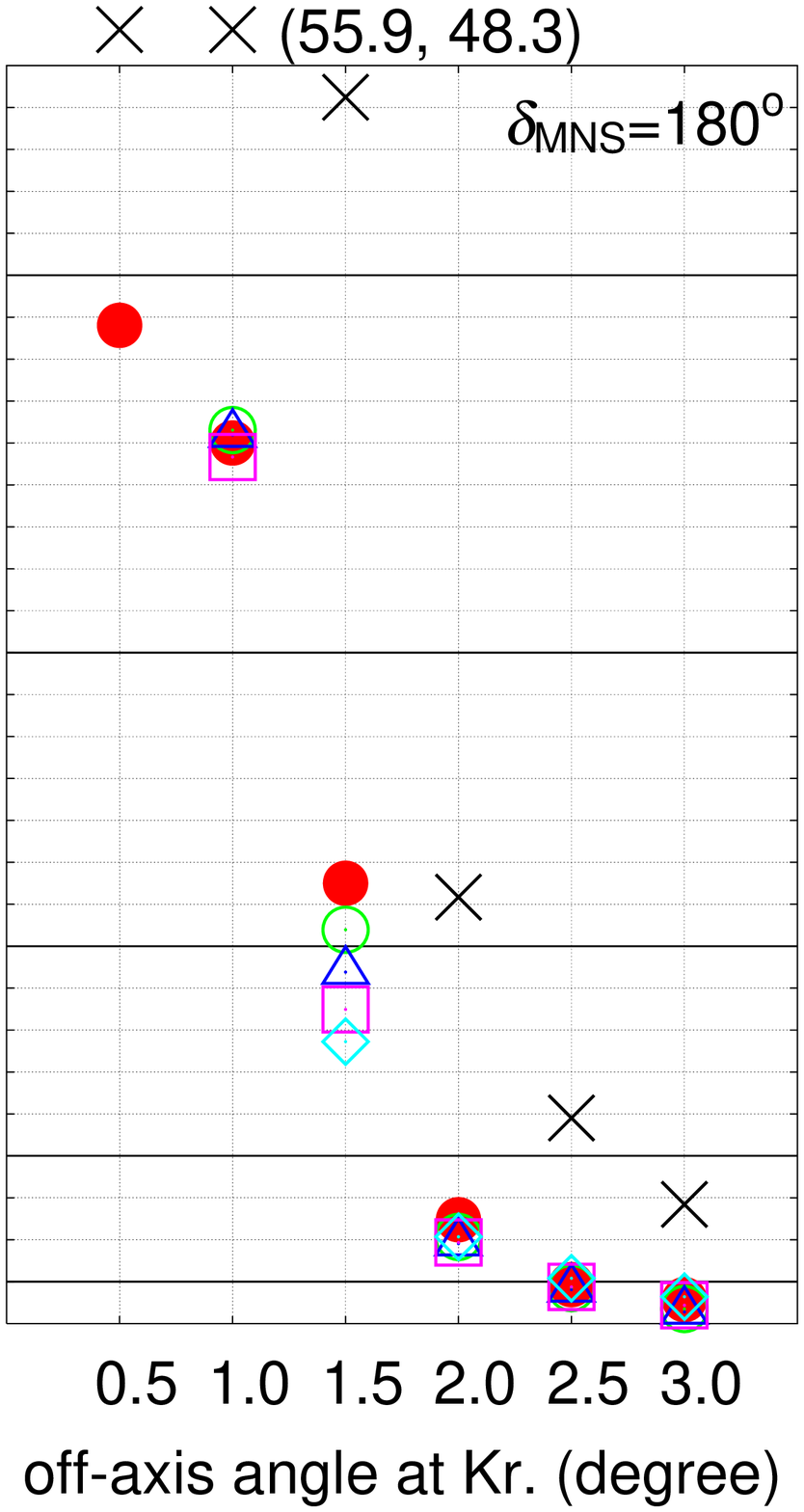}
 \includegraphics[width=0.24\textwidth]{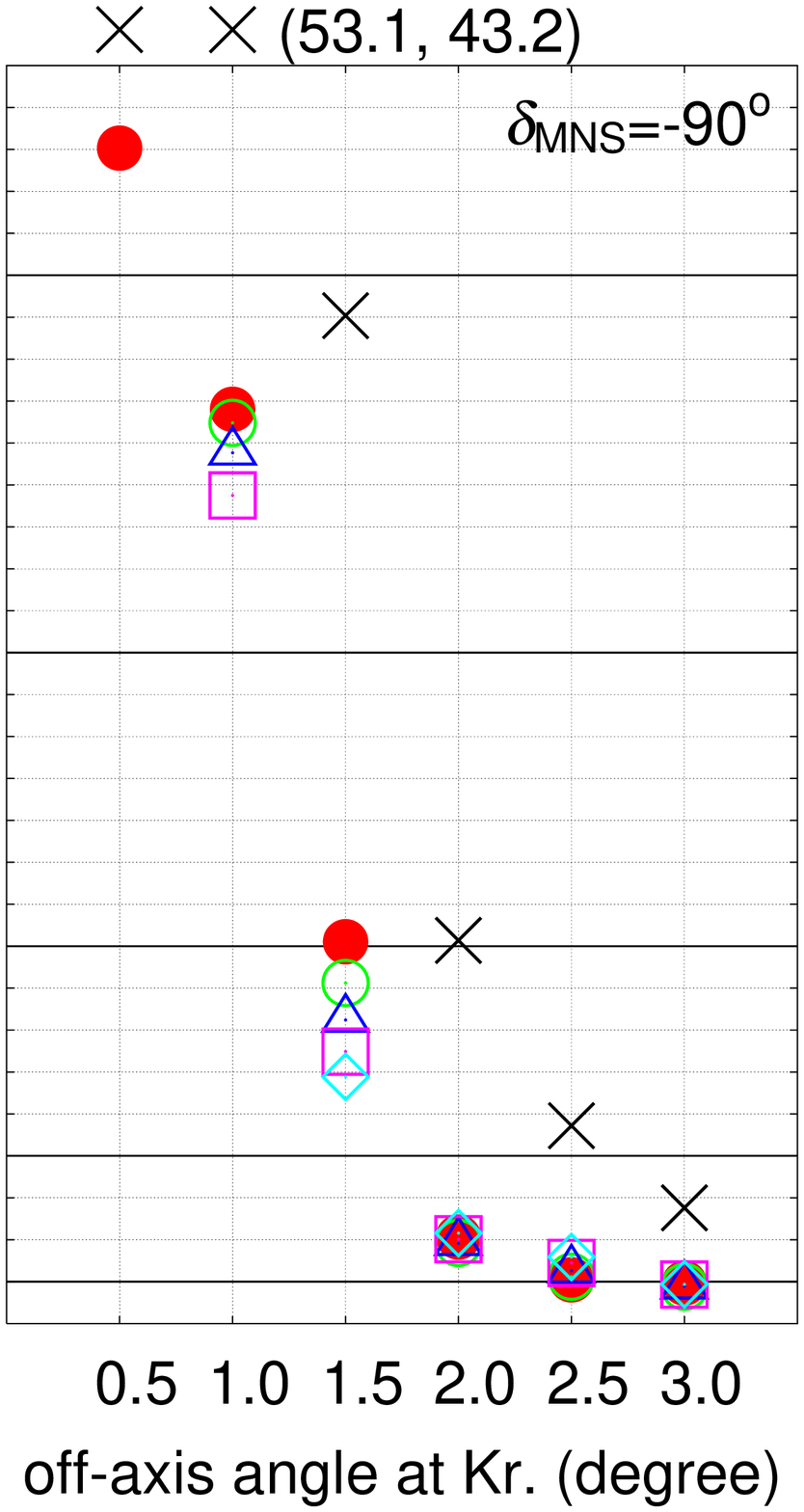}

\bigskip

(b) inverted hierarchy (OAB:3.0@SK)

 \includegraphics[width=0.24\textwidth]{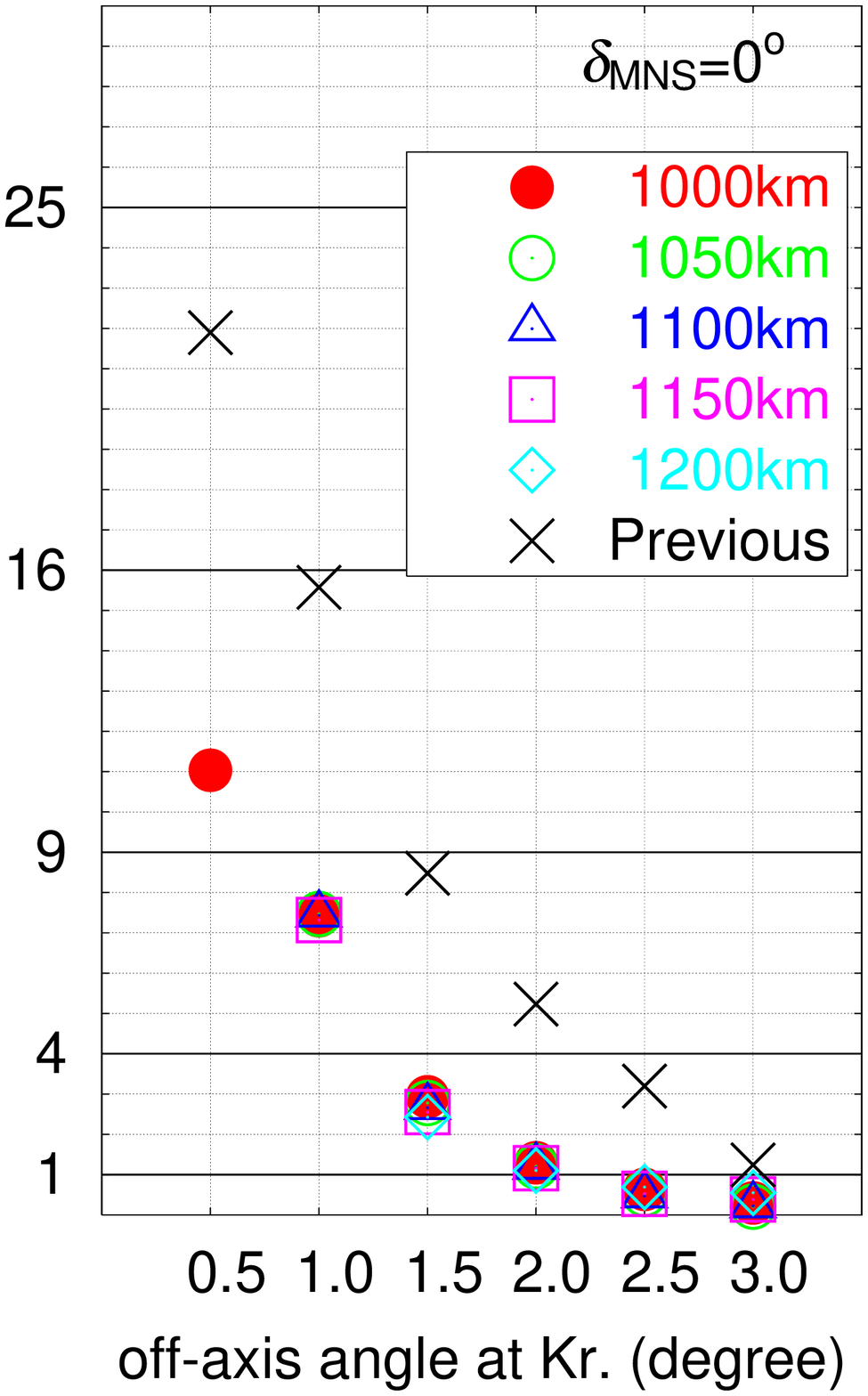}
 \includegraphics[width=0.24\textwidth]{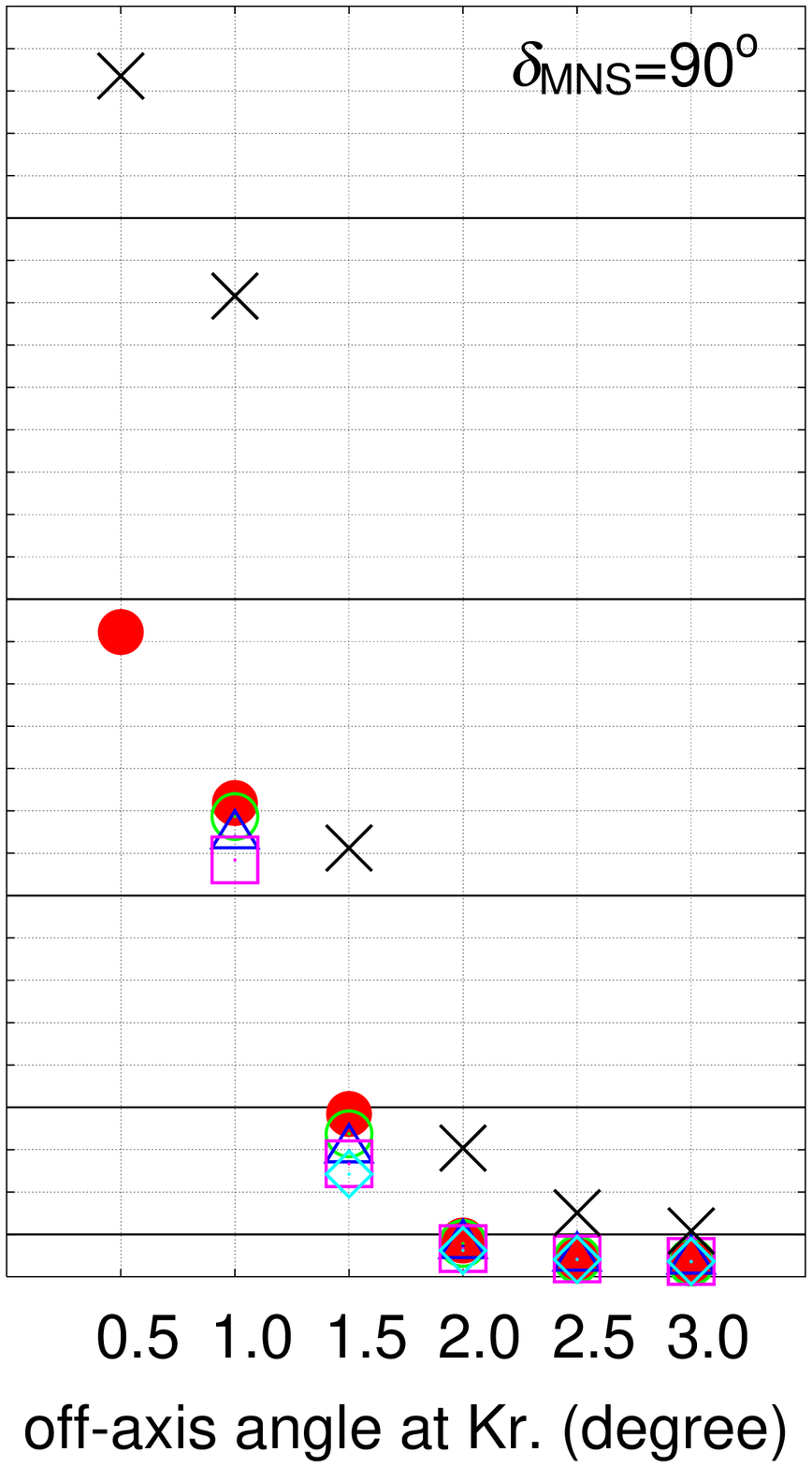}
 \includegraphics[width=0.24\textwidth]{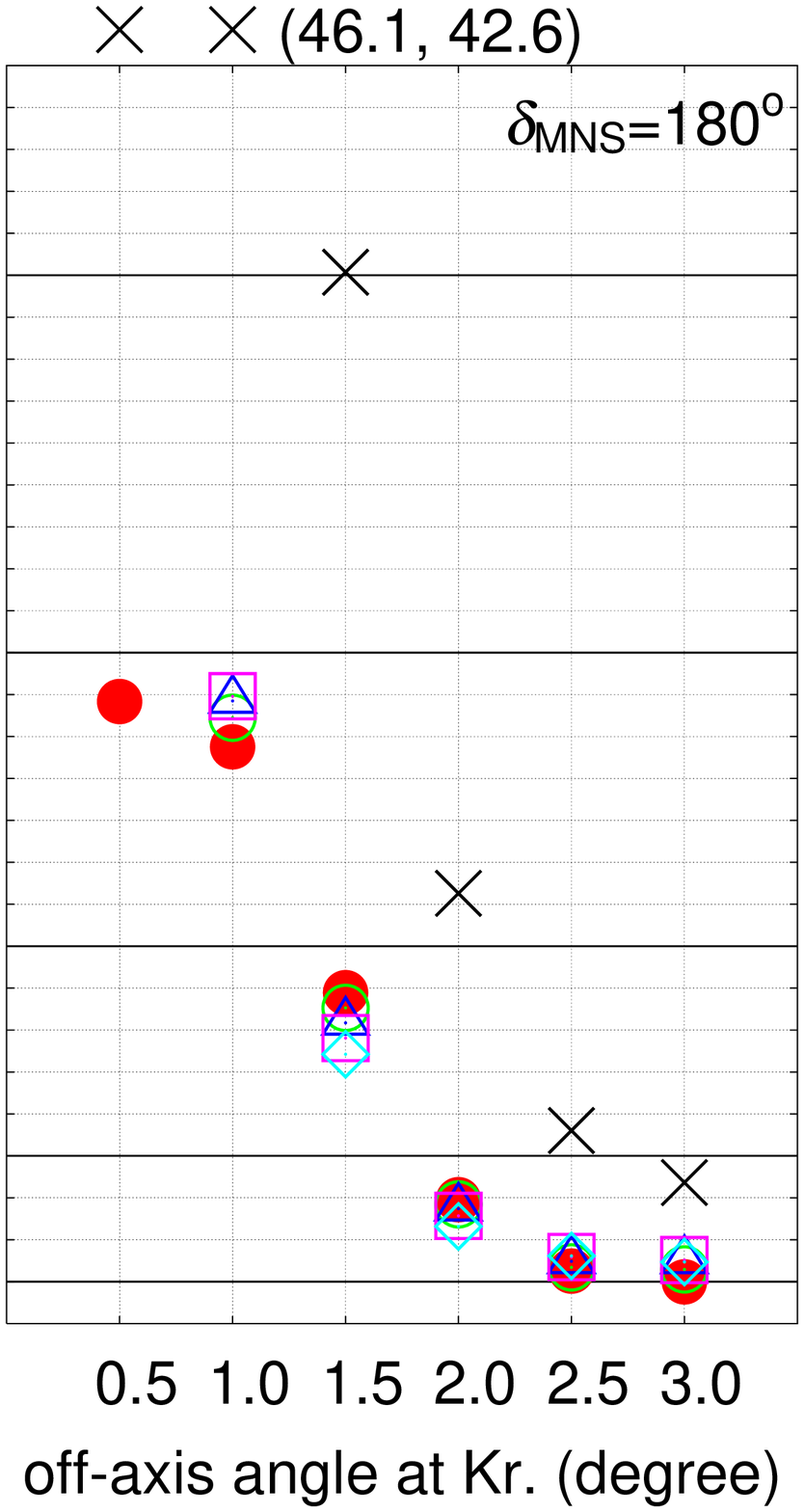}
 \includegraphics[width=0.24\textwidth]{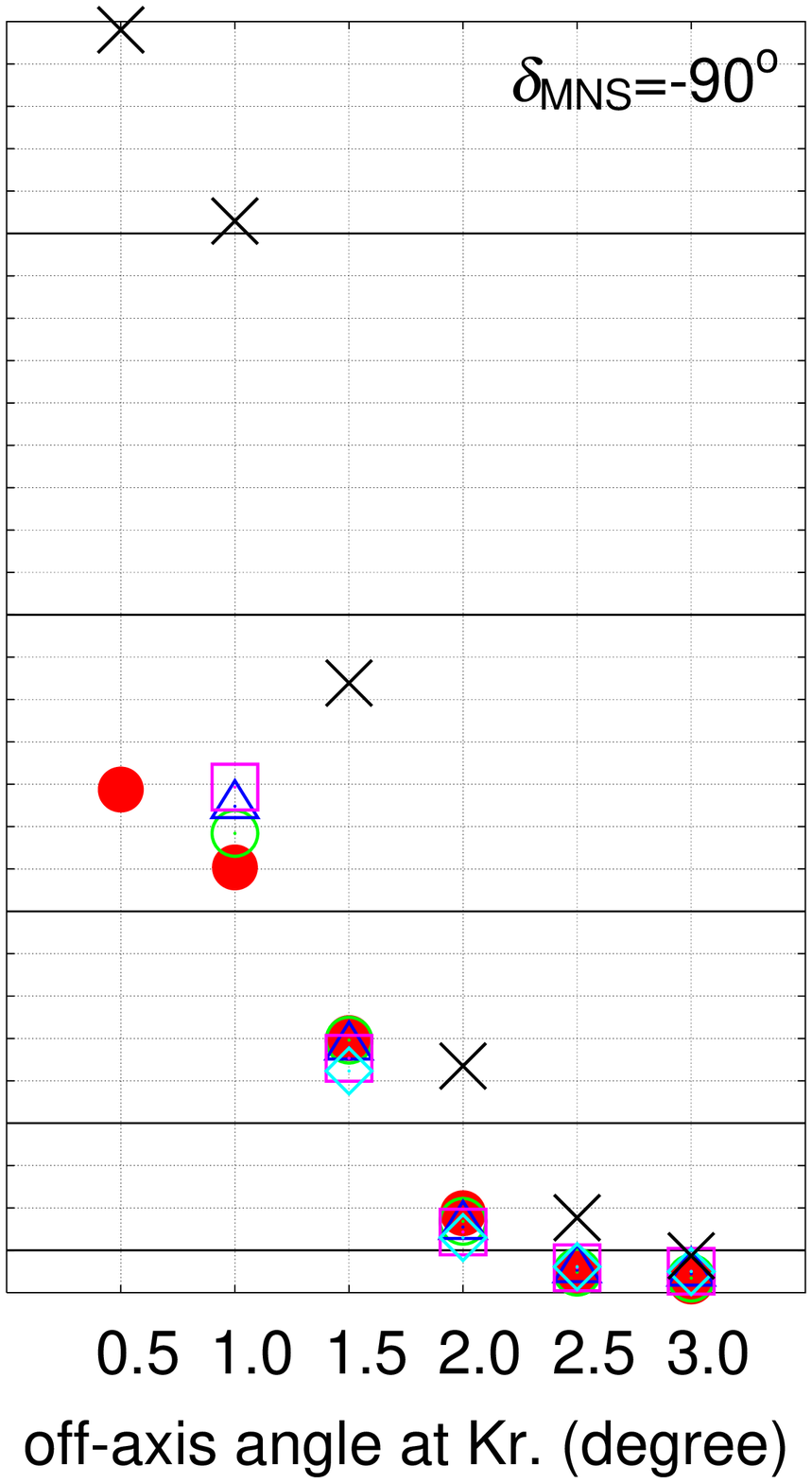}
 \caption{
(a):
Minimum $\Delta\chi^2$ of the T2KK experiment
as a function of the off-axis angle and the baseline length ($L$)
of a far detector from J-PARC,
after $\numt{5}{21}$ POT exposure 
of the $3.0^\circ$ OAB at SK
with a water \cerenkov detector of 100~kton fiducial volume in
Korea.
The normal hierarchy is assumed in generating the events
and the inverted hierarchy is assumed in the fit.
The solid-circle,
open-circle,
open-triangle,
open-square,
and
open-diamond,
shows $\Delta \chi^2_{\rm min}$
for $L=1000$~km, 1050~km, 1100~km, 1150~km, and 1200~km,
respectively.
We take $\srct{2}=0.1$ for all figures
and $\dmns=0^\circ$, $90^\circ$,
$180^\circ$, and $-90^\circ$, 
from the left to the right plots.
All the other input parameters 
are listed in 
eqs.~(\ref{eq:input_physet})-(\ref{eq:eff_and_missID}).
The cross symbols show the highest $\Delta \chi^2_{\rm min}$ value
of the previous study in ref.~\cite{HOS2}.
(b): The same as (a), but the inverted hierarchy is 
assumed in generating the events and 
the normal hierarchy is assumed in the fit.
}
\label{fig:POSnor30}
\end{figure}

%-----------------------------------------------------------------
We show in Fig.~\ref{fig:POSnor30} the minimum $\Delta\chi^2$ expected
for the T2KK experiment after $\numt{5}{21}$ POT exposure
as a function of the off-axis angle and the baseline length ($L$)
of the far detector in Korea,
when the off-axis angle is $3.0^\circ$ at SK.
Figure~\ref{fig:POSnor30}(a) shows the results
when the normal hierarchy is assumed in generating
the events and the inverted hierarchy is assumed in the
fit.
The opposite case, the results when the events are generated for
the inverted hierarchy and the normal hierarchy is assumed in the fit
are shown in Fig.~\ref{fig:POSnor30}(b).
The solid-circle, open-circle, open-triangle, open-square,
and open-diamond,
give the minimum $\Delta \chi^2$ for
the baseline length $L=1000$~km, 1050~km, 1100~km, 1150~km, and
1200~km, respectively.
The results depend strongly on the input values of
$\srct{2}$ and $\dmns$:
$\srct{2}=0.1$ is assumed for all the plots and $\dmns$
is $0^\circ$, $90^\circ$, $180^\circ$, and $-90^\circ$,
from the left to the right plots.
All the other input parameters are listed in
eqs.~(\ref{eq:input_physet})-(\ref{eq:eff_and_missID}).
In each plot, we show by the cross symbols 
the highest $\Delta \chi^2_{\rm min}$ values of
the previous study in ref.~\cite{HOS2}.
%---
When they are higher then 30,
the cross symbols are given on top of the frame
and their values are shown in parentheses.

All the plots in Fig.~\ref{fig:POSnor30} confirm the trend observed in
the previous studies in ref.~\cite{HOS1,HOS2}
that the sensitivity to the neutrino mass hierarchy is highest 
when the off-axis angle at a far detector is smallest 
and that there is little dependence on
the baseline length between $1000$~km and $1200$~km.
This is essentially because the first oscillation maximum in the
$\nu_\mu$-to-$\nu_e$ transition probability occurs at around
$E_\nu=2$~GeV in Korea,
which can be observed via the wide-band beam of small off-axis angle
but not with the narrow-band beam with $\gsim2.0^\circ$ off-axis angle
\cite{HOS1,HOS2}.
It is re-assuring that the mass hierarchy pattern can still be
determined at $3\sigma$ level just by adding a $100$~kton level water
\cerenkov detector at a right place (off-axis angle $\lsim 1^\circ$)
in Korea during the T2K experimental period ($\numt{5}{21}$POT),
even after the realistic estimation for the reconstructed energy
resolution and the background from single $\pi^0$ production
via neutral current are taken into account.
%

%-----------------------------------------------------------
Unfortunately,
the reduction of the $\Delta \chi^2_{\rm min}$ values from the
previous results are most significant at lower off-axis angles
$(\lsim 1^\circ)$
where the mass hierarchy discrimination power of the T2KK experiment
is highest.
This is because the high-energy tail of the wide-band beam
that gives the high sensitivity to the mass hierarchy also
gives rise to the higher rate of the single $\pi^0$ events via the
neutral currents, as shown in Fig.~\ref{fig:pi0pt}(a).
This results in the larger $\pi^0$ background to the
$\nu_\mu$-to-$\nu_e$ oscillation signal at a far detector;
see Figs.~\ref{fig:eventKr}(b) and (d).
At the most favorable location of $0.5^\circ$ OAB at $L=1000$~km,
the reduction in $\Delta \chi^2_{\rm min}$ is as large as $40\%$
to $60\%$, depending on $\dmns$ and the hierarchy.
We also note that the $\dmns$-dependence of the sensitivity
to the mass hierarchy is somewhat smaller than that of the previous
analysis:
For instance, the reduction of $\Delta \chi^2_{\rm min}$ value is
largest for $\dmns=180^\circ$ in Fig.~\ref{fig:POSnor30}, 
where the highest $\Delta \chi^2_{\rm min}$ value was reported
in ref.~\cite{HOS2}.
This is because the contribution proportional to $\cos\dmns$ in
the ``phase-shift'' term $B^e$ in eq.~(\ref{eq:Be}) is 
made less effective in discriminating the hierarchy by
the smearing in $\Erec$ due to the nucleon Fermi motion and
the finite detector resolutions, which have not been taken into
account in ref.~\cite{HOS1,HOS2}.

%--------------------------------------------------------------------
\begin{table}
 \centering
 \begin{tabular}{|l||l|l|l|l|}
  \cline{2-5}
  \multicolumn{1}{c}{}&\multicolumn{4}{|c|}{$\dmns^{\rm input}$}\\
  \hline
  parameters & ~~~$0^\circ$ 
             & ~~~$90^\circ$
             & ~~$180^\circ$
             & ~~$-90^\circ$ \\
  \hline
  \hline
  $\srct{2}$& ~0.74
         & ~0.18
         & ~0.90
	 & ~{\bf 1.8}\\
  $\ssun{2}$& ~0.024
         & -0.010
         &~0.052
	 &~0.12\\
  $\dm12$& ~0.14
         & ~0.067
         & ~0.35
	 & ~0.48\\
  \hline
  $f_\rho^{\rm SK}$& ~0.090
         & ~0.10
         & ~0.083
	 & ~0.061\\
  $f_\rho^{\rm Kr}$& -0.67
         & -0.55
         & -0.86
	 & {\bf -1.0}\\
  \hline
  $f_{\nu_\mu}^{\rm SK}$& -0.31
         & -0.28
         & -0.25
	 & -0.21\\
  $f_{\bar{\nu}_\mu}^{\rm SK}$ & ~0.032
         &~0.036
         &~0.036
	 &~0.027\\
  $f_{\nu_e}^{\rm SK}$ & -0.050
         &-0.067
         &-0.077
	 &-0.056\\
  $f_{\bar{\nu}_e}^{\rm SK}$ & -0.0013
         &-0.0026
         &~0.0044
	 &-0.0038\\
  \hline
  $f_{\nu_\mu}^{\rm Kr}$& ~0.14
         &~0.086
         &~0.13
	 &~0.18\\
  $f_{\bar{\nu}_\mu}^{\rm Kr}$ & ~0.0034
         &~0.015
         &~0.011
	 &~0.0063\\
  $f_{\nu_e}^{\rm Kr}$ & ~0.068
         &~0.068
         &~0.078
	 &~0.084\\
  $f_{\bar{\nu}_e}^{\rm Kr}$ & ~0.0052
         &~0.0042
         &~0.0042
	 &~0.0038\\
  \hline
  $f_{\nu}^{\rm CCQE}$ & -0.16
         &-0.20
         &-0.14
	 &-0.029\\
  $f_{\bar{\nu}}^{\rm CCQE}$ & ~0.032
         &~0.041
         &~0.039
	 &~0.031\\
  $f_{\nu}^{\rm Res}$ & ~0.13
         &~0.085
         &~0.099
	 &~0.11\\
  $f_{\bar{\nu}}^{\rm Res}$ & ~0.043
         &~0.075
         &~0.055
	 &~0.031\\
  $f_{\pi^0}$ & -0.13
         &-0.10
         &~0.047
	 &~0.12\\
  \hline
  $f_V^{\rm SK}$ & -0.33
         &~0.32
         &~0.30
	 &~0.24\\
  $f_V^{\rm Kr}$ & ~0.22
         &~0.17
         &~0.22
	 &~0.27\\
  \hline
  $\effe$ & ~0.48
         & ~0.11
         & ~0.61
	 &~{\bf 1.2}\\
  $\effm$ & -0.12
         & -0.066
         &-0.14
	 &-0.23\\
  $P_{e/\mu}$& ~0.48
         &~0.71
         &~{\bf 1.3}
	 &~{\bf 1.2}\\
  \hline
  \hline
  $\chi^2_{\rm para}$+$\chi^2_{\rm sys}$
   &~1.8
   &~1.2
   &~4.0
   &~7.6\\
\hline
($\chi^2_{\rm para}$+$\chi^2_{\rm sys}$)/$\Delta\chi^2_{\rm min}$
&~0.13
&~0.091
&~0.17
&~0.27\\
\hline
\end{tabular}
\caption{%
The pull factors of the parameters that characterize
systematic errors at $\Delta \chi^2_{\rm min}$ 
for
$3.0^\circ$ OAB at SK and $0.5^\circ$
OAB at $L=1000$~km,
when the normal hierarchy is assumed in generating the events and
the inverted hierarchy is assumed in the fit.
The model parameters are taken as in
eqs.~(\ref{eq:input_physet})-(\ref{eq:eff_and_missID})
for $\srct{2}=0.1$ and $\dmns=0^\circ$, $90^\circ$, $180^\circ$,
and $-90^\circ$.
The pull factors whose magnitudes are larger than unity are shown
by bold face letters.
The bottom lines give the squared sum of all the pull factors
and its fraction in the total $\Delta \chi^2_{\rm min}$.
}
\label{tab:pull_nor}
\end{table}
%--------------------------------------------------------------------

In Table~\ref{tab:pull_nor},
we list the pull factors of all the parameters for systematic errors
at $\Delta \chi^2_{\rm min}$,
for
$3.0^\circ$ OAB at SK and $0.5^\circ$ OAB at
$L=1000$~km,
for the normal hierarchy and 
for all the four $\dmns$ values in
Fig.~\ref{fig:POSnor30}(a).
%
%----------------------------------------
It is clearly seen that the pull factors for $\srct{2}$, 
$f_\rho^{\rm Kr}$, $\effe$, and $P_{e/\mu}$
are most significant.
The $\srct{2}^{\rm fit}$ is shifted upwards in order to
compensate for the small event numbers expected for
the inverted hierarchy.
The matter density between J-PARC and Korea 
is reduced to make the matter effect in the wrong sign small.
On the other hand,
$\rho_{\rm SK}$ is slightly shifted in the positive direction,
because it is the difference in the matter effects along the two
baselines that is sensitive to the mass hierarchy.
The positive pull factors of $\effe$ and $P_{e/\mu}$ also increase
the number of $e$-like events at a far detector in Korea.
Reduction of these errors, in particular that of $\srct{2}$ by
the next-generation reactor experiments,
should hence improve the sensitivity of the T2KK experiment on the
neutrino mass hierarchy.
On the other hand, the fraction of the systematic errors in the total
$\Delta \chi^2_{\rm min}$ is not large for a 100~kton detector with
$\numt{5}{21}$POT, as shown in the bottom line of
Table~\ref{tab:pull_nor}.
Therefore, a larger detector and/or higher beam power will improve the
sensitivity of the experiment.
%

%--------------------------------------------------------------------
\begin{table}
 \centering
 \begin{tabular}{|c|l||r|r|r|r|}
  \cline{3-6}
  \multicolumn{2}{c}{}&\multicolumn{4}{|c|}{$\dmns^{\rm input}$}\\
  \hline
  &analysis condition & $0^\circ$ 
  & $90^\circ$
  & $180^\circ$
  & $-90^\circ$ \\
  \hline
  \hline
  %org setup
  (0)&previous results~\cite{HOS2}
      & 22.9
  & 30.7
  & 55.9
  & 53.1
  \\
  \hline
  %hiEff
  (1)&$\rho_{\rm SK}/\rho_{\rm Kr}=2.6/3.0\times(1\pm0.06)$ g/cm$^3$,
      $\effm=(100^{+0}_{-1})\%$
      & 22.8
  & 30.3
  & 54.0 
  & 50.5
  \\
  %no
  (2)&$\effe=(90\pm5)\%$
      & 20.4
  & 26.8
  & 47.4
  & 42.7
  \\
  %woALL
  (3)& $\Erec$ for event energy with detector resolution
      & 18.3
  & 23.3
  & 39.8
  & 37.1
  \\
  %woPiRes
  (4)& $P_{e/\mu}=(1\pm1)\%$ & 17.4
      & 19.8
	  & 31.7
	      & 31.5
		  \\
  %woRes
  (5)& $\pi^0$ background & 11.1
      & 10.3
	  & 20.7
	      & 23.2
		  \\
  %final
  (6)& non-CCQE ``resonance'' events
  & 14.2
      & 12.7
	  & 23.8
	      & 28.0
		  \\
  \hline
 \end{tabular}
 \caption{%
Changes in $\Delta \chi^2_{\rm min}$ with various assumptions on the
systematic effects of the T2KK experiment,
for the combination of
$3.0^\circ$ OAB at SK and $0.5^\circ$ OAB at $L=1000$~km,
when the normal hierarchy is assumed in generating the events and
the inverted hierarchy in the fit.
We take $\srct{2}=0.1$ and $\dmns=0^\circ$, $90^\circ$, $180^\circ$,
and $-90^\circ$ as inputs and all the others are as in
eqs.~(\ref{eq:input_physet})-(\ref{eq:eff_and_missID}).
The top (0) row gives the previous results of ref.~\cite{HOS2},
and each row gives the results after changing the conditions
one by one.
The bottom (6) row gives our results shown in Fig.~\ref{fig:POSnor30}(a)
}
\label{tab:alteration1}
\end{table}
%-----------------------------------------------------------------

%-----------------------------------------------------------------
% Position select 2.5@SK
%-----------------------
\begin{figure}[t]
 \centering
(a) normal hierarchy (OAB:2.5@SK)

 \includegraphics[width=0.24\textwidth]{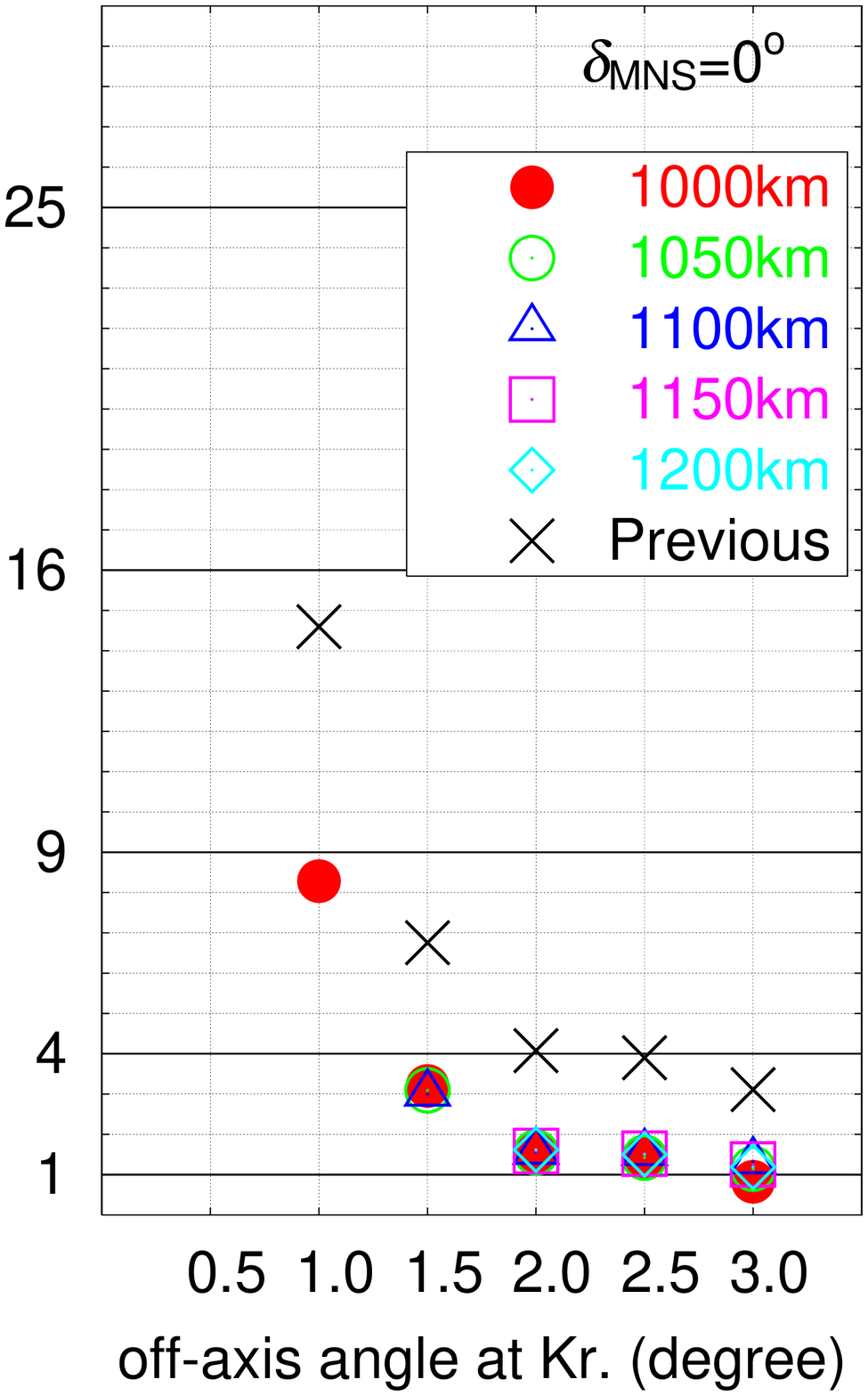}
 \includegraphics[width=0.24\textwidth]{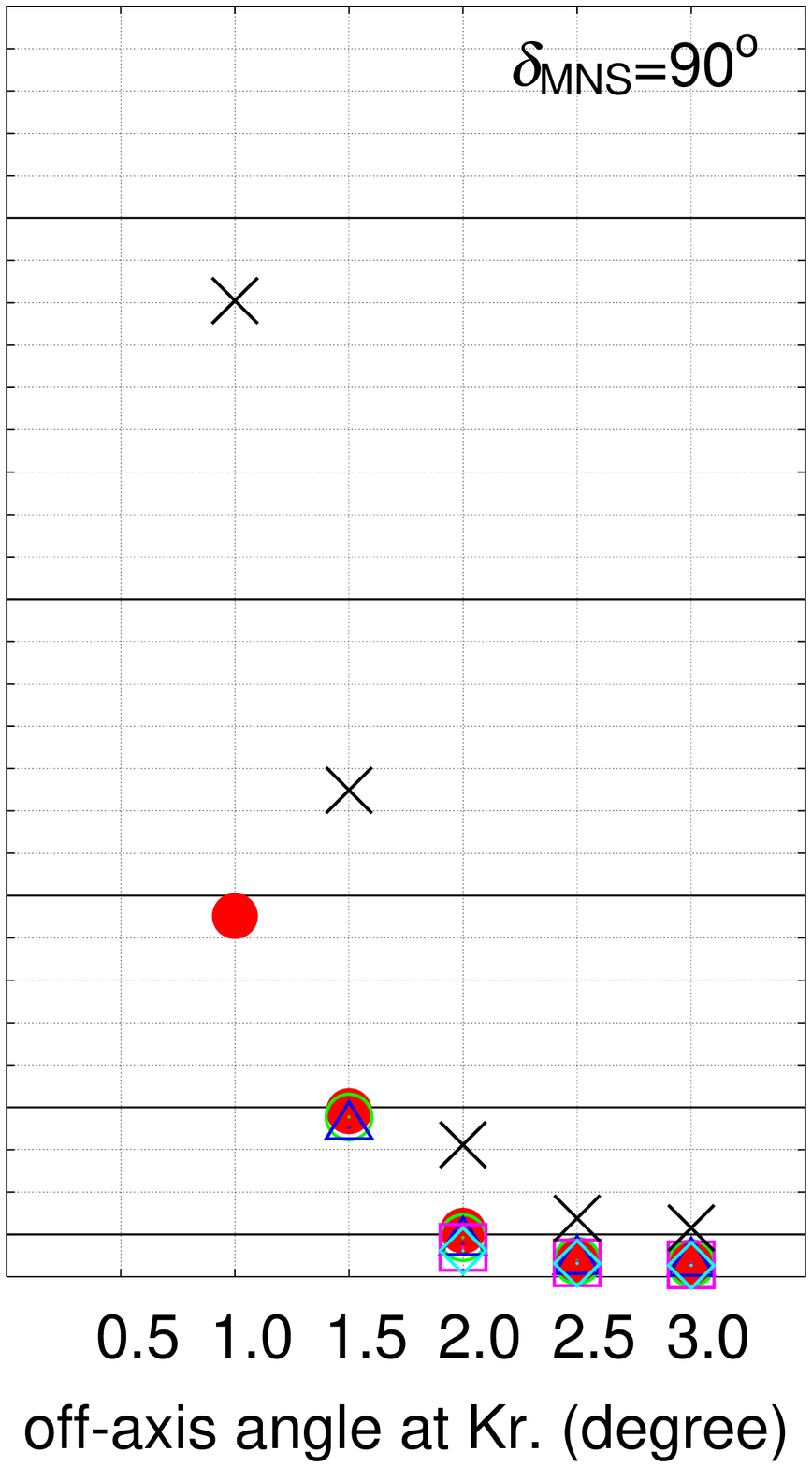}
 \includegraphics[width=0.24\textwidth]{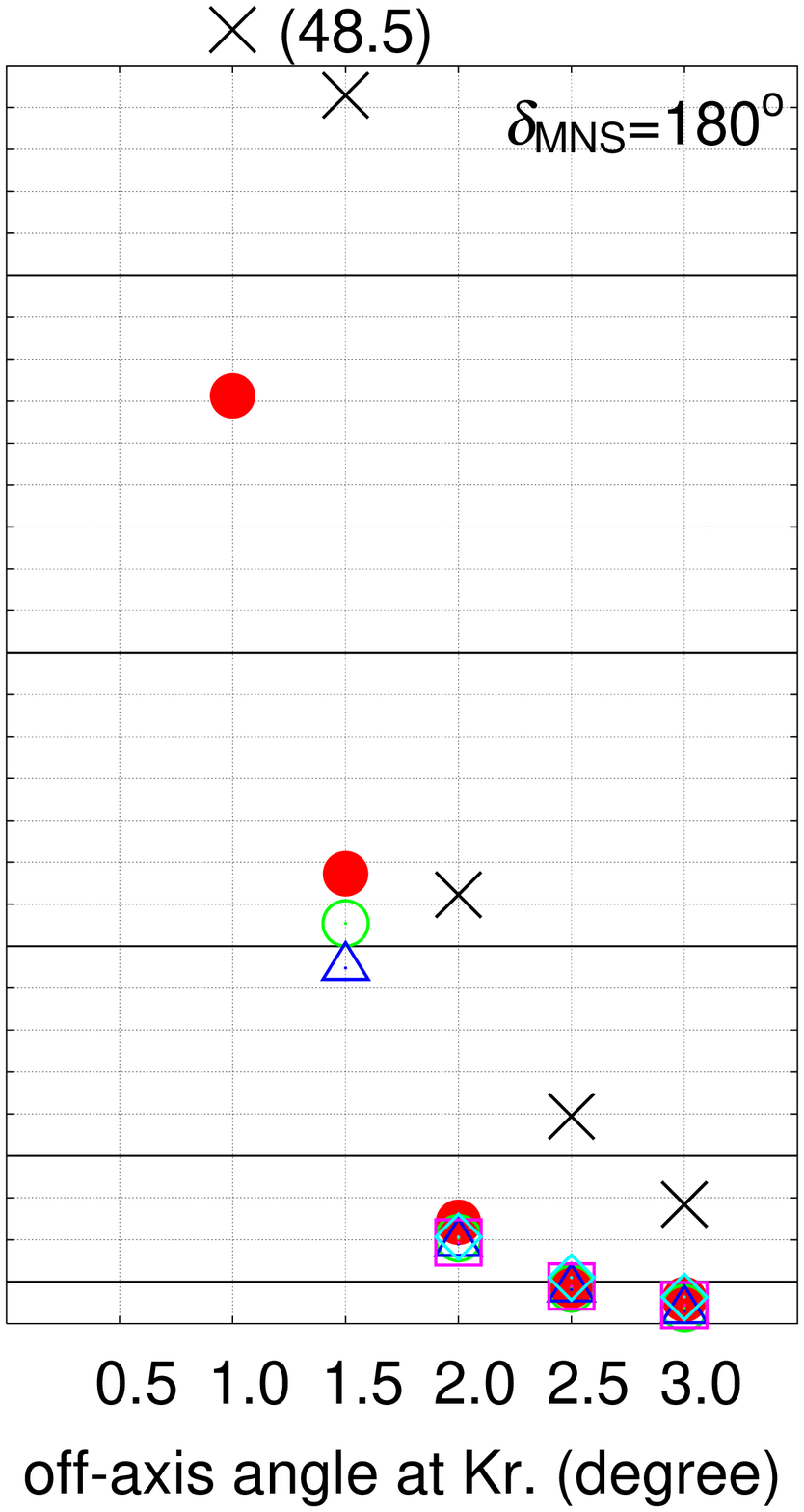}
 \includegraphics[width=0.24\textwidth]{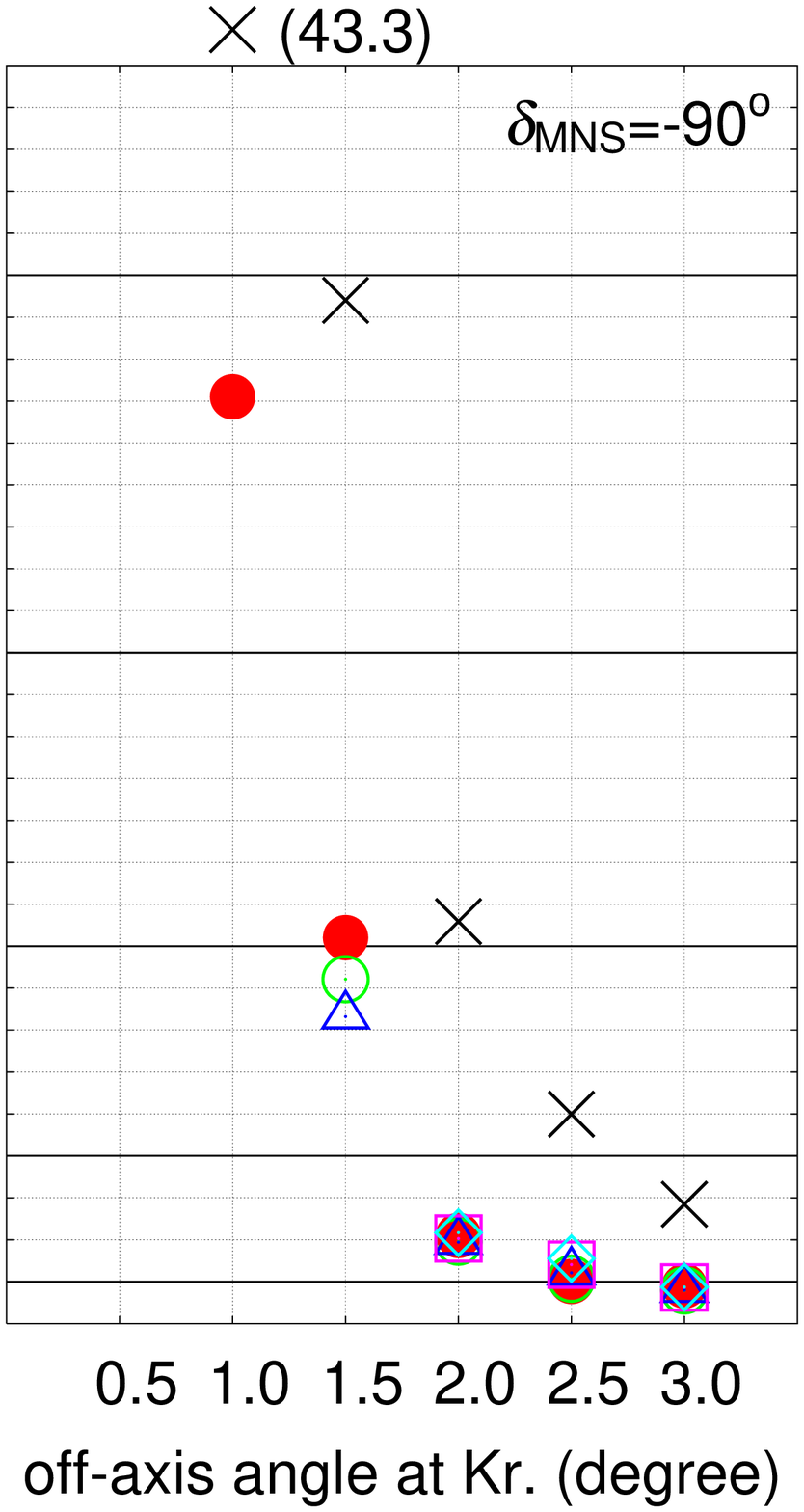}

\bigskip

(b) inverted hierarchy (OAB:2.5@SK)

 \includegraphics[width=0.24\textwidth]{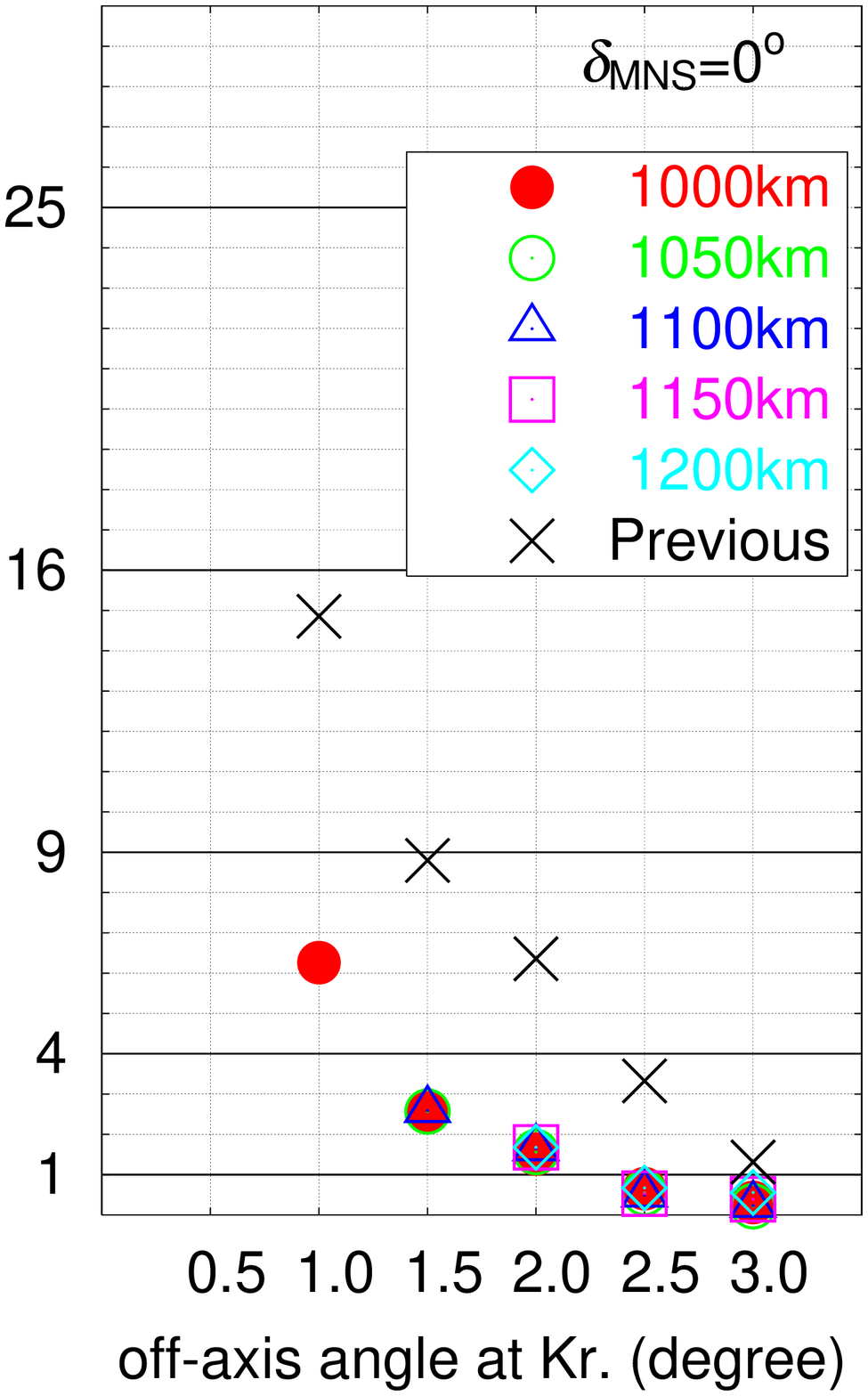}
 \includegraphics[width=0.24\textwidth]{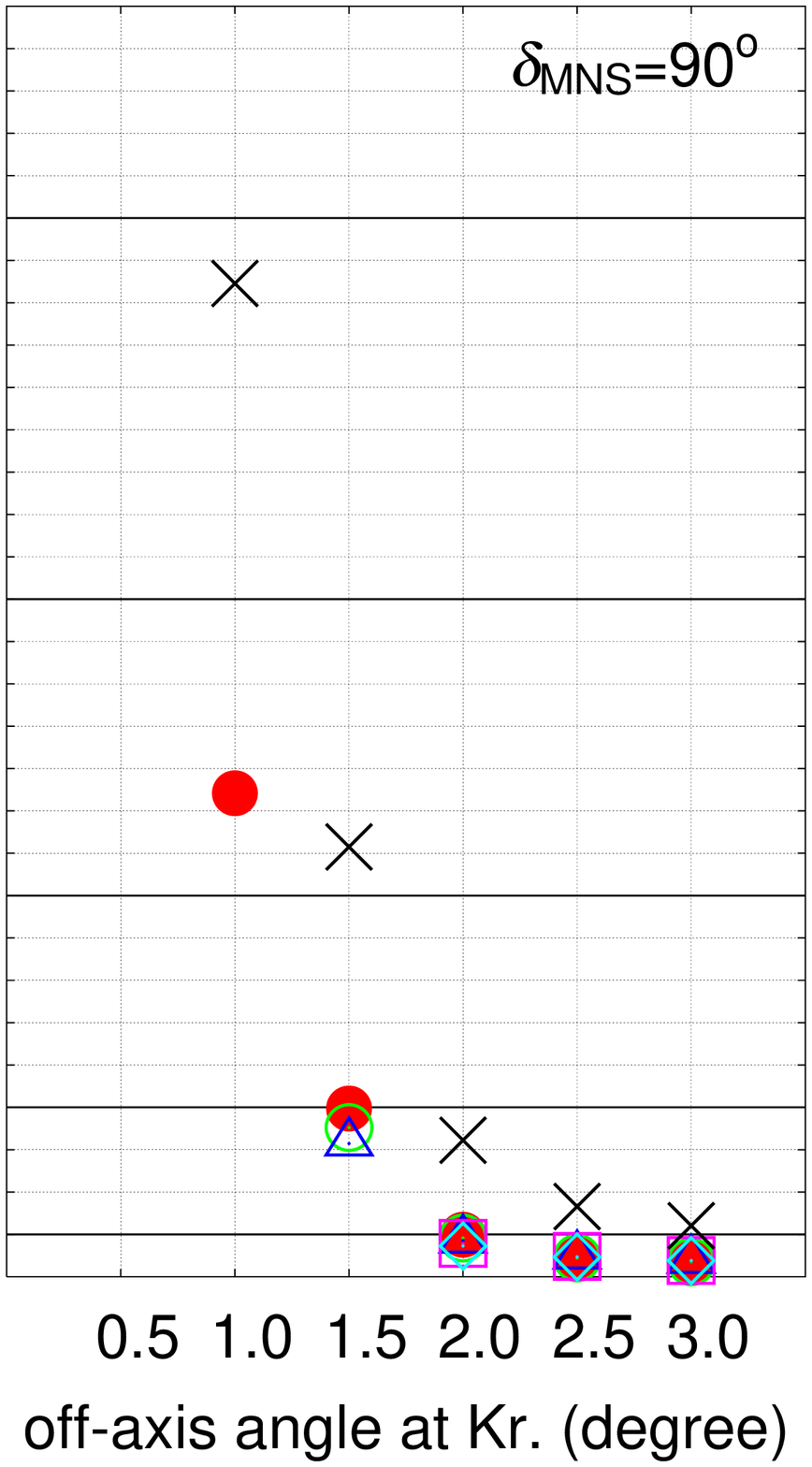}
 \includegraphics[width=0.24\textwidth]{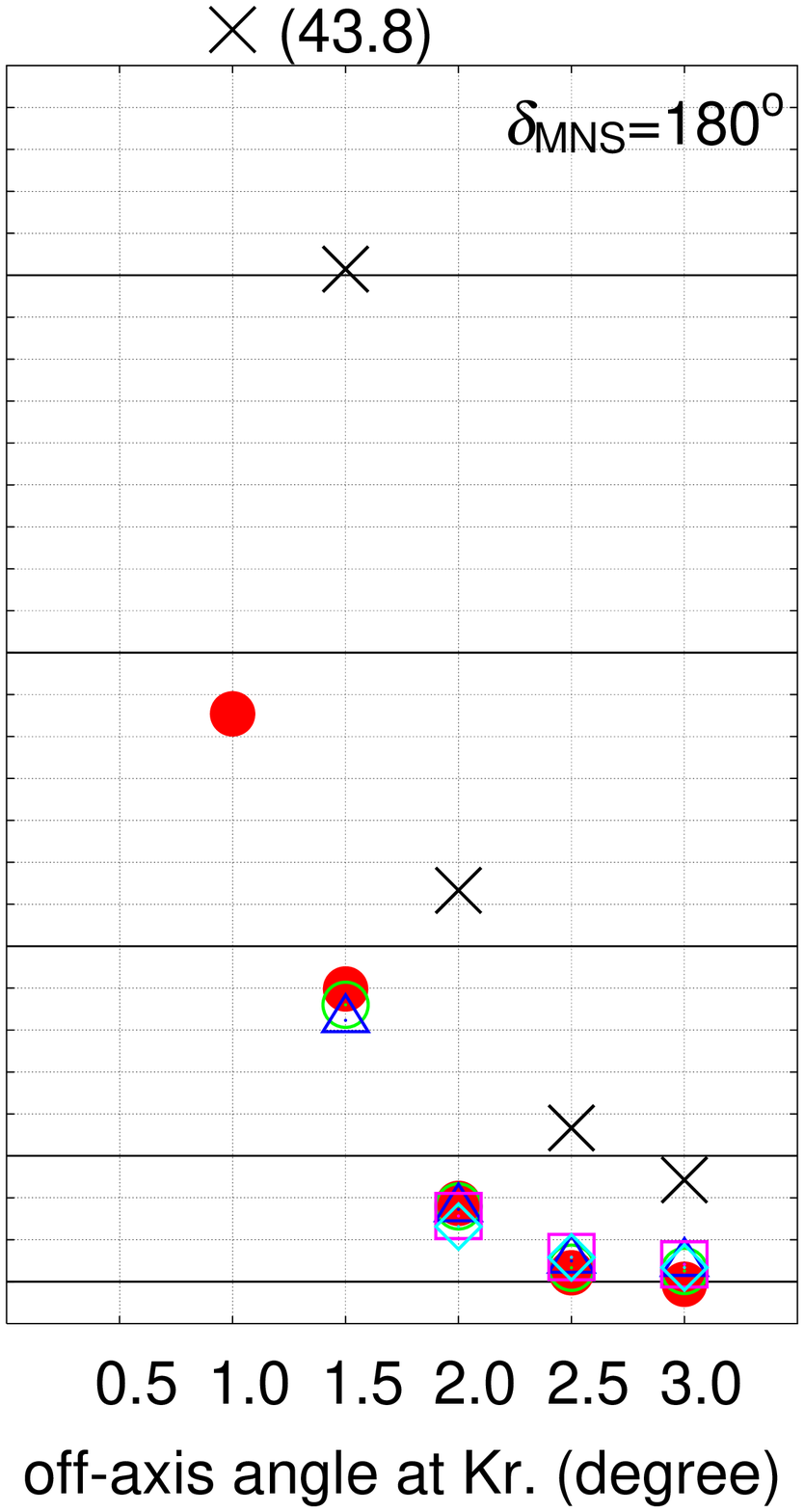}
 \includegraphics[width=0.24\textwidth]{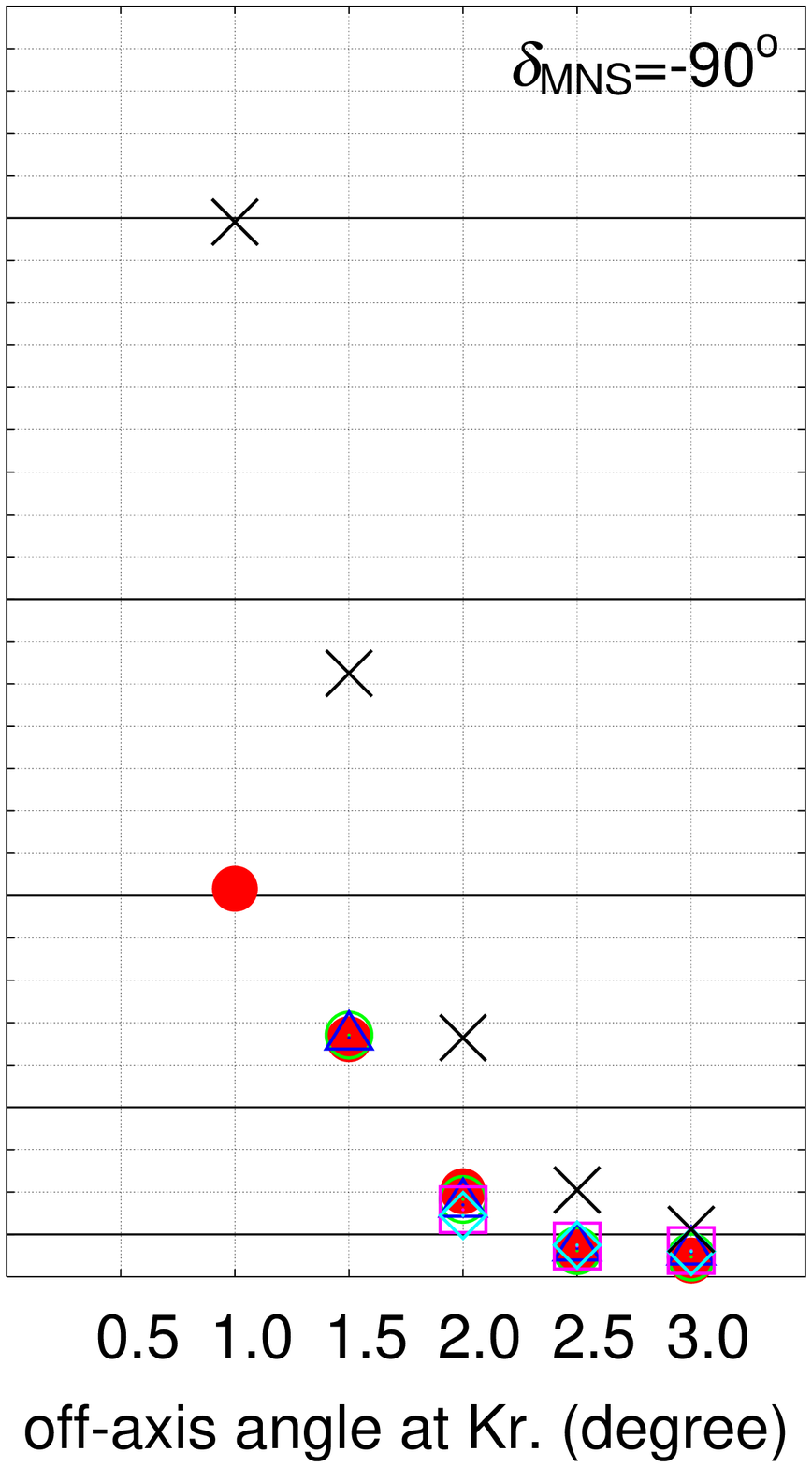}
 \caption{The same as Figure~\ref{fig:POSnor30},
but with $2.5^\circ$ OAB at SK.}
\label{fig:POSnor25}
\end{figure}
%-----------------------------------------------------------------

In Table~\ref{tab:alteration1},
we show how $\Delta \chi^2_{\rm min}$ changes from the values in
ref.~\cite{HOS2} by adding successively the effects introduced in this
analysis,
for the combination of $3.0^\circ$ OAB at SK and
$0.5^\circ$ OAB at $L=1000$~km,
when the normal hierarchy is assumed in generating the events
and the inverted hierarchy is assumed in the fit.
The first row (0) gives the results of the previous study in
ref.~\cite{HOS2}.
In the row (1), we change the average matter density along the T2K
baseline from 2.8 to 2.6 g/cm$^3$ and the error of $\rho_{\rm SK}$
and $\rho_{\rm Kr}$ are doubled from $3\%$ to $6\%$,
and we also introduced a $1\%$ error in the $\mu$ detection efficiency.
The $\Delta \chi^2$ values are slightly reduced for $\dmns=180^\circ$
and $-90^\circ$ cases,
mainly because of the increase in the matter density errors.
In the row (2), we further introduce the detection efficiency for the
$e$-like events, $\effe=(90\pm5)\%$, and
the $\Delta \chi^2_{\rm min}$ for all $\dmns$
decrease by about $10\%$ reflecting the $10\%$ decrease
of the signal events.
In the row (3),
we introduce smearing in $\Erec$ due to the nuclear Fermion motion and
realistic energy resolution of detectors.
Because the matter effects in the phase-shift term $B^e$ is 
diluted by the smearing,
the decrease in $\Delta \chi^2_{\rm min}$ is largest $\delta=180^\circ$;
see the term proportional to $\cos \dmns$ in eq.~(\ref{eq:Be}).
In the row (4),
we take into account the particle misidentification probability
$P_{e/\mu}=(1\pm1)\%$.
Since this change makes the fake $e$-like events around the dip of the
$\nu_\mu \to \nu_e$ transition probability,
the reduction in $\Delta \chi^2_{\rm min}$ is significant 
even for $1\%$ misidentification probability,
if its error is as large as $100\%$.
In the row (5),
the single $\pi^0$ events reduce the physics potential of
the T2KK experiment significantly,
because the $\nu_\mu \to \nu_e$ signal at small $\Erec$ is dominated by
the $\pi^0$ background at a far detector in Korea,
as shown in Fig.~\ref{fig:eventKr}.
In the bottom row (6),
we add the non-CCQE ``resonance'' events in the analysis.
These events make $\Delta \chi^2_{\rm min}$ large, 
because their magnitudes are also proportional to
the $\nu_\mu \to \nu_e$ transition probability.
%
%-----------------------------------------------------------------

In Fig.~\ref{fig:POSnor25}, 
we show the minimum $\Delta \chi^2$,
the mass hierarchy discrimination power of the T2KK experiment,
when the beam center is $2.5^\circ$ below the SK.
All the other contents of Fig.~\ref{fig:POSnor25} are the
same as those of Fig.~\ref{fig:POSnor30}.
Because of the geological constraint,
the $2.5^\circ$ OAB at SK cannot provide 
$0.5^\circ$ OAB in Korean peninsula~\cite{HOS1,HOS2}.
When the off-axis angle is $2.5^\circ$ at SK,
the optimum OAB for a far detector in Korea 
is $1.0^\circ$ at $L=1000$~km.
The value of $\Delta \chi^2_{\rm min}$ is not significantly different
between the $3.0^\circ$ OAB at SK and the $2.5^\circ$ OAB at SK,
when the off-axis angle in Korea is fixed as $1.0^\circ$.
It confirms our understanding that the energy profile or the hardness
of the neutrino beam observed at a far detector is essential for the
mass hierarchy discrimination.

%-----------------------------------------------------------

\subsection{Uncertainty of the $\pi^0$ background}
\label{sec:uncert-pi0-backgr}
\begin{figure}[t]
 \centering
 \includegraphics[scale=0.8]{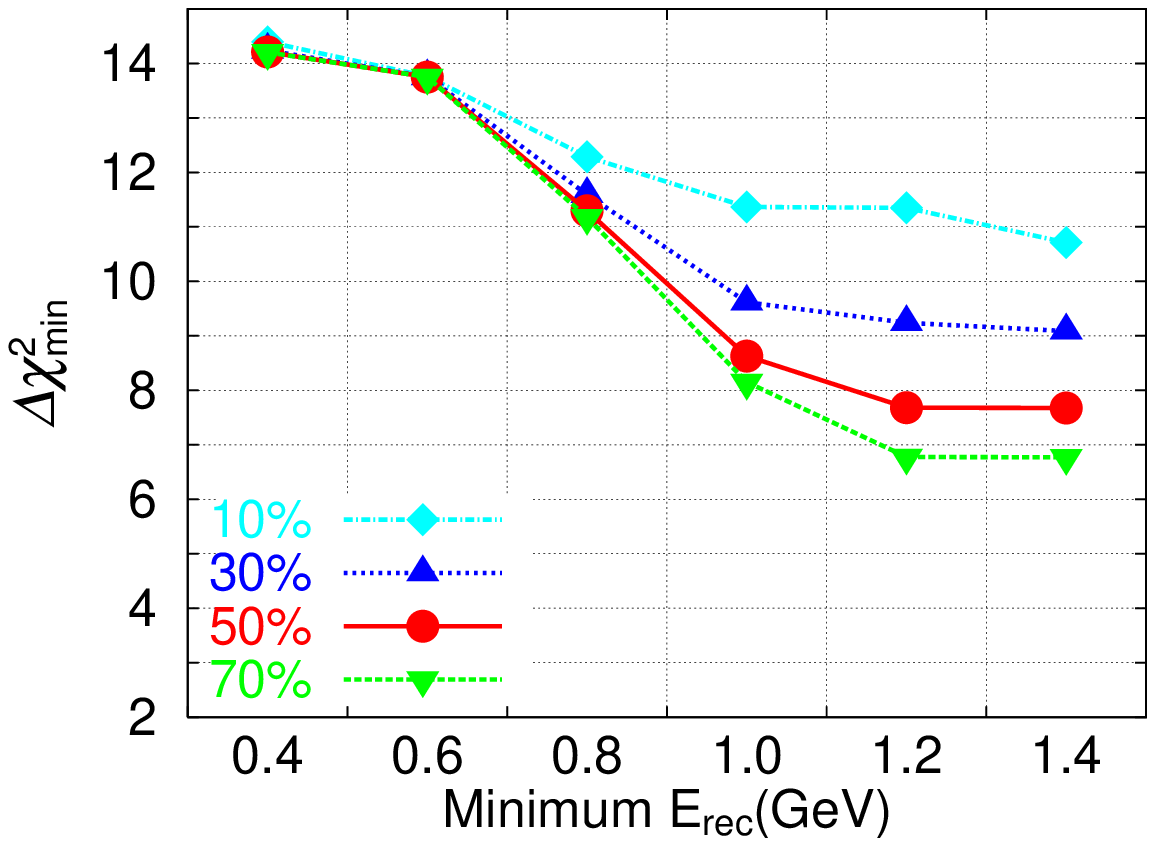}
\caption{
Minimum $\Delta\chi^2$ of the T2KK experiment
as a function of the minimum $\Erec$,
when only those $e$-like events with larger $\Erec$
values are retained in the analysis at a far detector.
The events are calculated for the $3.0^\circ$ OAB at SK
and 
a 100~kton water \cerenkov detector at $0.5^\circ$ OAB and
$L=1000$~km with $\numt{5}{21}$ POT exposure,
for $\srct{2}=0.1$ and $\dmns=0^\circ$ 
and the parameters of
eqs.~(\ref{eq:input_physet})-(\ref{eq:eff_and_missID}),
for the normal hierarchy, while the inverted hierarchy is assumed in
the fit.
The dashed-dotted line with diamonds,
the dotted line with up-triangles, 
the solid line with circles, 
and 
the dashed line with down-triangles
give $\Delta \chi^2_{\rm min}$
when the uncertainty in the $\pi^0$ background rate
$\Delta f_{\pi^0}$ is
$10\%$, $30\%$, $50\%$, and $70\%$, respectively.
}
\label{fig:PiSYS}
\end{figure}
%------------
In this subsection,
we examine the impacts of the $\pi^0$ background in more detail.
In our analysis, we adopt the following uncertainties for the relevant
cross sections
\begin{eqnarray}%list of uncertainties
f_{\beta}^{\rm CCQE} = 1 \pm0.03\,, \hspace*{5ex}
f_{\beta}^{\rm Res} = 1 \pm 0.2\,,\hspace*{5ex}
f_{\pi^0} = 1 \pm 0.5\,,
\label{eq:d_sigma}
\end{eqnarray}
where $\beta=\nu$ and $\bar{\nu}$; see eq.~(\ref{eq:sys_chi2}).
The $3\%$ error in the CCQE cross sections should be achieved
in the near future, 
whereas there is a possibility that the non-CCQE
``resonance'' cross sections
and the neutral current single $\pi^0$ production cross
section can be measured more accurately than $20\%$ and
$50\%$, respectively, assumed in this analysis.
We therefore repeat the fit by varying 
$\Delta f_\beta^{\rm Res}$ between $10\%$ and $30\%$,
and
$\Delta f_{\pi^0}$ between $10\%$ and $70\%$.
We find little impacts of those variations on the magnitude
of $\Delta \chi^2_{\rm min}$, which conform with the small
pull factors for these parameters in Table~\ref{tab:pull_nor}.
It turns out that the uncertainty in the non-CCQE cross
section does not affect the mass hierarchy sensitivity of
the T2KK experiment 
because  it tends to cancel in the ratio of the $\mu$-like
and $e$-like events.

%-----------
% pi0 effect
%-----------
In case of the $\pi^0$ background to the $e$-like events,
however,
the smallness of the impacts of varying $\Delta f_{\pi^0}$
between $10\%$ and $70\%$ is striking, and we examine the cause
carefully.
In Fig.~\ref{fig:PiSYS}, 
we show $\Delta\chi^2_{\rm min}$ of the T2KK experiment 
as a function of the lowest $\Erec$ above which the $e$-like events
are counted at the far detector in Korea.
All the other conditions and the input parameters are the same as
those of Fig.~\ref{fig:POSnor30}(a) and Table~\ref{tab:alteration1},
for $\dmns=0^\circ$.
The dash-dotted line with diamonds,
the dotted line with upper triangles, 
the solid line with circles, 
and 
the dashed line with lower triangles
are obtained with the $\pi^0$ background normalization error of
$\Delta f_{\pi^0}=10\%$, $30\%$, $50\%$, and $70\%$, respectively.

%-----
It is clearly seen that there is little dependence on the error
$\Delta f_{\pi^0}$ 
when we use all the data with $\Erec \geq0.4$~GeV as has been assumed
in our analysis.
As the $\Erec$ threshold is increased, however, the reduction in
$\Delta \chi^2_{\rm min}$ becomes significant as $\Delta f_{\pi^0}$
increases.
This is because the normalization of the $\pi^0$ background 
can be determined by the $e$-like event rate at low $\Erec$
where the $\pi^0$ background dominates the oscillation signal;
see Fig.~\ref{fig:eventKr}(b) and (d).
This suggests strongly that we should understand not only the overall
normalization of the $\pi^0$ background but also the energy and
angular distribution of singly produced $\pi^0$'s in the neutral
current events as well as the momentum dependence of the error of the
$\pi^0$-to-$e$ misidentification probability $P_{e/\pi^0}$,
whose parameterization is given in Fig.~\ref{fig:pi0pt}(b).
Detailed studies of the normalization and the shape of the $\pi^0$
background
should be the most important task before the physics case of the T2KK
experiment can be established.

\subsection{Dependence of the OAB at SK}
\label{sec:dependence-oab-sk}

In Figs.~\ref{fig:POSnor30} and \ref{fig:POSnor25},
we find that the best location of the far detector to determine the
neutrino mass hierarchy is at $L=1000$~km away from the J-PARC,
where $0.5^\circ$ OAB can be observed for the $3.0^\circ$ OAB
at SK (Fig.~\ref{fig:POSnor30}), 
or 
$1.0^\circ$ OAB for the $2.5^\circ$ OAB at SK (Fig.~\ref{fig:POSnor25}).
In this subsection, we compare carefully the two combinations since
they can be interchanged,
or the $(1.0 - \theta)^\circ$ OAB can be observed
for the $(2.5+\theta)^\circ$ OAB at SK, 
simply by adjusting the beam direction at J-PARC 
(up to $|\theta|\lsim0.5^\circ$) for a fixed 
far detector location along the baseline at $L\simeq 1000$~km.

%------------------------------------------------------------------
\begin{figure}[t]
 \centering
 \includegraphics[height=0.165\textheight]{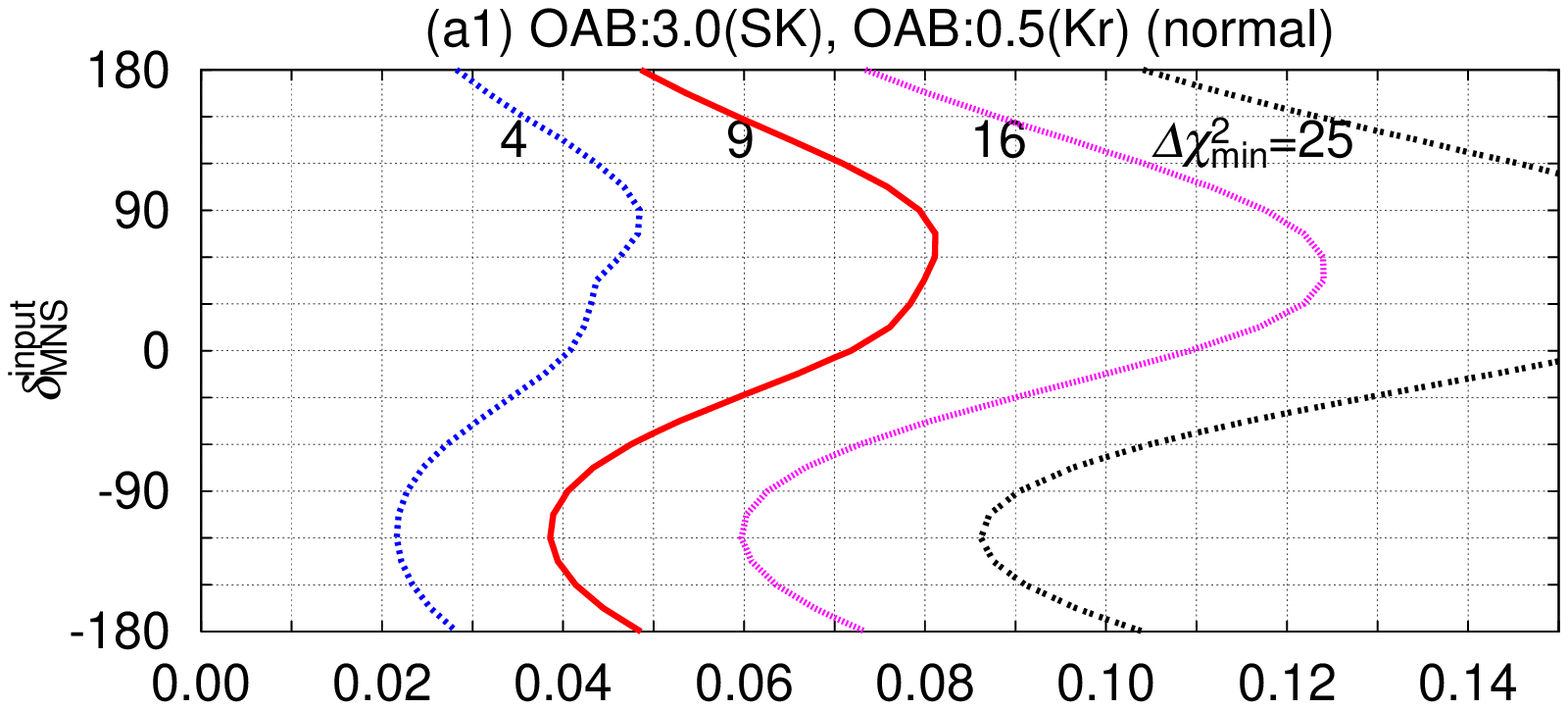}
 \includegraphics[height=0.165\textheight]{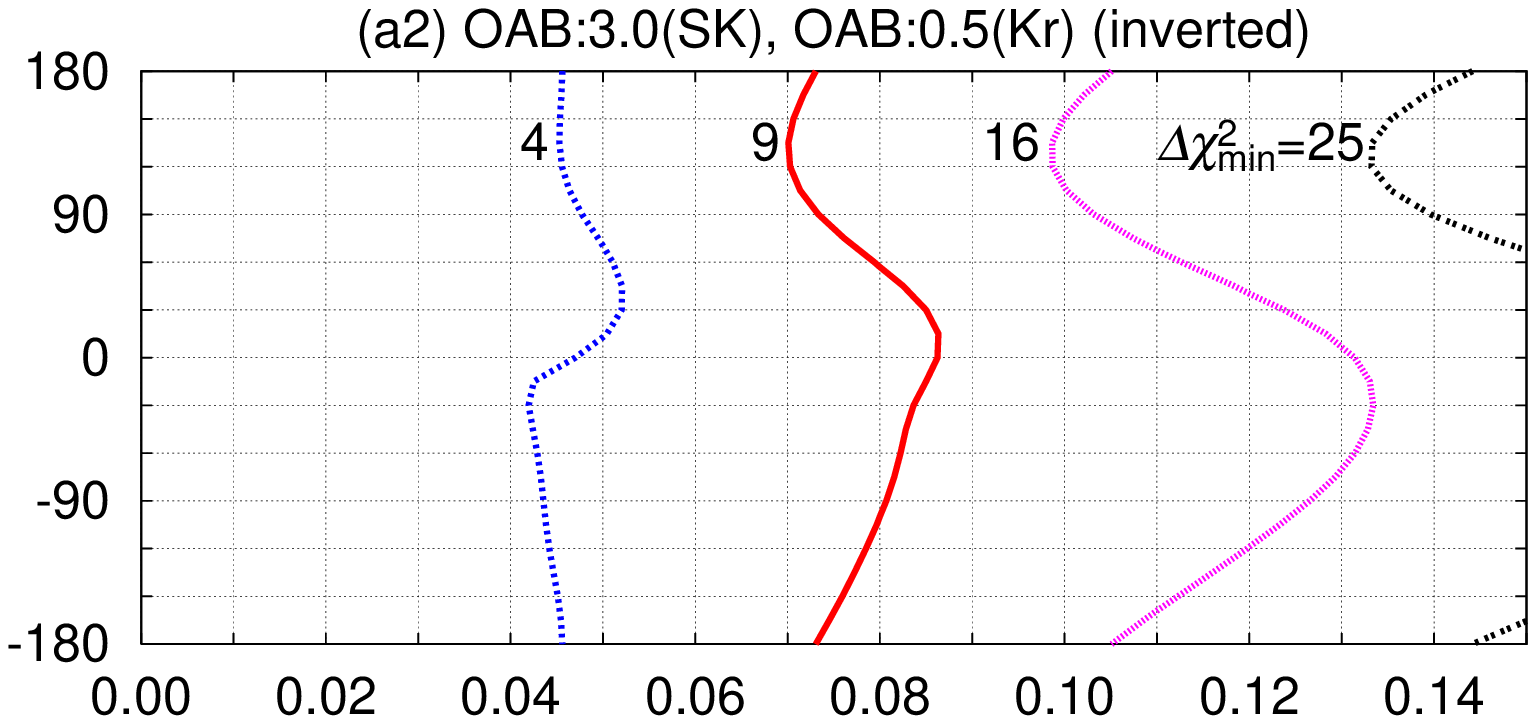}

\vspace*{-2ex}

 \includegraphics[height=0.165\textheight]{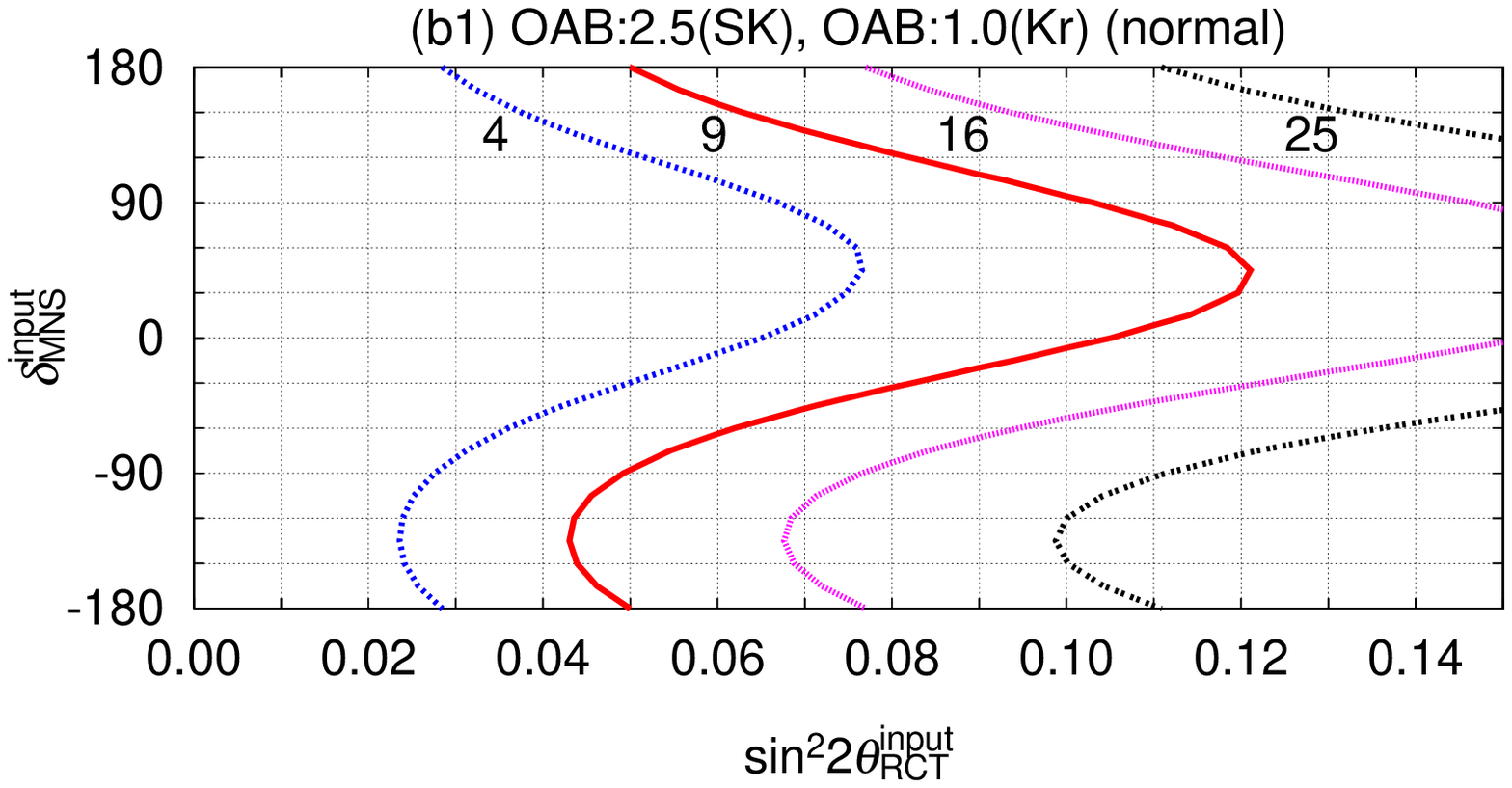}
 \includegraphics[height=0.165\textheight]{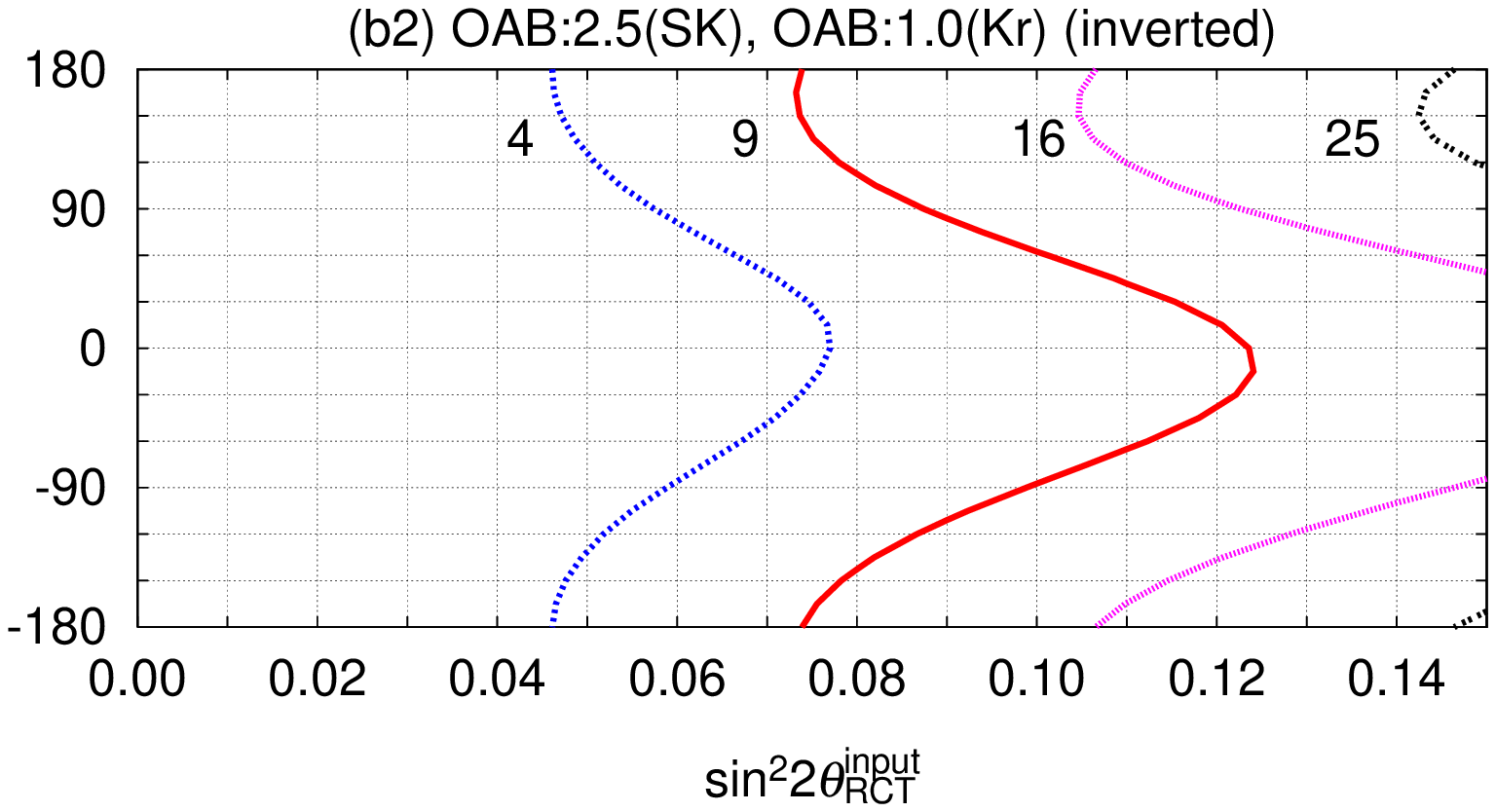}
 \caption{
 The $\Delta \chi^2_{\rm min}$ contour plot for the capability of the
T2KK experiment to determine the neutrino mass hierarchy on the
$\srct{2}^{\rm input}$ and $\dmns^{\rm input}$ plain.
(a1) and (b1) are for the normal hierarchy, while 
(a2) and (b2) are for the inverted hierarchy.
The OAB combination for (a1) and (a2) is 
$3.0^\circ$ OAB at SK and 
$0.5^\circ$ OAB at $L=1000$~km,
$2.5^\circ$ OAB at SK and 
$1.0^\circ$ OAB at $L=1000$~km
is for (b1) and (b2).
All the input parameters other than $\sin^22\theta_{\rct}^{\rm input}$
and $\dmns^{\rm input}$ are the same as those in
Figs.~\ref{fig:POSnor30} and \ref{fig:POSnor25}.
}
\label{fig:contOAB}
\end{figure}
%----------------------------------------------------------

In Fig.~\ref{fig:contOAB}, 
we show the contours for 
$\Delta \chi^2_{\rm min}=4$, $9$, $16$, $25$ in the plane of 
$\srct{2}^{\rm input}$ and $\dmns^{\rm input}$.
The wrong hierarchy can be excluded with the $n$-$\sigma$ confidence
level,
if the true values of $\srct{2}$ and $\dmns$ lie in the right-hand side
of the $\Delta \chi^2_{\rm min}=n^2$ contour.
The upper figures (a1) and (a2) are for $3.0^\circ$ OAB at SK
with $0.5^\circ$ OAB at $L=1000$~km,
and 
the lower figures (b1) and (b2) are for the $2.5^\circ$ OAB at SK with
$1.0^\circ$ OAB also at $L=1000$~km.

%----
It is clearly seen from the figures that 
the mass hierarchy can be determined better by
the combination of $3.0^\circ$ OAB at SK and
$0.5^\circ$ OAB at $L=1000$~km than the combination of
$2.5^\circ$ and $1.0^\circ$ for all the input values of
$\srct{2}$ and $\dmns$ and for both hierarchy patterns.
For instance,
by comparing the figures (a1) and (b1)
we find that the normal hierarchy can be established at $3\sigma$
level, $\Delta \chi^2_{\rm min}>9$,
when $\srct{2}\gsim 0.08$ $(0.12)$ for the combination of
$3.0^\circ$ and $0.5^\circ$ ($2.5^\circ$ and $1.0^\circ$).
Likewise, 
from the figures (a2) and (b2),
the inverted hierarchy can be established when
$\srct{2}\gsim 0.09$ $(0.12)$ for the combination of
$3.0^\circ$ and $0.5^\circ$ ($2.5^\circ$ and $1.0^\circ$).
The difference is significant when $|\dmns|\lsim90^\circ$
where it is difficult to determine the mass hierarchy.
On the other hand, we find little dependence on the off-axis angle
between $3.0^\circ$ and $2.5^\circ$ at SK when $\dmns\simeq
180^\circ$,
where the mass hierarchy can be determined with 
relative ease.
%---

%---
The reason for the strong dependence on the off-axis angle when 
$|\dmns|\lsim 90^\circ$ can be explained by the hardness of the 
$0.5^\circ$ OAB that provides sufficient flux at the 
$\nu_\mu \to \nu_e$ oscillation maximum around $E_\nu \sim 2.0$~GeV.
It is essentially the mass hierarchy dependence of the amplitude shift
term, $A^e$, in eqs.~(\ref{eq:Pme2}) and (\ref{eq:Ae}),
which contribute to the determination,
and the hardness of the $0.5^\circ$ OAB helps enhancing the signal.
When $\dmns\simeq 180^\circ$, 
in addition to the amplitude shift term,
the phase-shift term $B^e$, in eq.~(\ref{eq:Pme2}), becomes significant 
because the leading term and the sub-leading term in eq.~(\ref{eq:Be})
adds up to make $|B^e|$ large at $\cos \dmns \simeq -1$.
The mass hierarchy dependence due to the phase shift term $B^e$ turns
out to give significant difference in the $\nu_\mu \to \nu_e$
transition probability at lower $E_\nu$~\cite{HOS1,HOS2},
and the downward shift of the flux maximum $E_\nu$
in the $1.0^\circ$ OAB can be compensated for.

%---
In the absence of a concrete evidence that the nature chooses
$\cos \dmns \simeq 180^\circ$, it is clear that the effort to make the
off-axis angle at the far detector as small as possible should be
valuable.
The sensitivity difference between $0.5^\circ$ OAB and $1.0^\circ$ OAB
in Fig.~\ref{fig:contOAB} corresponds to about a factor of two
difference in the product of the fiducial volume of the far detector
and the POT,
the beam power times the running period.

\subsection{Impacts on the $|\dm13|$ measurement}
\label{sec:dm13}

In this subsection,
we comment on the implication of the mass hierarchy uncertainty
in the measurement of the absolute value of the larger mass-squared
difference.

%---
\begin{figure}
 \centering
 \includegraphics[width=0.49\textwidth]{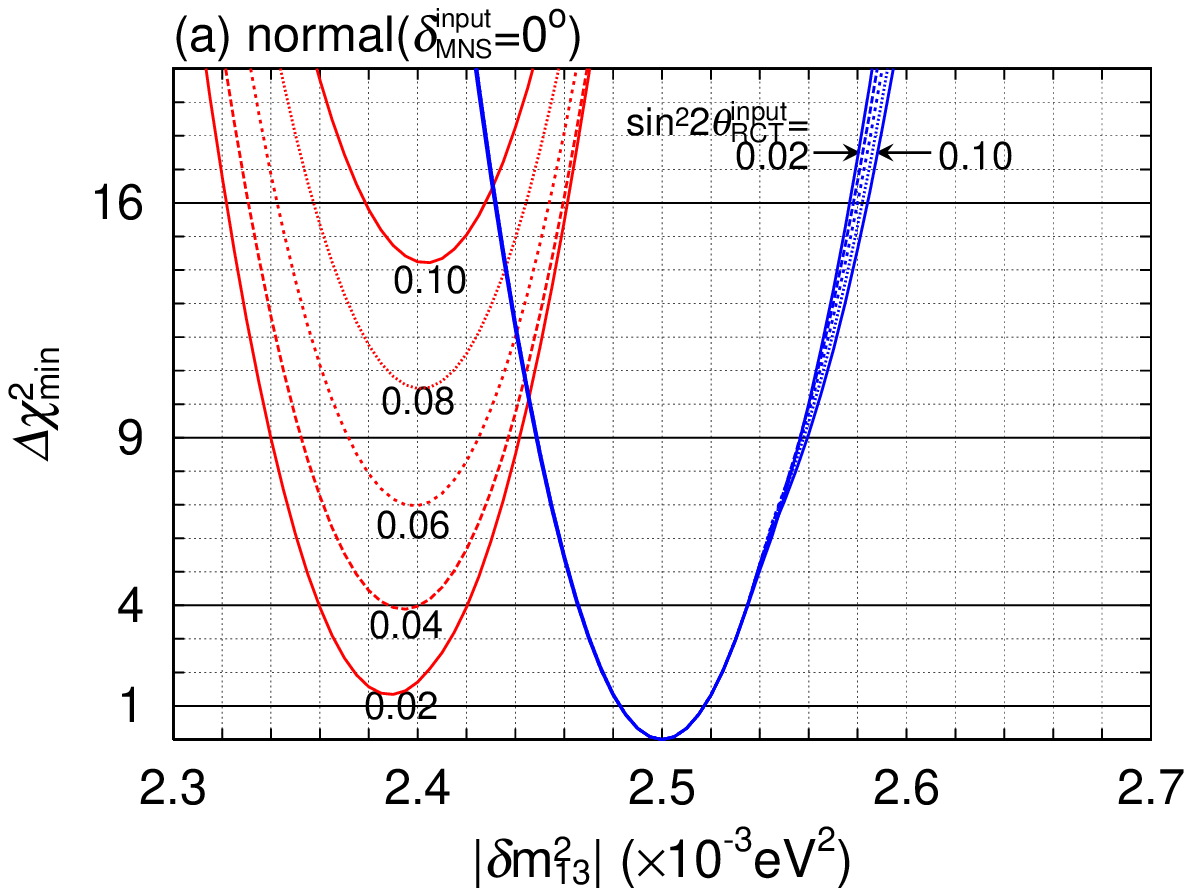}
 \includegraphics[width=0.49\textwidth]{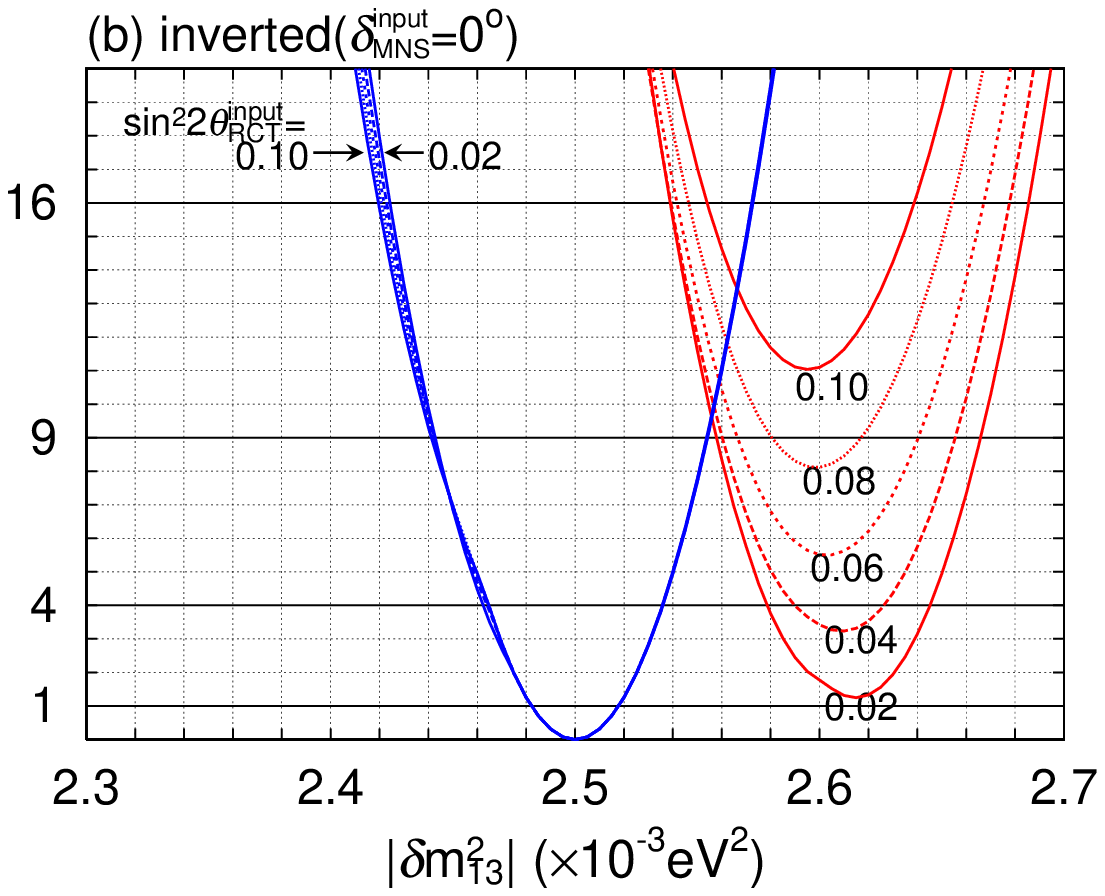}
 \caption{%
$\Delta\chi^2_{\rm min}$ of the T2KK experiment as a function of 
$|\dm13|$ for $3.0^\circ$ OAB at SK and $0.5^\circ$ OAB at
$L=1000$~km.
(a): The normal hierarchy case at $\dmns^{\rm input}=0^\circ$.
The solid line, long-dashed, short-dashed, dotted,
and the solid line again, show the results for
$\srct{2}^{\rm input}=0.02$, $0.04$, $0.06$,
$0.08$, and $0.1$, respectively.
All the other input parameters are those in
eqs.~(\ref{eq:input_physet})-(\ref{eq:eff_and_missID}).
The blue lines, which are almost degenerate, are obtained
when the right hierarchy is chosen in the fit,
whereas the red lines are obtained with the wrong hierarchy.
(b): The inverted hierarchy case.
}
\label{fig:dm13}
\end{figure}
%---

In Fig.~\ref{fig:dm13},
we show the minimum $\Delta\chi^2$ of the T2KK
experiment as a function of $|\dm13|$
with the optimum OAB combination
of $3.0^\circ$ at SK and $0.5^\circ$ at $L=1000$~km.
Fig.~\ref{fig:dm13}(a) is for the normal hierarchy and 
Fig.~\ref{fig:dm13}(b) is for the inverted hierarchy.
The five curves are for
$\srct{2}^{\rm input}=0.02$, $0.04$, $0.06$, $0.08$, and $0.1$,
which are denoted by the solid, long-dashed, short-dashed,
dotted, and the solid line again, respectively.
The CP phase is fixed at $\dmns=0^\circ$ and 
all the other parameters are those of
eqs.~(\ref{eq:input_physet})-(\ref{eq:eff_and_missID}).
In both cases there is a set of five curves with 
$\Delta \chi^2_{\rm min}=0^\circ$ at $|\dm13|=\numt{2.5}{-3}$ eV$^2$,
the input value.
All the five curves are almost degenerate in the set, which exhibits
the insensitivity of the $\nu_\mu \to \nu_\mu$ survival probability on
$\srct{2}$; see eqs.~(\ref{eq:Pmm}) and (\ref{eq:ABmu}).
On the other hand, there is another set of five curves
with $\Delta \chi^2_{\rm min}$
at $\numt{0.1}{-3}$ eV$^2$ smaller (larger) than the input value when
the mass hierarchy is normal (inverted).
These curves with $\srct{2}$ dependent $\Delta \chi^2_{\rm min}$
are obtained
when the opposite hierarchy is assumed in the fit.

%---
The larger mass-squared difference is determined 
from the T2KK experiment correctly as
\begin{equation}
 \left|\dm13\right|=\numt{\left(2.5 \pm 0.02 \right)}{-3}\mbox{{eV$^2$}}\,,
\label{eq:dm13input}
\end{equation}
if we know the mass hierarchy pattern.
However, if we do not know the mass hierarchy pattern,
the other solution
\begin{eqnarray}
 \left|\dm13\right|
&\simeq&\numt{(2.4\pm0.02)}{-3}\mbox{{eV$^2$~~~~(for the normal
hierarchy)}}\,, \nn\\
&\simeq&\numt{(2.6\pm0.02)}{-3}\mbox{{eV$^2$~~~~(for the inverted
hierarchy)}}\,, 
\label{eq:dm13T2KK}
\end{eqnarray}
appears for every $\srct{2}^{\rm input}$.
The wrong solution~(\ref{eq:dm13T2KK}) are about $3.5\sigma$ away from
the correct solution~(\ref{eq:dm13input}).
The difference of $\numt{\mp0.1}{-3}$~eV$^2$ in the mean value can be
explained by the phase shift term $B^\mu$
in the $\nu_\mu \to \nu_\mu$ survival probability;
see eqs.~(\ref{eq:Pmm}) and (\ref{eq:Bmu}).
From the peak location at 
\begin{equation}
\dfrac{\Delta_{13}}{2}+B^\mu=\dfrac{\pi}{2}\,,
\label{eq:dD13}
\end{equation}
the location of the solution with the wrong hierarchy can be
estimated as
\begin{eqnarray}
&&\hspace*{-10ex}
\left|\left(\dm13\right)^{\rm fit}\right|
-
\left|\left(\dm13\right)^{\rm input}\right|
\nn\\
&\simeq&
-\dfrac
{\left(\dm13\right)^{\rm input}}
{\left|\left(\dm13\right)^{\rm input}\right|} 
\left(0.11-0.023
\left(
\dfrac{\srct{2}^{\rm input}}{0.10}\right)^{1/2}\cos\dmns^{\rm input}
\right)
\times 10^{-3} \mbox{{eV$^2$}}\,.
\label{eq:dBmu3}
\end{eqnarray}
The magnitude of the difference is almost $\numt{0.09}{-3}$~eV$^2$
for $\srct{2}^{\rm input}=0.1$ at $\cos\dmns^{\rm input}=1$,
and it grows to $\numt{0.11}{-3}$~eV$^2$ as $\srct{2}$ decreases,
as can be observed from the figures.
This result suggests that the absolute value of 
the larger mass-squared difference
cannot be determined uniquely,
if the mass hierarchy pattern is not known.
Because the T2KK experiment can determine
the mass hierarchy from the $\nu_\mu \to \nu_e$ transition rates for
sufficiently large $\srct{2}^{\rm input}$,
the fake $|\dm13|$ can be excluded for larger $\srct{2}^{\rm input}$
as shown by the $\Delta \chi^2_{\rm min}$ values of the wrong solutions
in Figs.~\ref{fig:dm13}(a) and (b),
which grow with increasing $\srct{2}^{\rm input}$.
If we do not make use of the $\nu_\mu \to \nu_e$ transition signal in
the fit, all the solutions with the wrong mass hierarchy has
$\Delta \chi^2_{\rm min} \simeq 0$,
indistinguishable from the correct solution.
%---

%---
Let us note in passing that the T2K experiment suffers from the same
uncertainty in the measurement of $|\dm13|$.
If we drop all the data from the far detector in the above analysis
with $\dm13=\numt{\pm 2.5}{-3}$~eV$^2$,
we find the fake solution with
\begin{eqnarray}
 \left|\dm13\right|
&\simeq&\numt{(2.4\pm0.04)}{-3}\mbox{{eV$^2$~~~~(for the normal
hierarchy)}}\,, \nn\\
&\simeq&\numt{(2.6\pm0.04)}{-3}\mbox{{eV$^2$~~~~(for the inverted
hierarchy)}}\,, 
\end{eqnarray}
instead of eq.~(\ref{eq:dm13T2KK}).
The $\Delta \chi^2_{\rm min}$ values 
for the wrong solutions are indistinguishable from zero
for all the $\srct{2}$ input values.
The difference of about $\numt{0.1}{-3}$~eV$^2$ between the 
correct and the wrong solutions remains the same, because the
formulae~(\ref{eq:dD13})
and (\ref{eq:dBmu3}) are valid near the oscillation maximum at all
baseline length $L$ as long as the earth matter effect remains a
small perturbation as in eqs.~(\ref{eq:Prob2}) and (\ref{eq:ABmu}).
Since the two solutions are about $2\sigma$ away, 
the experiment should present two values of $|\dm13|$ until the mass
hierarchy is determined.

% ======================================================================== %
\section{CP phase}
\label{sec:6}
In this section, 
we study the capability of the T2KK experiment for measuring the
leptonic CP phase $\dmns$ with the optimum OAB combination,
$3.0^\circ$ OAB at SK and $0.5^\circ$ OAB at $L=1000$~km,
with $\numt{5}{21}$ POT exposure.

%---------------------
\begin{figure}
 \centering
  \includegraphics[height=0.23\textheight]{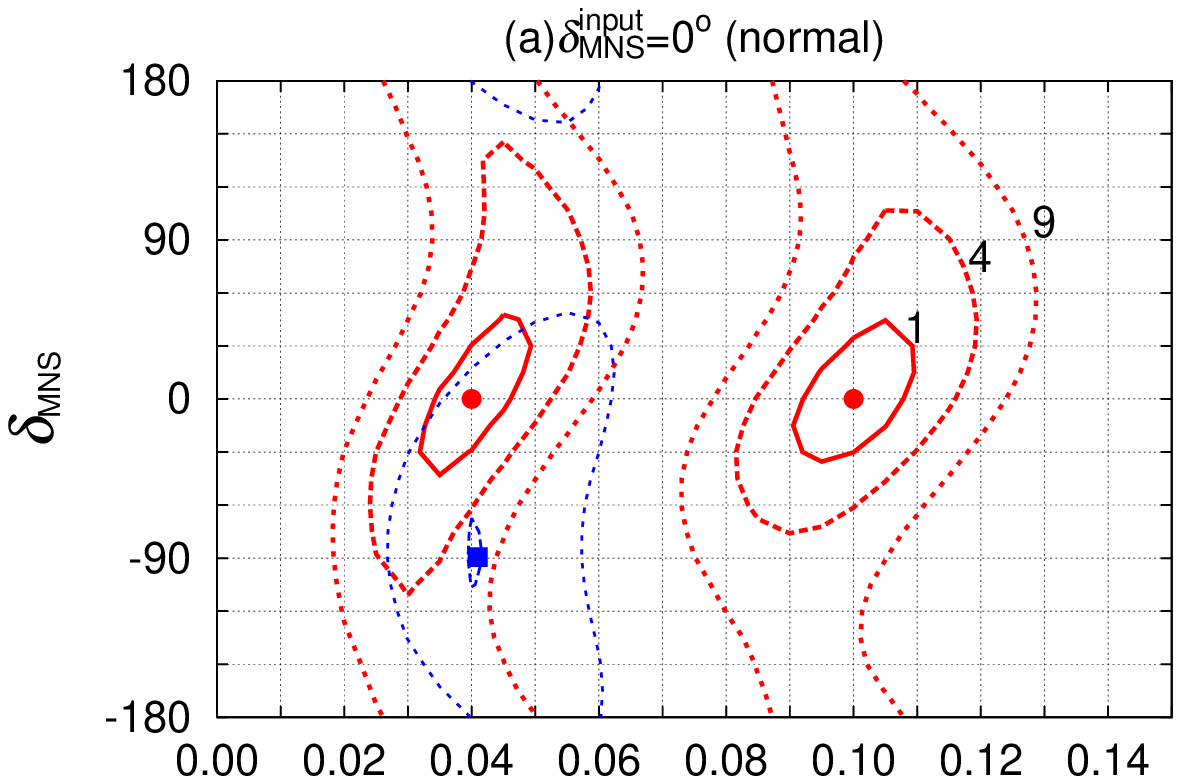}
 \hspace{-5ex}
  \includegraphics[height=0.23\textheight]{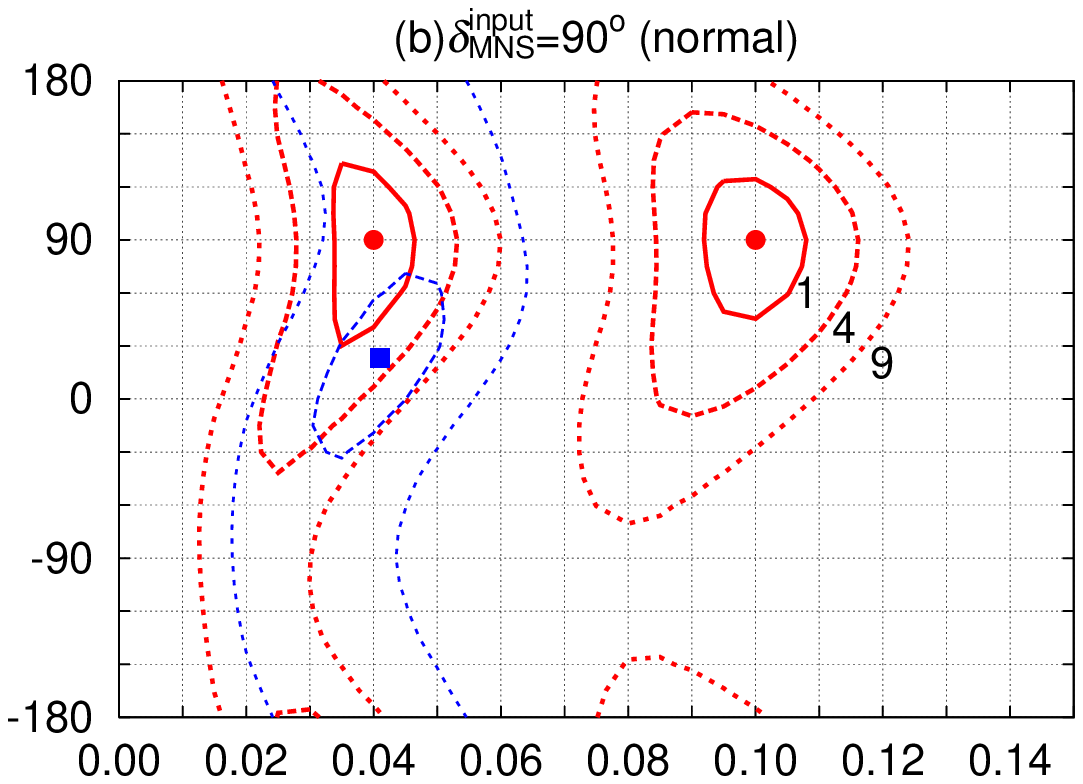}

  \includegraphics[height=0.23\textheight]{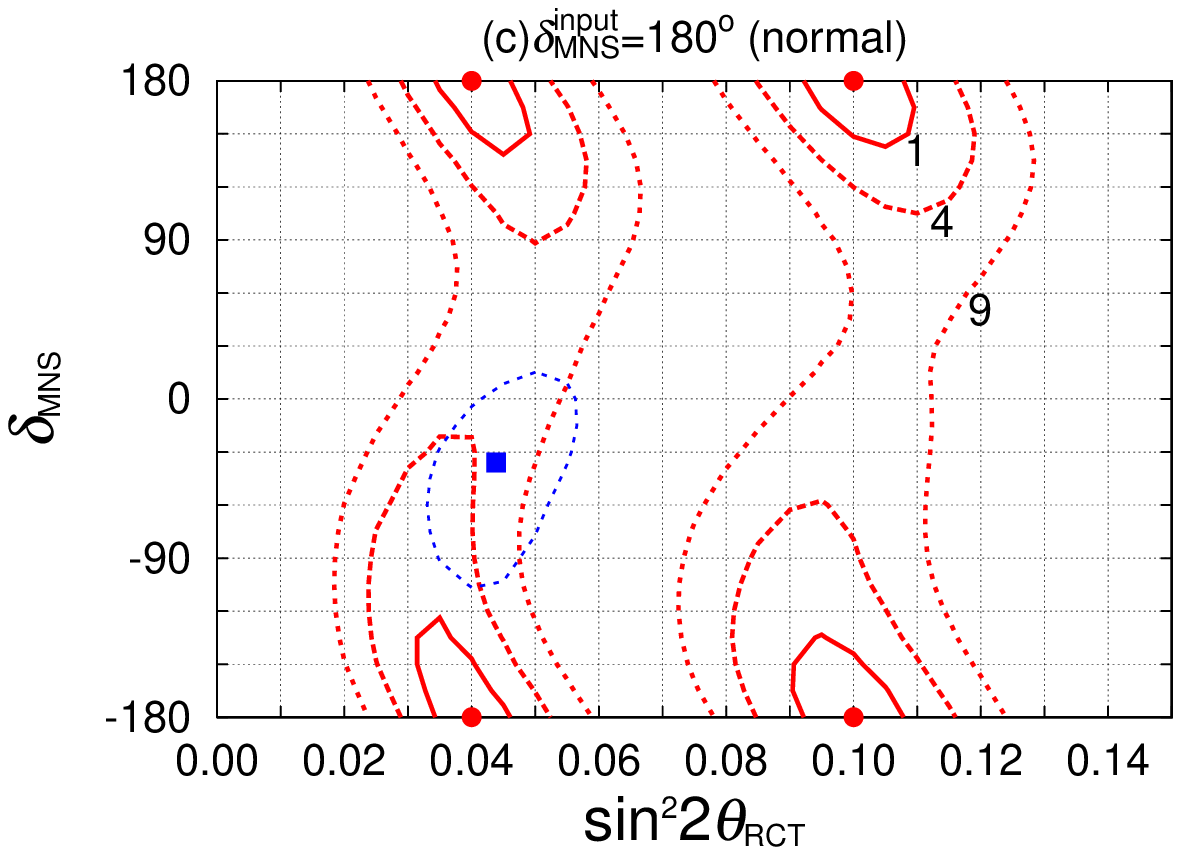}
 \hspace{-5ex}
  \includegraphics[height=0.23\textheight]{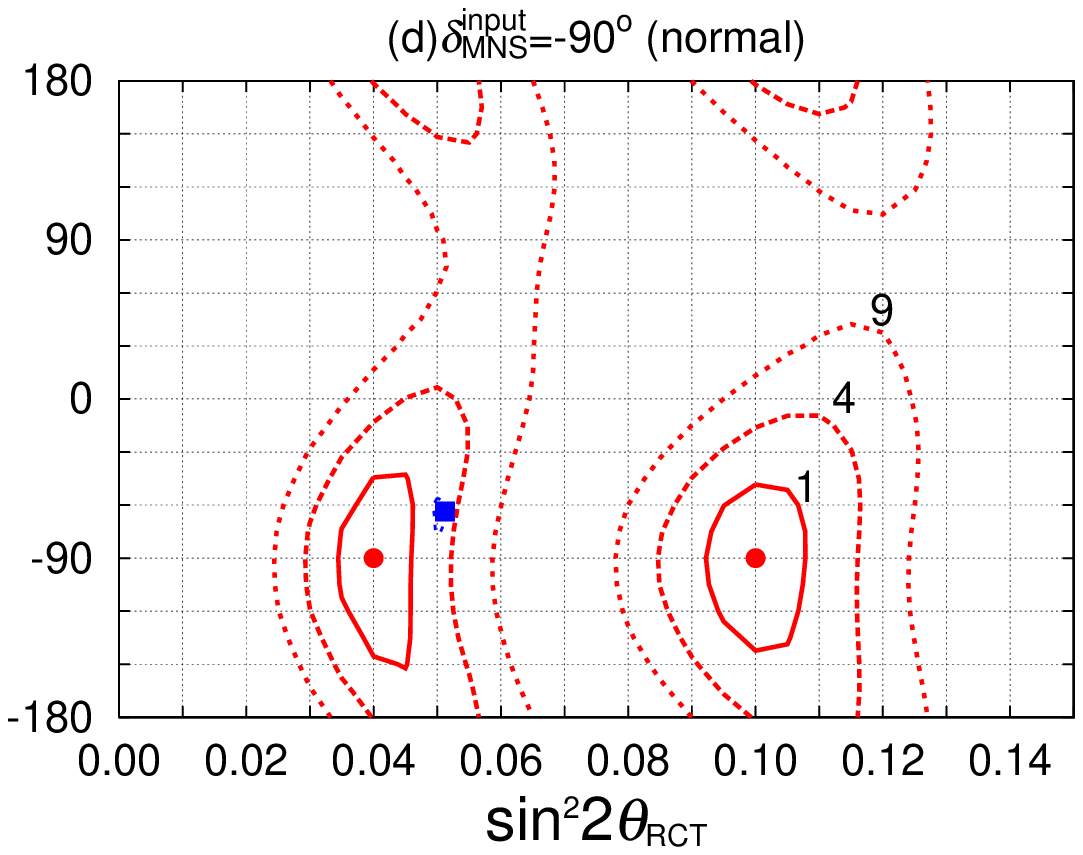}
  \caption{
The $\Delta \chi^2$ contour plot for the T2KK experiment in the
plane of $\srct{2}$ and $\dmns$ when the mass hierarchy is normal.
Allowed regions in the plane of $\srct{2}$ and $\dmns$ are shown for
the combination of $3.0^\circ$ OAB at SK and $0.5^\circ$ OAB 
at $L=1000$~km
with $\numt{5}{21}$ POT.
The input value of $\srct{2}$ is 0.10 and 0.04 for 
$\dmns=0^\circ$ (a), $\dmns=90^\circ$ (b),
$\dmns=180^\circ$ (c), and $\dmns=-90^\circ$ (d),
and the other input parameters are listed in 
eqs.~(\ref{eq:input_physet})-(\ref{eq:eff_and_missID}).
The input points are indicated as the solid blobs.
The contours for
$\Delta\chi^2=1$, $4$, and $9$
are shown by the solid, dashed,
and dotted lines, respectively.
The thick red contours are obtained when the right hierarchy 
is assumed in the fit, whereas the thin blue contours with the local
minimum by the solid square show the results when the opposite mass
hierarchy is assumed in the fit.
}
\label{fig:CPRCT.nor}
\end{figure}
%---------------------

%---
In Fig.~\ref{fig:CPRCT.nor},
we show $\Delta \chi^2$ contours in the plane
of $\srct{2}$ and $\dmns$ 
when the mass hierarchy is normal ($m^2_3-m^2_1>0$).
The input values are $\srct{2}^{\rm input}=0.10$ and 0.04, 
and $\dmns^{\rm input}=0^\circ$, $90^\circ$, $180^\circ$, and
$-90^\circ$ in Figs.~(a), (b), (c), and (d), respectively.
The other input parameters are those in
eqs.~(\ref{eq:input_physet})-(\ref{eq:eff_and_missID}).
The contours for $\Delta\chi^2=1$, $4$, and $9$ 
are shown by the solid, dashed, and dotted lines, respectively.
The thick red lines show the $\Delta \chi^2$ contours when
the right hierarchy is chosen in the fit,
whereas 
the thin blue lines show the results 
when the wrong hierarchy is assumed in the fit.
The solid blobs in each figure denote the input points
($\Delta \chi^2=0$)
and
the solid squares are the local minima for the fit
with the wrong hierarchy.
%--

%---------------------
\begin{figure}[t]
 \centering
  \includegraphics[height=0.23\textheight]{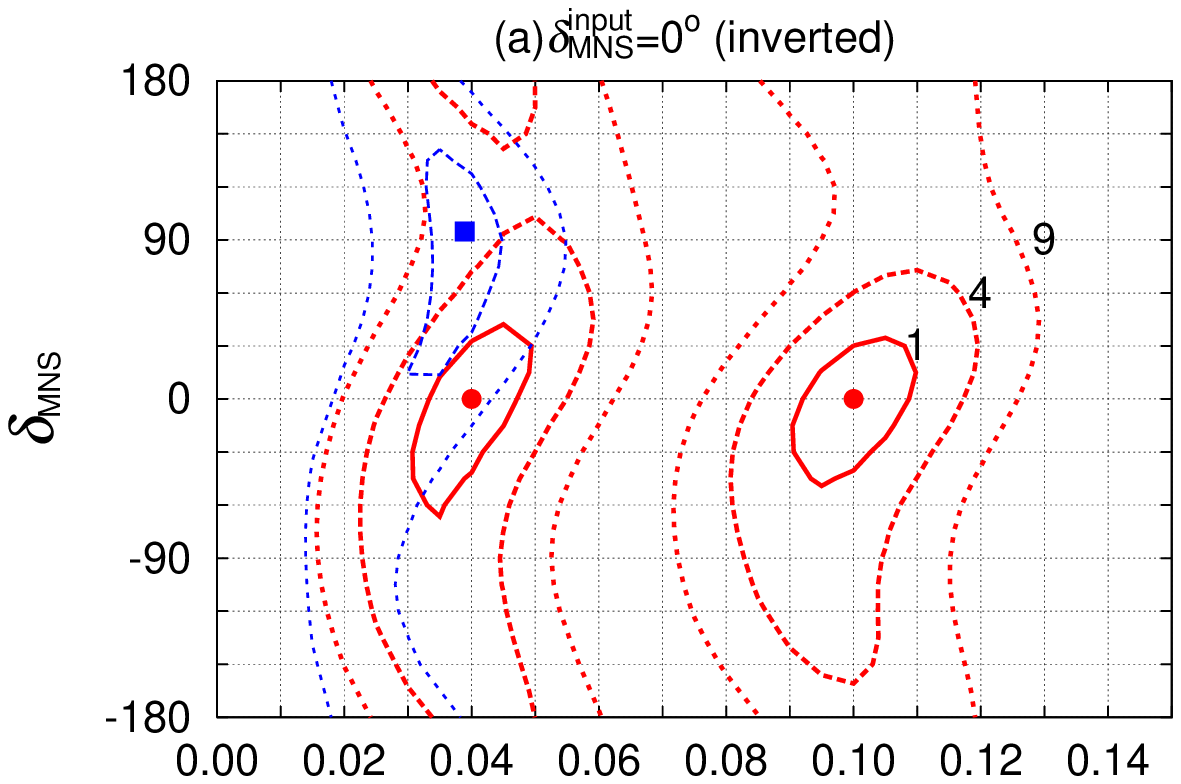}
 \hspace{-5ex}
  \includegraphics[height=0.23\textheight]{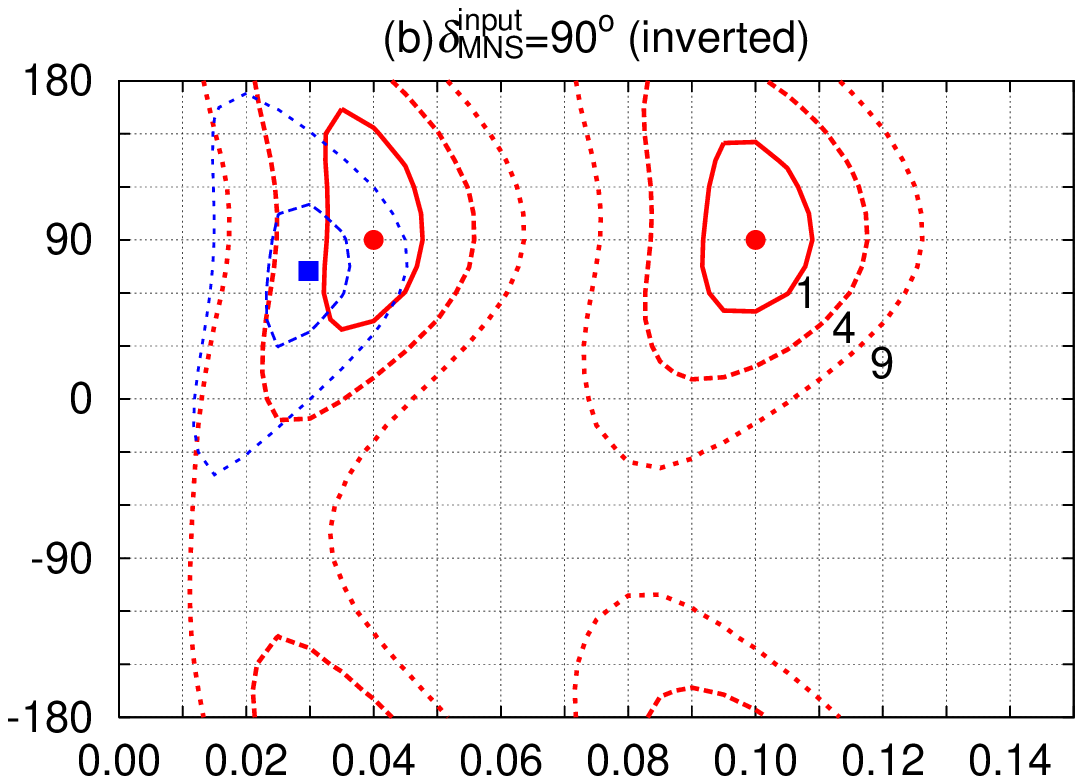}

  \includegraphics[height=0.23\textheight]{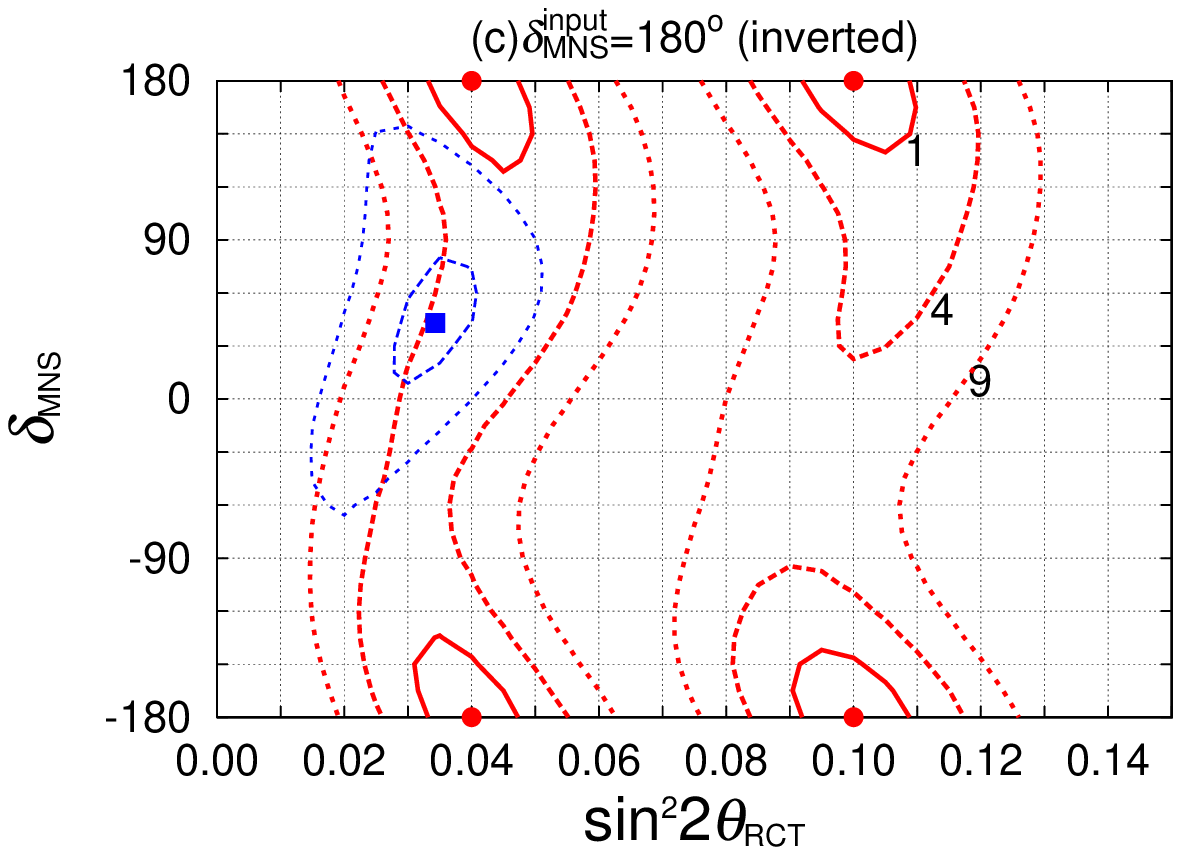}
 \hspace{-5ex}
  \includegraphics[height=0.23\textheight]{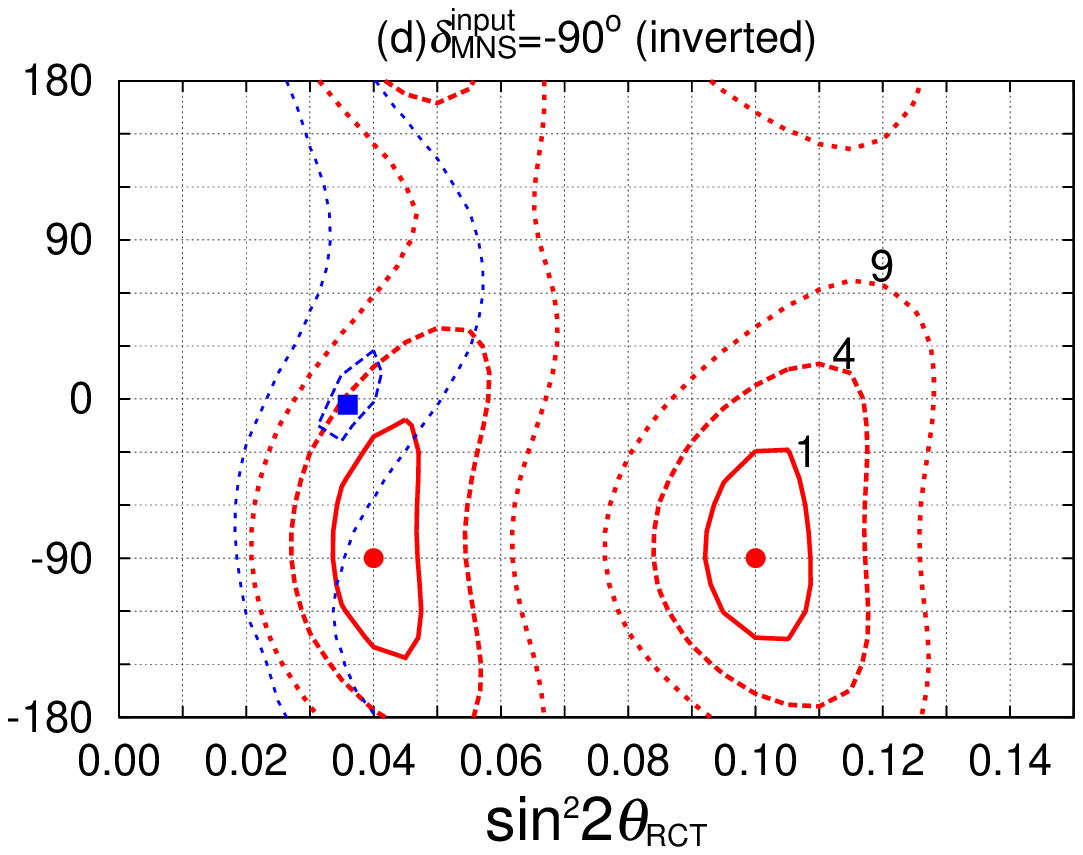}
  \caption{The same as Fig.~\ref{fig:CPRCT.nor},
but when the mass hierarchy is inverted ($m^2_3-m^2_1<0$).}
\label{fig:CPRCT.inv}
\end{figure}
%---------------------

We find from the figures that $\dmns$ can be constrained to about
$\pm 45^\circ$ at $1\sigma$ level for all the input value of
$\dmns^{\rm input}$ and $\srct{2}$.
The insensitivity of the measurement error of
$\dmns$ on $\srct{2}$~\cite{HOS1,HOS2,supernova}
persists.
However, at $3\sigma$ level, the contour closes only for 
$\dmns^{\rm input}=\pm90^\circ$ at $\srct{2}^{\rm input}=0.1$;
(b) and (d).
Moreover, there appears a shadow island where the inverted hierarchy
is assumed in the fit, for all the four $\dmns^{\rm input}$ cases
at $\srct{2}^{\rm input}=0.04$.
The shadow contours cover the whole $\dmns$ region
when $\dmns^{\rm input}=90^\circ$,
where the number of the $\nu_\mu \to \nu_e$ signal events is the smallest
among the four $\dmns^{\rm input}$ cases.
%-----

These observations are in sharp contrast with the previous ones,
shown in \eg Fig.~8 of ref.~\cite{HOS2},
where it has been shown that the $\dmns$ can be constrained to
about $\pm30^\circ$ for all the four input $\dmns$ values at
$\srct{2}^{\rm input} \gsim 0.02$ and
that the shadow islands from the wrong hierarchy solution are small and
they appear only for $\dmns^{\rm input}=90^\circ$ at
$\srct{2}^{\rm input} = 0.04$
and
for $\dmns^{\rm input}=0^\circ$, $90^\circ$, and $180^\circ$
at $\srct{2}^{\rm input} = 0.02$.
We find that both the reduction of the sensitivity from
$\pm 30^\circ$ to $\pm 45^\circ$ and the appearance of the big shadow
islands are mainly due to the $\pi^0$ background for the $e$-like
events,
while the smearing effects due to nuclear Fermi motion and the
detector resolution also contribute at the sub-leading level.

%-----
In Fig.~\ref{fig:CPRCT.inv},
we show the same contour plots as in Fig.~\ref{fig:CPRCT.nor},
but for the inverted hierarchy case.
We find that the $1\sigma$ constraints on $\dmns$
are slightly worse than those of the normal hierarchy case in
Fig.~\ref{fig:CPRCT.nor}:
The $1\sigma$ error remains at about $\pm45^\circ$ for 
$\dmns^{\rm input}=0^\circ$ and $180^\circ$,
but it grows to about $\pm 60^\circ$ or larger for
$\dmns^{\rm input}=\pm 90^\circ$.
As in the case of the normal hierarchy,
$\dmns$ can be constrained at $3\sigma$ level only for 
$\dmns^{\rm input}=\pm90^\circ$ at $\srct2^{\rm input}=0.1$.
The $2\sigma$-level shadow islands appear for all the four
$\dmns^{\rm input}$ cases at $\srct{2}^{\rm input}=0.04$, 
which is consistent with the observation of Fig.~\ref{fig:contOAB}(a2)
where all $\dmns^{\rm input}$ points lie below the 
$\Delta \chi_{\rm min}^2=4$ contours.
The $3\sigma$ contours of the wrong solutions,
denoted by the thin blue dotted lines,
cover the whole $\dmns$ region for $\dmns^{\rm input}=0^\circ$
and $-90^\circ$ at $\srct{2}^{\rm input}=0.04$.
The significant loss of the sensitivity to $\dmns$ as compared to 
Fig.~9 of ref.~\cite{HOS2} can also be explained by
the $\pi^0$ background to the $\nu_\mu \to \nu_e$ oscillation signal.
%-------

In summary, the capability of the T2KK experiment to measure the CP
phase of the lepton flavor mixing (MNS) matrix is significantly
worsened by the $\pi^0$ background in both normal and inverted
hierarchy cases.
This is because the large $\pi^0$ background to the $\nu_\mu \to \nu_e$
oscillation signal at the far detector, 
as shown in Figs.~\ref{fig:eventKr}(b) and (d), reduce significantly
the sensitivity to the amplitude-shift term $A^e$ and the phase-shift
term $B^e$ which have contributions proportional to $\sin \dmns$ and
$\cos \dmns$, respectively~\cite{HOS1,HOS2}.
These terms proportional to $\Delta_{12}$ in eq.~(\ref{eq:ABCe})
can be measured by comparing the shifts at a near ($L\simeq 300$~km)
and a far ($L\simeq1000$~km) detectors~\cite{HOS1,HOS2,supernova}
without using the $\bar{\nu}_\mu$ beam.
Since the $\pi^0$ background worsens the measurements of $A^e$ and $B^e$
at the far detector, the sensitivity to $\dmns$ deteriorates significantly.
The use of $\bar{\nu}_\mu$ beam in addition to the $\nu_\mu$ beam
\cite{Kajita,supernova}
may be helpful in recovering the sensitivity,
since at least the detector-dependent errors of the $\pi^0$ background
events should be common for both beams.

% ======================================================================== %
\section{Summary and conclusion}
\label{sec:7}
In this paper, 
we elaborate the previous analyses of ref.~\cite{HOS1,HOS2}
on the physics potential of the T2KK experiment
by taking into account
the smearing of reconstructed neutrino energy 
due to the Fermi motion of the target nucleus
and 
the finite resolution of $e^\pm$ and $\mu^\pm$ momenta
in a water \cerenkov detector.
We also include
the events from the non-CCQE ``resonance'' events that survive
the CCQE event selection cut of eq.~(\ref{eq:criteriaCC}),
and
the contribution from the single $\pi^0$ production via the
neutral current interactions,
which mimic the $\nu_e$ appearance signal in a water
\cerenkov detector.

In order to estimate the reconstructed energy ($\Erec$) distribution
efficiently,
we introduce the smearing functions for the CCQE and 
non-CCQE ``resonance'' events
that map the incoming neutrino energy $E_\nu$ onto the 
reconstructed energy $\Erec$ 
by using the Mote Carlo event generator {\sf nuance}~\cite{nuance}.
The effect of the detector resolution for $e^\pm$ and $\mu^\pm$,
see Table~\ref{tab:resolution}, has also been taken into account.
The smearing functions for the CCQE events are given
in eq.~(\ref{eq:fit_ccqe0}) 
with eqs.~(\ref{eq:r_ccqe_mu})-(\ref{eq:dE_ccqe_mu}) for $\nu_\mu$,
and
eqs.~(\ref{eq:r_ccqe_e})-(\ref{eq:dE_ccqe_e})
for $\nu_e$ .
Those for non-CCQE ``resonance'' events are parameterized as in
eq.~(\ref{eq:fit_res3}) with
eqs.~(\ref{eq:sig_res_m3})-(\ref{eq:r_res_e3})
in the region of $0.55\leq E_\nu \leq 1.2$~GeV
and
eq.~(\ref{eq:fit_res4}) with
eqs.~(\ref{eq:sig_res_m4})-(\ref{eq:r_res_e4})
for $1.2$~GeV$< E_\nu<6.0$~GeV.
%---
For estimating the background from the single
$\pi^0$ production,
we generate single $\pi^0$ events from the NC interactions for each
off-axis beam (OAB) also by using {\sf nuance}~\cite{nuance},
and
parameterize the probability that a $\pi^0$ is misidentified as
an $e^\pm$-like event, $P_{e/\pi}$, 
in terms of the energy ratio and the opening angle of the two photons
for the $\pi^0$ decay-in-flight; see Fig.~\ref{fig:pi0pt}(b)
and eqs.~(\ref{eq:def_P}) and (\ref{eq:def_Ppie2}).
%---

%---
We study the sensitivity of the T2KK experiment on the neutrino
mass hierarchy by placing a water \cerenkov detector with 100~kton
fiducial volume at various location in Korea for the $3.0^\circ$ and 
$2.5^\circ$ OAB at SK.
The neutrino beam at an off-axis angle greater than about 0.5 (1.0)
can be observed in Korea, at the baseline length
$1000$~km$\lsim L \lsim 1200$~km,
for the $3.0^\circ$ OAB ($2.5^\circ$ OAB) at SK.
We find that the highest sensitivity is achieved for the combination
of $3.0^\circ$ OAB at SK and $0.5^\circ$ OAB at $L=1000$~km,
confirming the results of ref.~\cite{HOS1,HOS2}.
With $\numt{5}{21}$ POT, which is the planned exposure of the T2K
experiment,
the mass hierarchy can be determined at $3\sigma$ level if
$\srct{2}\gsim 0.08$ ($0.09$) for the above OAB combination,
when the neutrino mass hierarchy is normal (inverted).
For the combination of $2.5^\circ$ OAB at SK and $1.0^\circ$ OAB at
$L=1000$~km, the $3\sigma$ sensitivity is obtained for $\srct{2}\gsim0.12$
for both hierarchies; see Fig.~\ref{fig:contOAB} in 
section~\ref{sec:dependence-oab-sk}.
These figures show significant reduction of the sensitivity as compared
to the results of the previous studies, such as Fig.6 of ref.~\cite{HOS2},
which show that the neutrino mass hierarchy 
can be determined for $\srct{2}\gsim0.05 (0.06)$ at $3\sigma$,
when the hierarchy is normal (inverted),
with the same combinations of the OAB's, 
and with the same detector size and the POT.
We find that the main cause of the reduction in the sensitivity is
the background from the single $\pi^0$ production;
see Table~\ref{tab:alteration1} in
section~\ref{sec:best-combination}.
The smearing in the reconstructed energy has a significant effect when
$\dmns\simeq 180^\circ$, 
where the mass hierarchy dependent 
oscillation phase-shift term is large.
The contribution from the non-CCQE ``resonance'' events
help discriminating the mass hierarchy, 
because these events are also a part of the $\nu_\mu \to \nu_e$
oscillation signal.
%---

%---
We also examine the prospect of the CP phase measurement for the T2KK
experiment with the above OAB combination.
The sensitivity of the $\dmns$ measurement is also reduced 
significantly from that of the previous study in ref.~\cite{HOS2},
which found the $1\sigma$ error of about $\pm30^\circ$,
to about $\pm 45^\circ$ or even $\pm 60^\circ$ in some cases.
The main cause of the worsening of the error is again
the $\pi^0$ background for the $e$-like events
at the far detector that makes it difficult to measure the baseline
dependence of the $\nu_\mu \to \nu_e$ oscillation amplitude and the
phase:
$\sin \dmns$ is measured by the amplitude difference
and 
$\cos \dmns$ is measured by the phase difference~\cite{HOS1,HOS2}.

%---
The $\pi^0$ background reduces significantly
the physics potential for the mass hierarchy
determination and the CP phase measurement
of the T2KK experiment.
If we understand better the physics of the $\pi^0$ production and 
its decay signal inside the water \cerenkov detector,
the sensitivity of the experiment on these fundamental parameters
should be improved.
Detailed investigation of the normalization and the shape of the $\pi^0$
background should be one of the most important tasks 
to evaluate quantitatively the physics discovery potential of the T2KK
experiment.

% ======================================================================== %
\section*{Acknowledgments}
\label{sec:ack}

We would like to thank Y.~Hayato for useful discussions and comments
on the neutrino interactions and the background for the 
water \cerenkov detector.
We also thank our colleagues A.K.~Ichikawa, T.~Kobayashi, T.~Nakaya,
and K.~Nishikawa from whom we learn about the K2K and T2K experiments
and K-i. Senda for discussions on the matter profile along the T2K
and T2KK baselines.
We are also grateful to D.~Casper for providing us with the newest
{\sf nuance}~\cite{nuance} code and 
to C.V.~Andreopoulos for his help with {\sf genie}~\cite{genie}.
K.H. wishes to thank the Aspen Center for Physics
and the Phenomenology Institute at the University of Wisconsin
for their hospitality during his visits,
where he enjoyed stimulating discussions with V.~Barger, P.~Huber, and
S.~Petcov.
The work is supported in part by the Core University Program of JSPS,
and 
in part by the Grant in Aid for Scientific Research (No.18340060,
No.20039014) from MEXT, Japan.

% ======================================================================== %
\appendix
\setcounter{equation}{0}
\renewcommand{\theequation}{A\arabic{equation}}
\renewcommand{\thesection}{Appendix~\Alph{section}}
\section{Smearing functions $f^X_{\alpha}(\Erec;E_\nu)$}
\label{sec:appA}

In the appendix,
we show our parameterization of the smearing functions, 
$f^X_{\alpha}(\Erec;E_\nu)$,
which map the incoming neutrino energy, $E_\nu$, onto the
reconstructed energy, $\Erec$,
for the quasi-elastic events.
The superscript $X$ denotes the event type,
$X=$ CCQE for the CCQE events, or
$X=$ Res for the non-CCQE ``resonance'' events
that pass the CCQE selection criteria of eq.~(\ref{eq:criteriaCC}),
and the subscript $\alpha$ is for $\mu$ or $e$:
$\alpha=\mu$ for $\nu_\mu$ events
and
$\alpha=e$ for $\nu_e$ events.
These functions take account of 
the Fermi motion of the target nucleon
inside the oxygen nucleus
and
the finite energy-momentum resolutions of a muon and 
an electron in a water \cerenkov detector
listed in Table~\ref{tab:resolution}.

\subsection{CCQE events}
\label{sec:appA1}
The $\Erec$ distribution of the CCQE events,
which are generated by {\sf nuance}~\cite{nuance},
can be parameterized accurately by 3 Gaussians,
\begin{eqnarray}
 f^{\rm CCQE}_{\alpha}(\Erec;E_\nu) =
 \dfrac{1}{A^\alpha(E_\nu)}
\sum_{n=1}^{3}
r^\alpha_n(E_\nu) 
\exp
\left(-\dfrac{(\Erec-E_\nu+{\delta}E^\alpha_n(E_\nu))^2}
{2(\sigma^\alpha_n(E_\nu))^2}
\right)
\,,
\label{eq:fit_ccqeA0}
\end{eqnarray}
in the region of
\begin{eqnarray}
0.3\mbox{{~GeV}}\leq E_\nu \leq 6.0\mbox{{~GeV}}
\mbox{{~~for~~}}
0.4\mbox{{~GeV}}\leq \Erec \leq 5.0\mbox{{~GeV}}\,.
\label{eq:regionCCQE}
\end{eqnarray}
The index $\alpha$ takes $\mu$ for $\nu_\mu$
and $e$ for $\nu_e$ events,
and each function is normalized by
\begin{eqnarray}
 A^\alpha(E_\nu)
= \sqrt{2\pi}\sum_{n=1}^{3}
  r^\alpha_n(E_\nu)
     \sigma_n^\alpha(E_\nu)\,.
\label{eq:normalzationCCQE}
\end{eqnarray}
The variance $\sigma^\alpha_n$, 
the energy shift ${\delta}E^\alpha_n$,
and the normalization factors
$r^\alpha_{2,3}$
are functions of the incoming neutrino energy $E_\nu$,
with $r^{\alpha}_1 (E_\nu) = 1$.
The first and the second Gaussians account mainly
for the nuclear Fermi motion,
and we can set $\delta E_2^\mu=\delta E_1^\mu$.
The third Gaussian is necessary to account for the asymmetry in the
$\Erec-E_\nu$ distribution
such as the Fermi block effect at the low energies,
and the asymmetric momentum resolution effects at
high energies.
%---

%--
All the coefficients are parameterized compactly by using the variables
\begin{equation}
 x = E_\nu~\mbox{{[GeV]}}-1\,,
\hspace*{10ex}
 \xi = \sqrt{E_\nu~\mbox{{[GeV]}}}-1\,,
\label{eq:defx}
\end{equation}
which vanish at $E_\nu=1$~GeV.
The variance $\sigma^\mu_n$(MeV) 
of the three Gaussians 
\begin{subequations} %CCQE-sigma
\begin{eqnarray}
 \sigma^\mu_1 &=&
 39.7+68.5 \xi
\,,\\
 \sigma^\mu_2 &=&
    82.7
  - 50.2x
  + 259\xi\,,\\
\sigma^\mu_3 &=&
 197
+486x
-606\xi
+203x\xi\,,
\end{eqnarray} 
\label{eq:sig_ccqe_mu}
\end{subequations}
$\!\!\!\!$
and the energy shift terms $\delta E_n^\mu$ (MeV)
\begin{subequations} %CCQE_E
\begin{eqnarray}
 {\delta}E^\mu_1 =
 {\delta}E^\mu_2 &=& 
 35-2.5x(1-1.2x)/(1+x)\,,\\
 {\delta}E^\mu_3 &=& 
 \sigma_3^\mu
\left[
0.053
+0.033x
\left(
\dfrac{
1 + 5.27x - 8.67x^2 +  1.83x^3
}{1 + 3.65x + 4.35x^2}
\right)
\right]\,,
\end{eqnarray}
\label{eq:dE_ccqe_mu}
\end{subequations}
$\!\!\!\!$
are given in units of MeV.
The normalization factors are $r^\mu_1=1$ and
\begin{subequations} %CCQE_r
\begin{eqnarray}
 r^\mu_2&=& 
  1.1
 -0.96x
 +0.44x^2
 -0.076x^3
 +0.0047x^4
\,,\\
 r^\mu_3&=&
0.365-1.97\xi
(1-0.634\xi+0.464\xi^2+0.293\xi^3-0.342\xi^4)
/(1+1.23x)
\,.~~~~~
\end{eqnarray} 
\label{eq:r_ccqe_mu}
\end{subequations}

\vspace*{-3ex}
%--- variance
The first and second variances are determined mainly by 
the sum of the nuclear Fermi motion and
the momentum resolution of the water \cerenkov detector.
In the absence of the momentum resolution error, two Gaussians,
one with a constant variance of $\sim 60$~MeV and the other with
a larger variance of $\sim 190$~MeV at $E_\nu\sim1$~GeV which
decreases slowly with energy,
can account for the bulk of the Fermi motion effects on
the $\Erec-E_\nu$ distribution;
see Fig.~\ref{fig:event-select}.
It is the smearing effect due to the energy resolution which increases
the first two variances as $\sqrt{E_\nu}$ at high energies.
%
%---
The value of $\delta E_{1,2}^\mu$ does not depend on $E_\nu$ much,
because they are essentially determined by the nucleon and lepton
masses; see eq.~(\ref{eq:erec1}).
The third Gaussian has much larger variance than the first two,
and it accounts for the Fermi-blocking effect at small $E_\nu$
and 
the momentum resolution asymmetry at high energies.
Consequently, $r_3^\mu$ is significant only at low energies
($E_\nu<0.7$~GeV) and at high energies ($E_\nu\geq4$~GeV).

%%%-------
For the $\nu_e$ case,
the variance $\sigma^e_n$(MeV) is expressed as
\begin{subequations} %CCQE-sigma
\begin{eqnarray}
 \sigma^e_1&=&
55.5
-19.6x
+98.9\xi\,,\\
 \sigma^e_2&=&
125
-51.3x
+201\xi\,,\\
 \sigma^e_3&=&
  273
-102x
+1560\xi
+111x\xi\,,
\end{eqnarray} 
\label{eq:sig_ccqe_e}
\end{subequations}
$\!\!\!\!$
the shift term ${\delta}E_n^e$(MeV) is given as
\begin{subequations} %CCQE_E
\begin{eqnarray}
 {\delta}E^e_1 = 
 {\delta}E^e_2 &=& 
  40
-0.99x
+3.3x^2
-0.71x^3
-2.2x/(1+x)\,,\nn\\
 {\delta}E^e_3&=& 
\sigma_3^e
\left[
-0.16
+0.68x
-2.6\xi
+1.1x/(1+x)
\right]\,,
\end{eqnarray}
\label{eq:dE_ccqe_e}
\end{subequations}
$\!\!\!$
and 
the normalized factors are $r^e_1=1$ and
\begin{subequations} %CCQE_r
\begin{eqnarray}
r^e_2 &=& 
0.67
-0.58x
+0.58x^2
-0.16x^3
+0.019x^4\,,\\
 r^e_3 &=&
0.094
-0.040x
+0.031x^2
-0.016x^3
+0.0059x^4\,.
\end{eqnarray} 
\label{eq:r_ccqe_e}
\end{subequations}

%--------
\vspace*{-4ex}
The three variances in eq.~(\ref{eq:sig_ccqe_e})
behave similarly to those for $\nu_\mu$,
but $\sigma_n^e$ is larger than $\sigma_n^\mu$,
because the energy resolution of the $e$-like events
are worse than that of the $\mu$-like events;
see Table~\ref{tab:resolution} in section~\ref{sec:3}.
%
%-----
The energy shifts $\delta E_{1,2}^e$ behave similarly to
$\delta E_{1,2}^\mu$,
while
$\delta E_3^e$ differs significantly from
$\delta E_3^\mu$ at low energies,
because the asymmetry of the $\Erec-E_\nu$ distribution
in the sub-GeV region is sensitive to
the mass and the momentum resolution of the emitted charged lepton.
The normalizations $r_{2,3}^e$ behave similarly to $r_{2,3}^\mu$,
except 
at very low energies ($E_\nu \lsim 0.7$~GeV)
when the muon mass in not negligible
and 
at very high energies ($E_\nu \gsim 3.5$~GeV)
due to resolution effects.

\subsection{Nuclear Resonance Events}
\label{sec:appA2}
The $\Erec$ distribution generated by {\sf nuance}~\cite{nuance}
for the non-CCQE events that pass the CCQE selection criteria of
eq.~(\ref{eq:criteriaCC})
is also parameterized for $\nu_\mu$ or $\nu_e$.
We find that 3 Gaussians
\begin{eqnarray}
 f^{\rm res}_{\alpha}(\Erec;E_\nu\leq1.2\mbox{~GeV}) = 
\dfrac{1}{\hat{A}^\alpha(E_\nu)}
\sum_{n=1}^{3}
\hat{r}^\alpha_n(E_\nu) 
\exp
\left(-\dfrac{(\Erec-E_\nu+\delta\hat{E}_n^\alpha(E_\nu))^2}
     {2(\hat{\sigma}_n^\alpha(E_\nu))^2}
\right)
\,,
\label{eq:fit_resA3}
\end{eqnarray}
suffice in the region of
\begin{eqnarray}
0.55\mbox{{~GeV}}\leq E_\nu \leq 1.2\mbox{{~GeV}}
\mbox{{~~for~~}}
0.4\mbox{{~GeV}}\leq \Erec \leq 5.0\mbox{{~GeV}}\,,
\label{eq:Range-Res-1}
\end{eqnarray}
whereas 4 Gaussians
\begin{eqnarray}
f^{\rm res}_{\alpha}(\Erec;E_\nu>1.2\mbox{~GeV})
=
\dfrac{1}{\tilde{A}^\alpha(E_\nu)}
\sum_{n=1}^{4}
\tilde{r}^\alpha_n(E_\nu) 
\exp
\left(-\dfrac{(\Erec-E_\nu+{\delta}\tilde{E}^\alpha_n(E_\nu))^2}
 {2(\tilde{\sigma}_n^\alpha(E_\nu))^2}
\right)
\,,
\label{eq:fit_resA4}
\end{eqnarray}
are necessary in the region of
\begin{eqnarray}
1.2\mbox{{~GeV}} <    E_\nu \leq 6.0\mbox{{~~GeV}}
\mbox{{~~for~~}}
0.4\mbox{{~GeV}} \leq \Erec \leq 5.0\mbox{{~~GeV}}\,,
\label{eq:Range-Res-2}
\end{eqnarray}
because the number of contributing resonances grow at high energies.
Here again
$\alpha=\mu$ for $\nu_\mu$
and
$\alpha=e$ for $\nu_e$ events,
and the functions are normalized as in eq.~(\ref{eq:normalzationCCQE}).

By using the same variables $x$ and $\xi$ in eq.~(\ref{eq:defx}),
the variances $\hat{\sigma}_n^\mu$(MeV)
of the 3 Gaussians in eq.~(\ref{eq:fit_resA3})
are 
\begin{subequations} %sig_res_m3
\begin{eqnarray}
 \hat{\sigma}^\mu_1 &=& 
97.8
-3670 x
+7410 \xi
+1540 x\xi\,,\\
 \hat{\sigma}^\mu_2 &=& 
98
+390x
+500x^2
\,,\\
 \hat{\sigma}^\mu_3 &=& 
27
-22x
-33x^2
-480x^3
-1000x^4\,,
\end{eqnarray} 
\label{eq:sig_res_m3} 
\end{subequations}
$\!\!\!\!$
and the energy shift terms $\delta\hat{E}_n^\mu$(MeV) are
\begin{subequations} %dE_res_m3
\begin{eqnarray}
\delta\hat{E}^\mu_1 &=&
382
-8170x
+16200\xi
+3670x\xi\,,\\
\delta\hat{E}^\mu_2 &=&
579
-22600x
+45800\xi
+10800x\xi \,,\\
\delta\hat{E}^\mu_3 &=&
210
-12x
-7.8x^2
-170x^3
\,,
\end{eqnarray}
\label{eq:dE_res_m3} 
\end{subequations}
and the normalized factors are $\hat{r}_1^\mu=1$ and 
\begin{subequations} %r_res_m3
\begin{eqnarray}
\hat{r}^\mu_2 &=&
0.42
+ 0.16x 
- 1.9x^2
\,,\\
\hat{r}^\mu_3 &=&
0.1
-0.202x/\left(1+2.03x\right)\,.
\end{eqnarray} 
\label{eq:r_res_m3} 
\end{subequations}

%-----------------
\vspace*{-3ex}
The first Gaussian is mainly related to the $\Delta$-resonance.
The order of the first variance is similar to the sum of 
the width of the $\Delta(1232)$, $\Gamma_\Delta\simeq 60$~MeV~\cite{PDB},
the Fermi motion of the target, $\sigma_{\rm Fermi}\simeq 60$~MeV,
and the momentum resolution $\sigma_{\delta p/p}\sim 30$~MeV
at $E_\nu\simeq 1$~GeV.
The value of $\delta\hat{E}_1^{\mu,e}$ is roughly the distance
between the peak of the CCQE events and that of the $\Delta$ events,
which is about 400~MeV.
The second Gaussian with growing variance of about $100$~MeV
at $E_\nu=1$~GeV and with larger energy shift of $600$~MeV
accounts for contribution of N(1440) and higher resonances.
The third Gaussian is necessary to take account of 
the nuclear effects
and
the asymmetry from the momentum resolution.

%---
For the $\nu_e$ case, we find
\begin{subequations} %sig_res_e3
\begin{eqnarray}
 \hat{\sigma}^e_1 &=&
102
-2960x
+5940\xi
+1300x\xi \,,\\
 \hat{\sigma}^e_2 &=& 
114 
+370.0x
+244x^2
\,,\\
 \hat{\sigma}^e_3 &=& 
27.1
-717x
+1390\xi
+425x\xi
\,,
\end{eqnarray} 
\label{eq:sig_res_e3} 
\end{subequations}
$\!\!\!\!$
and
\begin{subequations} %dE_res_e3
\begin{eqnarray}
\delta\hat{E}^e_1 &=&
372
-9530x
+18900\xi
+4630x\xi\,,\\
\delta\hat{E}^e_2 &=&
580.0
-15600x
+31400\xi
+6070x\xi\,,\\
\delta\hat{E}^e_3 &=&
\hat{\sigma}^e_3\left(
7.6 + 5.3x - 9.4x^2  - 24x^3
\right)
\,,
\end{eqnarray}
\label{eq:dE_res_e3} 
\end{subequations}
$\!\!\!\!$
both in MeV units, $\hat{r}^e_1=1$ and
\begin{subequations} %r_res_e3
\begin{eqnarray}
 \hat{r}^e_2 &=&
0.37-5.8x+13\xi
\,,\\
 \hat{r}^e_3 &=&
 0.067-0.068x-0.37x^2-3.4x^3
\,.
\end{eqnarray} 
\label{eq:r_res_e3} 
\end{subequations}

There are no large difference between $\nu_e$ and $\nu_\mu$ for
all variances, energy shift terms, and the normalization factors.
The small differences are mainly due to the difference in the $e$
and $\mu$ momentum resolutions.

%%%-------------------------------------------------------------
In the high-energy region of eq.~(\ref{eq:Range-Res-2}),
we introduce variables $y$ and $\eta$
\begin{equation}
 y = E_\nu~\mbox{{[GeV]}}-2\,,
\hspace*{10ex}
\eta = \sqrt{E_\nu~\mbox{{[GeV]}}/2}-1\,,
\label{eq:defy}
\end{equation}
which vanish at $E_\nu=2$~GeV.
For $\nu_\mu$ $(\alpha=\mu)$,
the four variances are
\begin{subequations} %sig_res_m4
\begin{eqnarray}
 \tilde{\sigma}^\mu_1 &=& 
110 + 12y + y^2
\,,\\
 \tilde{\sigma}^\mu_2 &=& 
160 + 36y + 2.4y^2
\,,\\
 \tilde{\sigma}^\mu_3 &=& 
320-1240y+5640\eta+328y\eta
\,,\\
\tilde{\sigma}^\mu_4 &=& 
177+2580y-9190\eta-747y\eta
\,,
\end{eqnarray} 
\label{eq:sig_res_m4} 
\end{subequations}
$\!\!\!\!$
and the energy shift terms are
\begin{subequations} %dE_res_m4
\begin{eqnarray}
\delta\tilde{E}^\mu_1 &=&
350 - 1.2y + 1.4y^2
\,,\\
\delta\tilde{E}^\mu_2 &=&
530 + 20y - 4.9y^2 + 0.58y^3
\,,\\
\delta\tilde{E}^\mu_3 &=&
823
-\left(3550y - 14100\eta - 1840y\eta\right)
/\left(1+0.921y\right)
\,,\\
\delta\tilde{E}^\mu_4 &=&
 1500+990y-1500\eta-760y\eta
\,,
\end{eqnarray}
\label{eq:dE_res_m4} 
\end{subequations}
$\!\!\!\!$
both in MeV units. 
The normalization factors are $\tilde{r}_1^\mu=1$ and
\begin{subequations} %r_res_m4
\begin{eqnarray}
 \tilde{r}^\mu_2 &=&
0.31-0.18y + 0.22y^2 - 0.096y^3 + 0.015y^4
\,,\\
 \tilde{r}^\mu_3 &=&
0.194+0.316y-1.37\eta-0.122y\eta
\,,\\
 \tilde{r}^\mu_4 &=&
0.0369
-0.0728y
+0.254\eta
+0.0356y\eta\,.
\end{eqnarray} 
\label{eq:r_res_m4} 
\end{subequations}

%---
\vspace*{-3ex}
The first Gaussian is mainly related to the $\Delta$-resonance; 
$\tilde{\sigma}_1^{\mu}\simeq 100$~MeV and 
$\delta\tilde{E}_1^\mu \simeq 350$~MeV
at all energies,
similarly to $\sigma_1^\mu$ and $\delta E_1^\mu$ in
eqs.~(\ref{eq:sig_res_m3}) and (\ref{eq:dE_res_m3}).
%----
Because the number of the resonance modes 
which contribute to the second Gaussian
increases with $E_\nu$,
the second variance grows from
$\tilde{\sigma}_2^{\mu}\simeq 160$~MeV at $E_\nu=2$~GeV
to $240$~MeV at $E_\nu=4$~GeV.
$N(1440)$ dominates the second Gaussian,
and $\delta \tilde{E}_2^\mu$ does not grow much
from $530$MeV at $E_\nu=2$~GeV.
%------
The resonances of mass greater than $2$~GeV
contribute to the third Gaussians;
$\tilde{\sigma}_3^\mu\simeq 320$~MeV and 
$\delta\tilde{E}_3^\mu\simeq 830$~MeV
at $E_\nu=2$~GeV,
which grow to 
$450$~MeV and $900$~MeV, respectively, at $E_\nu=4$~GeV.
%----
The last Gaussian is necessary to reproduce the tail at
low energies.

%-------------------------------------------------------------------
For the $\nu_e$ events at $E_\nu>1.2$~GeV, we find
\begin{subequations} %sig_res_e4
\begin{eqnarray}
 \tilde{\sigma}^e_1(y) &=& 
110 + 14y - 0.50y^2
\,,\\
 \tilde{\sigma}^e_2(y) &=& 
181+2150y+207y^2-8440\eta-1890y\eta
\,,\\
 \tilde{\sigma}^e_3(y) &=& 
 334-1750y+7620\eta+556y\eta
\,,\\
 \tilde{\sigma}^e_4(y) &=& 
222 + 10300y  - 4030\eta-3570y\eta\,,
\end{eqnarray} 
\label{eq:sig_res_e4} 
\end{subequations}
$\!\!\!\!$
and
\begin{subequations} %dE_res_e4
\begin{eqnarray}
\delta\tilde{E}^e_1(y) &=&
360 - y - 1.6y^2  + 0.95y^3  - 0.026y^4
\,,\\
\delta\tilde{E}^e_2(y) &=&
510-61.1y+261\eta\,,\\
\delta\tilde{E}^e_3(y) &=&
796
-\left(
3260y + 115y^2 - 12900\eta - 2070y\eta
\right)/(1+0.94y)
\,,\\
\delta\tilde{E}^e_4(y) &=&
1500-690y+4600\eta\,,
\end{eqnarray}
\label{eq:dE_res_e4} 
\end{subequations}
$\!\!\!\!$
%%%---
both in MeV units. The normalization factors are $\tilde{r}^e_1=1$ and
\begin{subequations} %r_res_e4
\begin{eqnarray}
 \tilde{r}^e_2 &=&
0.28- 0.092y+ 0.23y^2- 0.11y^3+ 0.017y^4
\,,\\
\tilde{r}^e_3 &=&
0.196-0.487y+2.02\eta+0.148y\eta\,,\\
\tilde{r}^e_4 &=&
0.0298 + 0.05\eta-0.213\eta^2+0.23\eta^3\,.
\end{eqnarray} 
\label{eq:r_res_e4} 
\end{subequations}
$\!\!\!\!$
There is no big difference between
$\nu_\mu$ and $\nu_e$,
because these Gaussians account for the same resonance modes.

% ======================================================================== %

\end{document}